\documentclass[%
 reprint,
superscriptaddress,
 amsmath,amssymb,
 aps,
]{revtex4-1}

\usepackage[T1]{fontenc}
\usepackage[utf8]{inputenc}
\usepackage{graphicx}
\usepackage{dcolumn}
\usepackage{bm}
\usepackage{braket}
\usepackage{float}
\usepackage{multirow}
\usepackage{booktabs}
\usepackage{mathrsfs}
\usepackage{hyperref}

\begin{document}

\title{Impact of gate-voltage noise on silicon spin-qubit variational quantum eigensolvers}

\author{Xinning Wang}
\email{x2337wan@uwaterloo.ca}
\affiliation{Institute for Quantum Computing, University of Waterloo, Waterloo, Ontario N2L 3G1, Canada}
\affiliation{Department of Chemistry, University of Waterloo, Waterloo, Ontario N2L 3G1, Canada}

\author{Bohdan Khromets}
\affiliation{Institute for Quantum Computing, University of Waterloo, Waterloo, Ontario N2L 3G1, Canada}
\affiliation{Department of Physics and Astronomy, University of Waterloo, Waterloo, Ontario N2L 3G1, Canada}

\author{Zach Merino}
\affiliation{Institute for Quantum Computing, University of Waterloo, Waterloo, Ontario N2L 3G1, Canada}
\affiliation{Department of Physics and Astronomy, University of Waterloo, Waterloo, Ontario N2L 3G1, Canada}

\author{Jonathan Baugh}
\email{baugh@uwaterloo.ca}
\affiliation{Institute for Quantum Computing, University of Waterloo, Waterloo, Ontario N2L 3G1, Canada}
\affiliation{Department of Chemistry, University of Waterloo, Waterloo, Ontario N2L 3G1, Canada}

\date{\today}

\begin{abstract}
Quantum computers offer a route to outperform classical methods in tasks such as molecular simulation, motivating hybrid algorithms like the Variational Quantum Eigensolver (VQE) for near-term devices. Silicon spin qubits are a promising platform for scalable quantum computation, but their performance is limited by hardware imperfections---most notably charge-noise--induced potential fluctuations and static miscalibration of gate-electrode voltages---which degrade quantum gate fidelities and, ultimately, algorithmic accuracy. Here we develop a hardware--algorithm co-simulation framework for silicon quantum-dot processors that links 3D electrostatics to effective $g$-factors and exchange couplings, and propagates voltage-level noise through realistic control pulses. Using VQE for $\mathrm{H}_2$ ground-state energy estimation as a circuit-level testbed, we study both static scaling/offset errors on the gate-electrode voltages and stochastic fluctuations modeled as random-telegraph noise with tunable amplitudes and switching times. At the gate level, we show that exchange-based two-qubit gates are roughly an order of magnitude more sensitive to these types of noise than ESR-driven single-qubit rotations. Quantum process tomography and Kraus-operator analysis further distinguish coherent and incoherent contributions and quantify the fraction of error that is, in principle, correctable by a compensating unitary. Embedding these noise models into the VQE circuit, we identify regimes of miscalibration strength and noise switching time compatible with chemically accurate energy estimates, and discuss how statistical post-processing based on the full distribution of noisy energy estimates could further improve accuracy.

\begin{description}
    \item[Keywords] Variational Quantum Eigensolver, Spin Qubits, Silicon Quantum Dots, Charge Noise, Quantum Control, Quantum Process Tomography, Kraus Operators, Quantum Simulation
\end{description}
\end{abstract}

\maketitle

\section{\label{sec:level1} Introduction}

Since the fundamental ideas of quantum computation were first articulated in the 1980s and 1990s~\cite{steane1998quantum,feynman1982simulating}, the field has evolved into the present noisy intermediate-scale quantum (NISQ) era, in which medium-scale but non--fault-tolerant devices can already tackle classically challenging tasks~\cite{PreskillReview, Mi2024PhaseTransitionsRCS}. Contemporary quantum processors based on superconducting circuits, trapped ions, and neutral atoms~\cite{Ladd2010} now coherently manipulate and entangle more than one hundred physical qubits, and first demonstrations of error-corrected logical qubits with substantially enhanced coherence times have been reported~\cite{PreskillReview,GoogleWillow,IBMHeron,GoogleQEC,Liu2025CertifiedRandomness,Bluvstein2024LogicalProcessor,Manetsch2025TweezerArray}. 

In the longer term, the goal is to move beyond NISQ technology to fault-tolerant application-scale quantum (FASQ) machines capable of running deep quantum algorithms for tasks such as integer factorization and combinatorial optimization that offer compelling theoretical advantages over classical methods~\cite{harrow2009quantum,shor1994algorithms,cerezo2021variational}. Bridging the gap from today's NISQ devices to future FASQ architectures remains the central challenge, as hardware noise, limited coherence, and imperfect control still severely constrain the realizable quantum circuit depth and, hence, practical performance.

\begin{figure*}[htbp]
    \centering
    \includegraphics[width=0.75\linewidth]{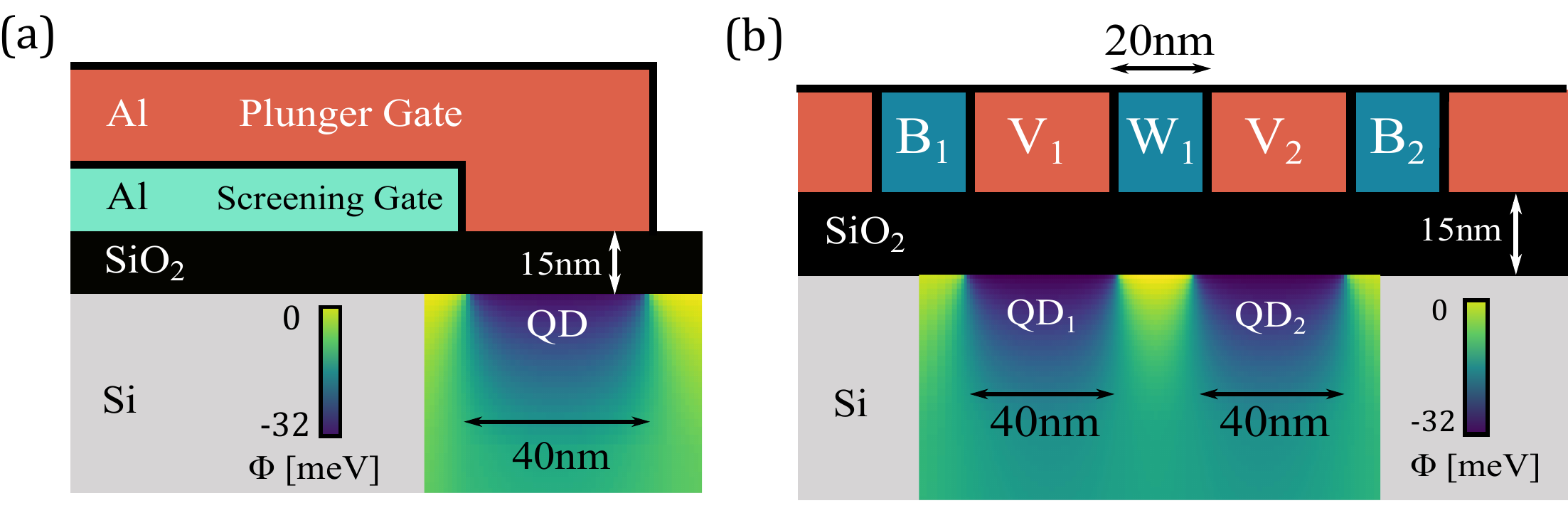}
    \caption{Schematic of a double-QD device used for the one and two-qubit logic gate simulations: (a) side view, (b) front view.
    Electrons, accumulated at the Si/SiO${}_2$ interface, are laterally confined in QDs under the positive plunger gate voltages $V_1$ and $V_2$. The confining potential is further controlled by a grounded screening gate and the tunnel gate voltage $W_1$. Barrier gates $B_1, B_2$ are kept at 0 V throughout the simulations. The electrostatic potential $\Phi(\vec{r})$ shown here was calculated for $V_1=V_2=0.2\,\text{V}$, $W_1=-0.1\,\text{V}$. }
 
    \label{fig:device_single}
\end{figure*}

Within the NISQ regime, variational quantum algorithms (VQAs) such as the Variational Quantum Eigensolver (VQE)~\cite{Peruzzo2014} are leading candidates for problem-specific near-term quantum advantage. They are particularly attractive for quantum chemistry~\cite{cao2019quantum} and materials science~\cite{Mustafa2022}, where shallow, parameterized circuits optimized in a hybrid quantum--classical loop can exploit modest circuit depths and exhibit some resilience to noise~\cite{fedorov2022vqe}. At the same time, their performance is tightly constrained by the detailed noise characteristics, connectivity, and calibration of the underlying hardware~\cite{bharti2021noisy,PreskillReview}, so that realizing robust, reproducible advantage with VQAs remains a key open challenge on the path toward fully fault-tolerant FASQ machines.\\
\indent Silicon-based spin qubits~\cite{maurand2016cmos} have emerged as a compelling platform for scalable processors, combining long coherence times~\cite{veldhorst2015two,stano2022review} with compatibility with industrial semiconductor fabrication and high integration density~\cite{zwerver2022qubits}. Recent experiments with silicon quantum dots have demonstrated high-fidelity multi-qubit operations~\cite{philips2022universal,tanttu2024errors,mills2022twoqubit} and proof-of-principle quantum algorithms~\cite{watson2018programmable,thorvaldson2025grover}, underscoring their promise for larger-scale architectures. At the same time, charge noise remains a central obstacle, limiting gate fidelities and compromising device stability~\cite{paqueletwuetz2023reducing}. It arises from microscopic defects (interface traps, lattice damage, bulk impurities) and shows strongly device-dependent spectra~\cite{culcer2009dephasing,connors2022charge}, so its characterization and mitigation are key prerequisites for advancing silicon spin qubits toward FASQ-capable systems.\\
\indent Despite advances in generic error-mitigation techniques, the interplay between device-specific noise processes and algorithmic performance remains insufficiently quantified, particularly for silicon spin qubits. Most prior analyses of VQE and its mitigation strategies rely on abstract, platform-agnostic noise models---such as simple depolarizing channels or zero-noise extrapolation~\cite{kim2023scalable, Fontana2021VQENoiseResilience}---which obscure the impact of concrete control imperfections like gate-voltage miscalibration and charge-noise-induced potential fluctuations. Related modeling efforts have addressed the connection between semiconductor spin-qubit hardware and effective Hamiltonians, as well as the impact of charge noise on gate-level performance~\cite{PhysRevB.67.121301,PhysRevB.108.045305}, but generally do so without explicitly resolving the structure of the underlying noise process and stop short of assessing their consequences at the algorithmic level.\\
\indent Here we develop a hardware--algorithm co-analysis framework tailored to silicon quantum-dot processors. We consider a silicon metal-oxide-semiconductor (MOS) architecture with a global microwave magnetic field for electron spin resonance and single-qubit addressability via Stark-shifted electronic $g$-factors~\cite{Veldhorst2014AddressableQubit}. Within this setting, we explicitly model two dominant classes of voltage-level imperfections on the gate electrodes: (i) static miscalibration of gate voltages and (ii) dynamically fluctuating charge noise, emulated as independent random telegraph noise (RTN) processes on each gate voltage with tunable amplitudes and switching times. By varying these RTN time scales and amplitudes, we capture how different effective noise spectra map onto logical error channels.\\
\indent In the remainder of this work, we first assess the impact of these voltage-level imperfections at the gate level in a simulated two-qubit device. For both single- and two-qubit operations, we compute average gate fidelities and perform full quantum process tomography (QPT)~\cite{o2004quantum,mohseni2008quantum} to reconstruct the corresponding noise channels; a Kraus-operator analysis then separates coherent from incoherent contributions and quantifies the fraction of error that is, in principle, correctable via targeted calibration and control refinements. We next embed the same noise models into a four-qubit simulation of VQE-based ground-state energy estimation for the hydrogen molecule, sampling over distributions of gate-voltage scaling and offset parameters as well as RTN amplitudes and switching times to identify regimes in which chemical accuracy can still be achieved. Throughout, we focus on voltage- and charge-noise--equivalent mechanisms and neglect, for example, hyperfine and spin--orbit--mediated relaxation; within this scope, the combined gate- and algorithm-level analysis yields qualitative and semi-quantitative design rules that link noise strength and temporal correlations to VQE performance, providing guidance for calibration targets and noise engineering as silicon spin-qubit platforms mature.

\section{Methods}
\label{sec:methods} 
\begin{figure*}[th]
    \centering

    \begin{minipage}[t]{0.32\textwidth}
        \centering
        \makebox[0pt][l]{\raisebox{21.5\height}{\textbf{(a)}}}%
        \includegraphics[width=\linewidth]{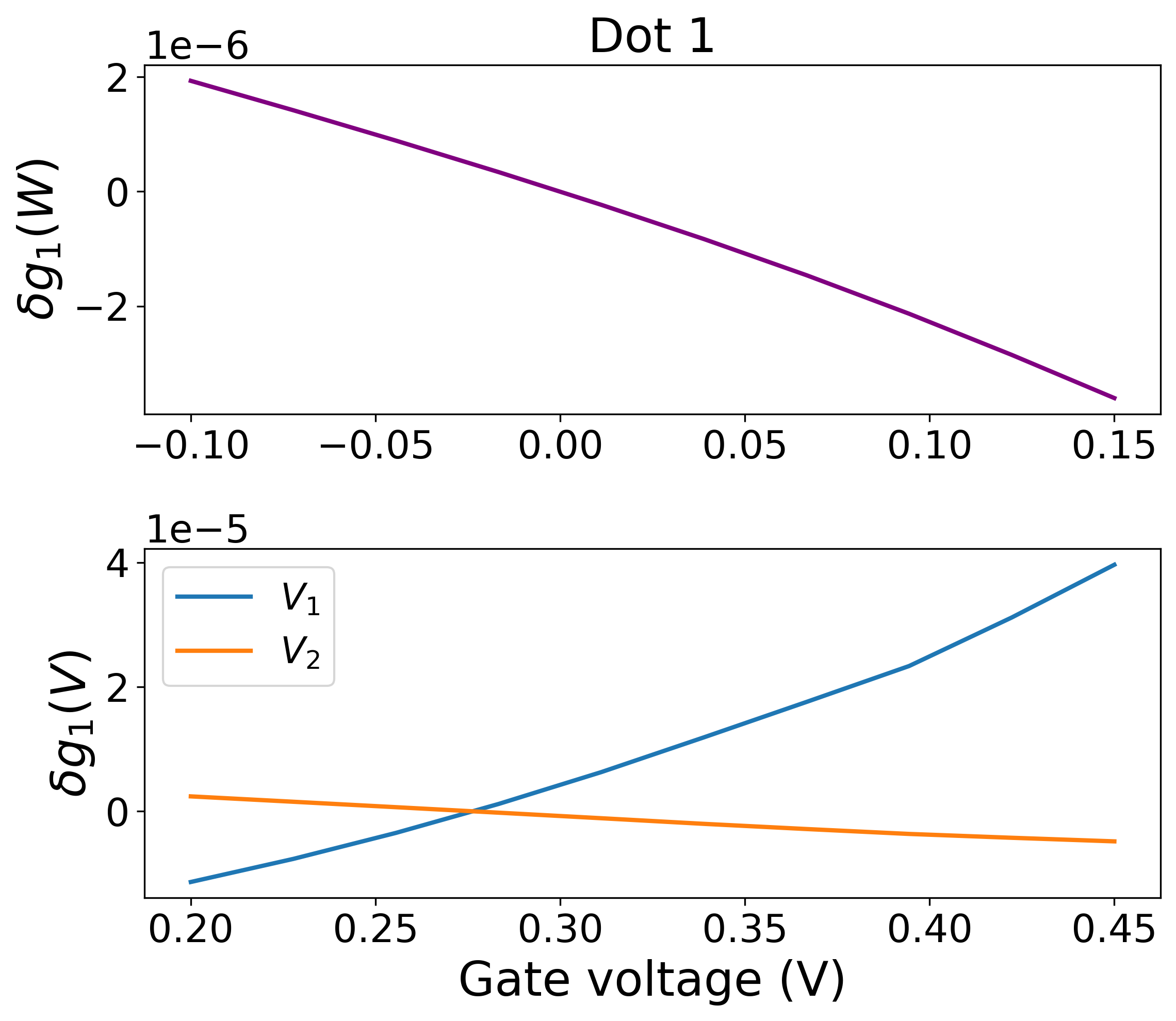}
    \end{minipage}\hfill
    \begin{minipage}[t]{0.32\textwidth}
        \centering
        \includegraphics[width=\linewidth]{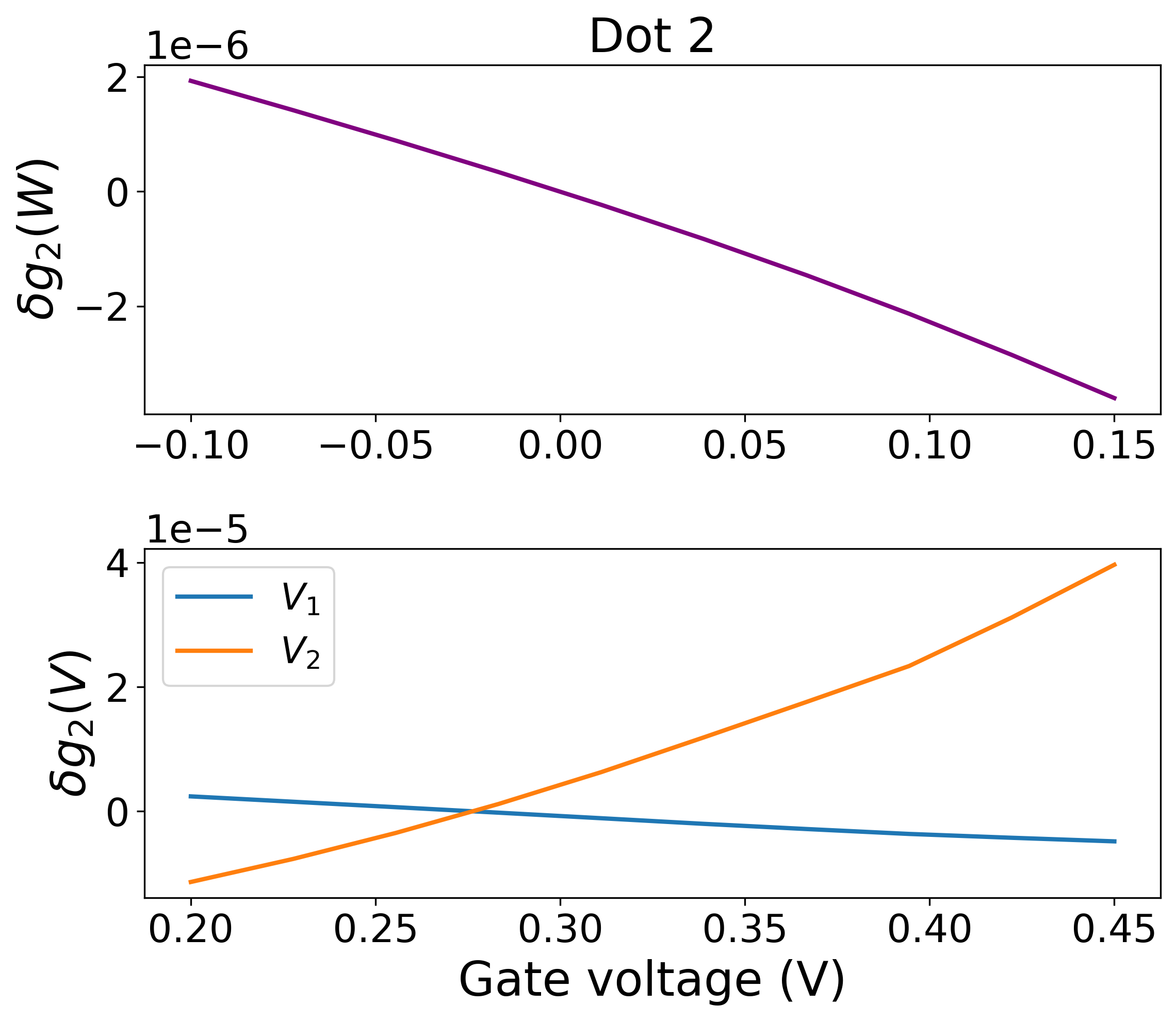}
    \end{minipage}\hfill
    \begin{minipage}[t]{0.32\textwidth}
        \centering
        \makebox[0pt][l]{\raisebox{21.5\height}{\textbf{(b)}}}%
        \includegraphics[width=\linewidth]{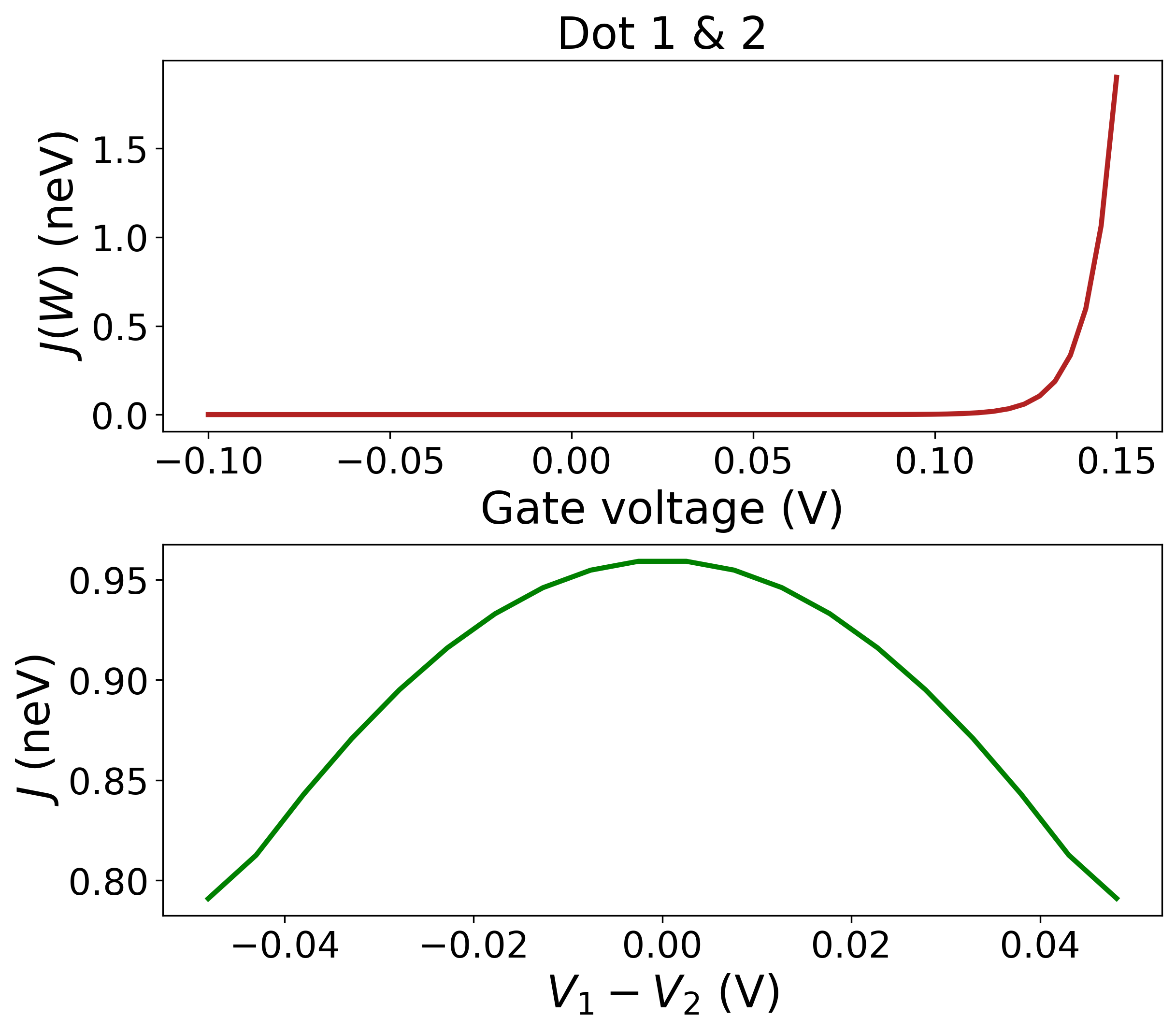}
    \end{minipage}

    \caption{
    Voltage dependence of effective spin Hamiltonian parameters in a double-QD device. (a) Gate-induced variations of the electron $g$-factors for dot~1 and dot~2 as functions of the plunger gate voltages $V_1$ and $V_2$, and the tunnel gate voltage $W_1$.
    All quantities in panel~(a) are shown relative to a reference (idling) voltage point
    $(V_{1}, V_{2}, W_{1}) = (0.2765~\mathrm{V},\, 0.2765~\mathrm{V},\, 0~\mathrm{V})$.
    (b) Exchange coupling $J$ as a function of the tunnel gate voltage $W_1$ and the inter-dot detuning $V_1 - V_2$.
    In panel~(b), the exchange coupling is evaluated relative to a reference voltage point
    $(V_1, V_2, W_1) = (0.2765~\mathrm{V},\, 0.2765~\mathrm{V},\, 0.1~\mathrm{V})$,
    highlighting the exponential dependence of the inter-dot exchange on tunnel gate voltage.  
    }
    
    \label{fig:dg_J_two_column}
\end{figure*}

\subsection{Spin qubit simulator}
Throughout this work we consider a specific class of silicon spin-qubit devices: silicon MOS quantum dots without micromagnets, in which all spins experience a global ESR field and individual addressability is provided by small Stark-shift-induced changes in the in-plane $g$-factor. Within this architecture, we model spin dynamics in realistic two- and four-dot devices using a full-stack simulator that links gate-electrode voltages to an effective spin Hamiltonian.\\
\indent Three-dimensional electrostatic simulations are performed with a finite-difference Poisson solver~\cite{Birner2007_nextnano}, using the realistic gate geometry and applied electrode voltages as electrostatic boundary conditions. The resulting potentials are sampled on a two-dimensional slice near the Si/SiO$_2$ interface, where the electron wave function is localized (see Fig.~\ref{fig:device_single}). Because electrons in Si MOS quantum dots are strongly confined in the vertical direction, with vertical confinement lengths much smaller than the lateral dot dimensions and interdot spacing, we use an effective-potential approximation: a horizontal slice of the 3D electrostatic potential,
\[
\varphi(x,y)=\Phi(x,y,z_0),
\]
taken at the expected position $z_0$ of maximum vertical envelope density, is used as the confinement potential for a two-dimensional Schr\"odinger equation. For a grid of gate-voltage configurations, these effective 2D potentials are interpolated to obtain a smooth dependence $\varphi(\vec{V})$ on the control-voltage vector $\vec{V}$, and are used as inputs to a 2D single-particle Schr\"odinger solver to compute ground-state wavefunctions for individual dots and dot pairs. This should therefore be regarded as an effective-potential approach rather than a fully self-consistent Poisson--Schr\"odinger simulation with charge-density feedback into the electrostatic potential.\\
\indent From these wavefunctions we extract effective parameters: local $g$-factors via Stark-shift models and interdot exchange couplings via Heitler--London / Hund--Mulliken approximations~\cite{Burkard_1999,Pedersen_2007}. Figure~\ref{fig:dg_J_two_column} shows the resulting dependencies of $\delta g$ and $J$ on plunger and tunnel voltages: in the operating regime, $\delta g$ is approximately linear in the voltages, whereas $J(V)$ is steep and close to exponential. These voltage-dependent parameters define an effective spin Hamiltonian with tunable Zeeman and exchange terms; its explicit form is given in Appendix~\ref{appendix:spin_hamiltonian}.\\
\indent Control inputs, including time-dependent global ESR fields and local gate voltages, are mapped to the corresponding Hamiltonian parameters, and the spin dynamics are simulated via Lindblad master-equation evolution with optional relaxation and dephasing channels. A custom pulse-design routine constructs gate-voltage and ESR envelopes for single- and two-qubit gates by enforcing a time-ordered Hamiltonian of the form $H(t) = H_0 S(t)/T$~\cite{10.1007/978-3-031-84869-8_13}, where $H_0$ generates the target gate, $S(t)$ is a smooth shape function, and $T$ is the gate duration. The reverse mapping from desired $g_i(t)$ and $J_{i}(t)$ to $\vec{V}(t)$ solves the coupled voltage equations using the full device model, so cross-talk between electrodes is automatically included in the pulse design. The limited $g$-factor tunability in Si MOS ($\sim 10^{-4}$~\cite{Veldhorst2014AddressableQubit}) sets single-qubit gate times in the microsecond regime; in this work we choose a common duration $t_g = 10~\mu\mathrm{s}$ for all gates, including exchange and CZ gates, and treat this noiseless control as the baseline onto which voltage miscalibration and charge-noise--equivalent fluctuations are later superimposed.

\subsection{\label{sec:citeref}Emulated charge noise}

\begin{figure}[htbp]
    \centering
    \includegraphics[width=0.45\textwidth]{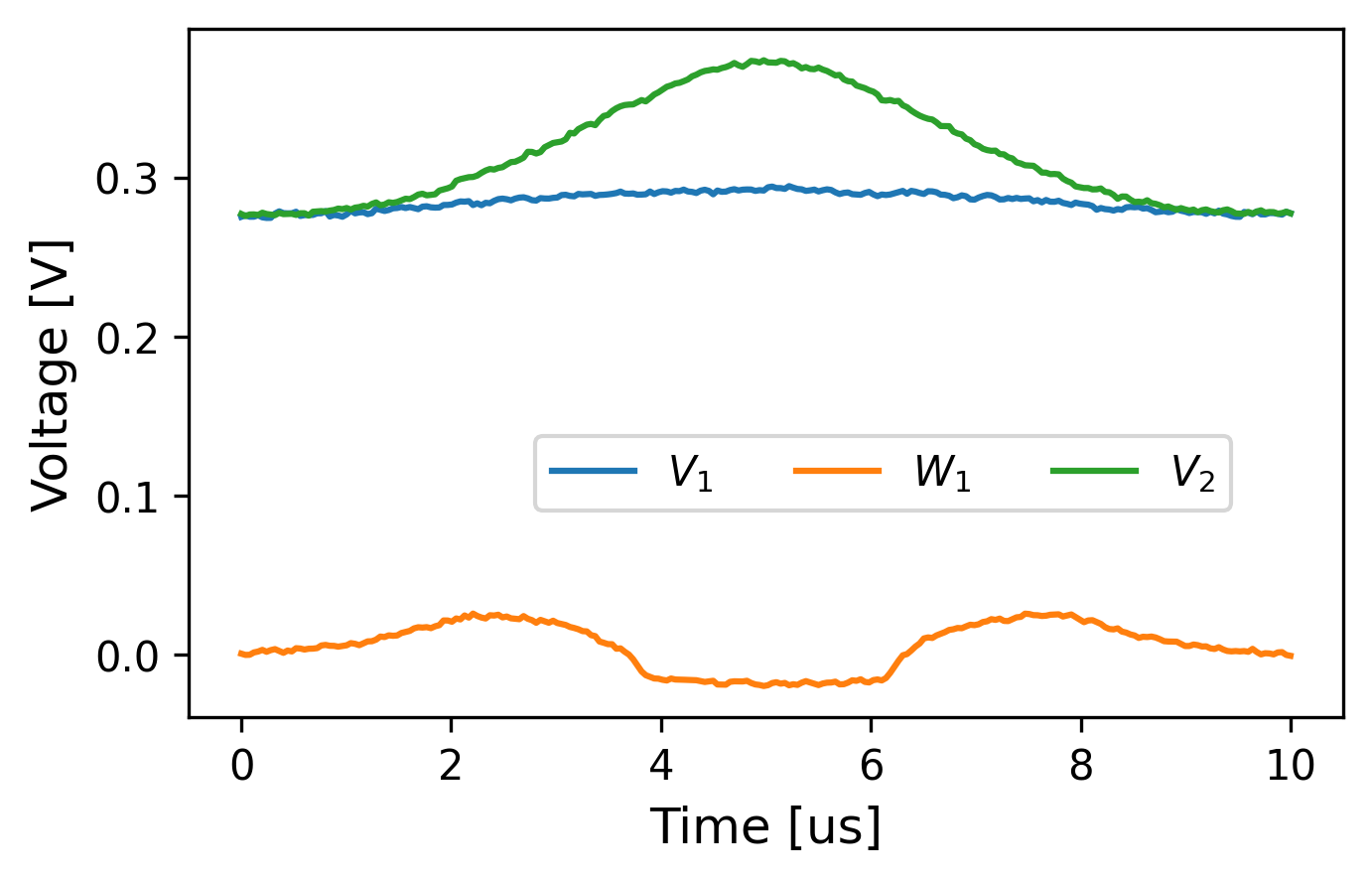}
    \caption{Representative noisy voltage control pulse used to implement an $R_X(\pi/2)$ gate on qubit~1 in the two-qubit simulation, while qubit~2 undergoes a simultaneous identity operation. The controls $V_1$ and $V_2$ denote plunger-gate voltages and $W_1$ denotes the interdot barrier-gate voltage. The ideal waveform realizes the target gate operation, whereas the superimposed fluctuations correspond to sampled random telegraph voltage noise. These noisy control voltages are mapped directly onto the time-dependent Hamiltonian through the voltage-dependent parameters $g_i(\vec{V})$ and $J_i(\vec{V})$.
}
    \label{fig:noisy_voltage_pulse}
\end{figure}

\begin{figure}[htbp]
    \centering
    \begin{minipage}{0.45\textwidth}
        \centering
        \makebox[0pt][l]{\raisebox{47.0\height}{\textbf{(a)}}}%
        \includegraphics[width=\linewidth]{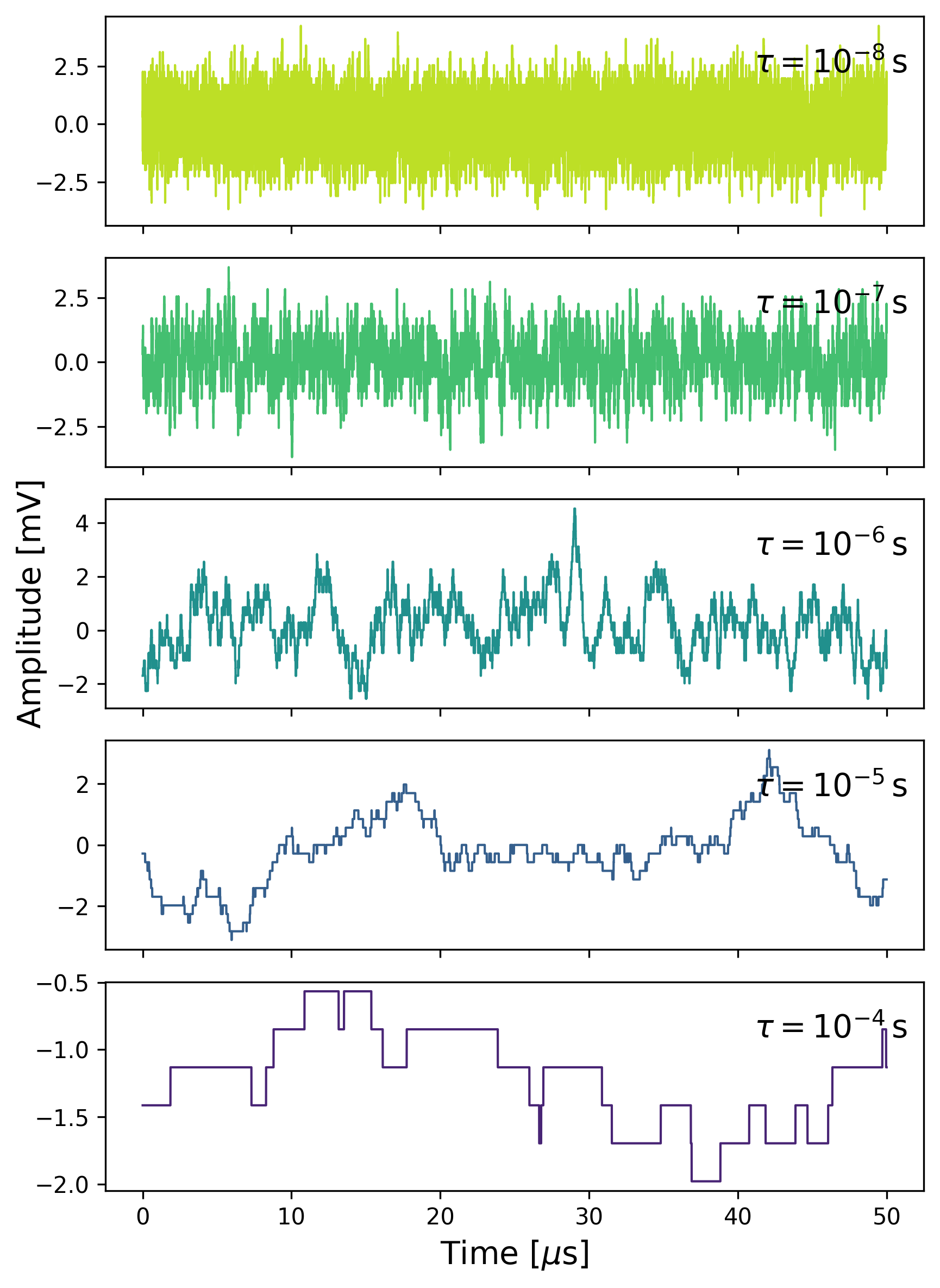}
    \end{minipage}
    \hfill
    \begin{minipage}{0.45\textwidth}
        \centering
        \makebox[0pt][l]{\raisebox{24.0\height}{\textbf{(b)}}}%
        \includegraphics[width=\linewidth]{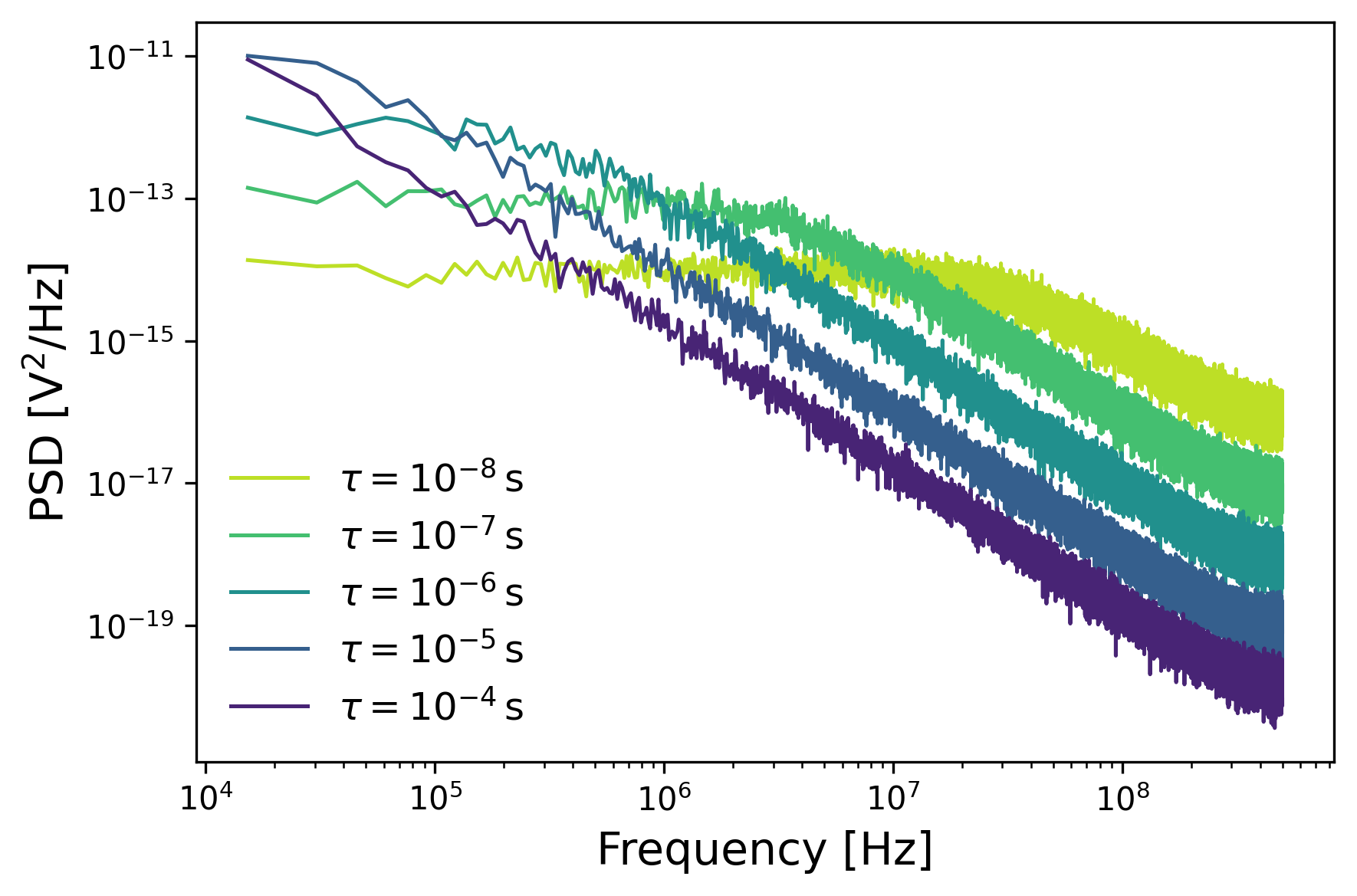}
    \end{minipage}

    \caption{(a) Simulated voltage-noise time series and (b) corresponding power spectral densities (PSDs) for five RTN switching times, $\tau \in \{10^{-8}, 10^{-7}, 10^{-6}, 10^{-5}, 10^{-4}\}\,\mathrm{s}$, at fixed RMS amplitude $A = 1~\mathrm{mV}$. For each value of $\tau$ (with $\tau = \langle \tau_0\rangle = \langle \tau_1\rangle$), the charge-noise signal on a given gate is constructed as a normalized sum of $N_\textsc{rtn}=50$ independent RTN realizations [Eq.~\eqref{eq:rtn_sum}]. The time series extend beyond the single-gate duration $t_g = 10\,\mu\mathrm{s}$ and span the timescale of a simulated variational single-shot experiment, $t_{\mathrm{exp}} = 130\,\mu\mathrm{s}$; only the first $50\,\mu\mathrm{s}$ are shown for clarity.}
    \label{fig:rtn_psd}
\end{figure}

Precise manipulation of electron spins in quantum dots relies on tight control of local electrostatic potentials, which is essential for realizing high-fidelity single- and two-qubit gates. These potentials, tuned by gate voltages, are susceptible to fluctuations from environmental sources such as charged defects, dielectric interface traps, and noise in the control electronics. Building on the simulator framework described above, we incorporate charge-noise effects into the model to investigate this dominant source of control error in spin-qubit systems.

In our simulations we do not resolve all microscopic spin--orbit--mediated channels individually. Instead, we use the voltage dependence of the effective parameters obtained from the electrostatic and single-particle simulations: the local $g$-factors $g_i(\vec{V})$ and exchange couplings $J_{i}(\vec{V})$. Charge noise is emulated as stochastic fluctuations of the gate voltages,
\[
    \vec{V}(t) = \vec{V}_0 + \delta \vec{V}(t),
\]
so that the Stark-shift derivatives $\partial g_i / \partial V_k$ and $\partial J_{i} / \partial V_k$ map gate-voltage noise into time-dependent $g_i(t)$ and $J_{i}(t)$. This captures the dominant longitudinal effect of low-frequency charge noise---fluctuations in qubit frequencies and exchange splittings in silicon spin qubits~\cite{culcer2009dephasing,connors2022charge}---while neglecting additional transverse and phonon-mediated mechanisms expected to play a subleading role on the $\sim 10~\mu\mathrm{s}$ gate timescales considered here~\cite{Burkard2023,yoneda2018coherence}.

Experimentally, the power spectral density (PSD) of charge noise in silicon quantum dots often exhibits a characteristic \(1/f^\alpha\) dependence over a broad frequency range---from mHz up to MHz and beyond~\cite{Kepa2023APL}. This behavior is consistent with a large ensemble of two-level fluctuators (TLFs), each representing a localized charge trap that randomly switches between two metastable charge configurations. When the TLF switching rates are distributed approximately log-uniformly, their combined spectrum approaches \(1/f^\alpha\) with \(\alpha \approx 1\).

To connect such measurements to voltage-noise amplitudes in our model, we extract an effective root-mean-square (RMS) gate-voltage noise from representative $1/f$ spectra. We consider a charge-noise energy spectrum of the form $S_{\varepsilon}(f) = D^2 / f$, where $D = 1.70~\mu\mathrm{eV}$ denotes the mean charge-noise amplitude at $1~\mathrm{Hz}$ extracted from the range of silicon quantum-dot devices reported in Ref.~\cite{Freeman2016ChargeNoise}. Integrating this spectrum over an experimentally relevant bandwidth $f_{\min} = 10^4~\mathrm{Hz}$ to $f_{\max} = 10^8~\mathrm{Hz}$ yields an RMS energy fluctuation
\begin{equation}
    \sigma_{\varepsilon}
    =
    \left[
    \int_{f_{\min}}^{f_{\max}} \frac{D^2}{f}\,df
    \right]^{1/2}
    = 5.16~\mu\mathrm{eV}.
\end{equation}
Using the plunger-gate lever arm $\alpha = \partial \mu / \partial V = 0.218~\mathrm{eV/V}$, obtained from our simulator, this energy fluctuation corresponds to an equivalent gate-voltage RMS noise
\begin{equation}
    \sigma_V = \sigma_{\varepsilon} / \alpha \approx 2.5\times10^{-5}~\mathrm{V}.
\end{equation}
This value serves as a physically motivated reference point for experimentally relevant gate-voltage noise levels; in the simulations below we explore a broader range of amplitudes to systematically assess gate performance across different noise regimes.

The electrostatic contribution from an individual TLF is modeled as a RTN process, i.e., a two-level signal that switches between $\pm A$ according to a Poisson process. For a stationary RTN process with zero mean, RMS amplitude $A$, and mean dwell time $\tau = \langle \tau_0 \rangle = \langle \tau_1 \rangle$ in each state, the corresponding PSD is the Lorentzian
\begin{equation}
    S_{\textsc{rtn}}(f)
    = \frac{A^2 \tau}{1 + (\pi f \tau)^2},
    \label{eq:single_rtn_psd}
\end{equation}
with a characteristic cutoff frequency $f_c = (2\pi\tau)^{-1}$~\cite{Gauthier2022RTN}. For $f \ll f_c$ the PSD saturates at a constant value (quasi-static fluctuations), while for $f \gg f_c$ it decays as $1/f^2$.

In practice we emulate charge noise on each gate electrode by superimposing a finite ensemble of independent RTN signals. For each electrode we generate $N_{\textsc{rtn}}$ statistically independent RTN sequences $\mathrm{rtn}_i(t)$ with identical amplitude and mean switching time $\tau$, and form their normalized sum
\begin{equation}
    \mathrm{RTN}_{\mathrm{avg}}(t)
    = \frac{1}{\sqrt{N_{\textsc{rtn}}}}
      \sum_{i=1}^{N_{\textsc{rtn}}} \mathrm{rtn}_i(t),
    \label{eq:rtn_sum}
\end{equation}
so that the composite noise has zero mean and RMS amplitude $A$, each plunger and barrier gate acquires its own statistically independent noisy voltage component with controlled amplitude $A$ and switching time $\tau$. Within a given simulation, all gate electrodes share the same RTN parameters $(A,\tau)$, while independent realizations are generated and applied to each gate.

To illustrate how stochastic voltage fluctuations enter the device-level control sequence, Fig.~\ref{fig:noisy_voltage_pulse} shows a representative noisy voltage sequence used for a two-qubit $RX(\pi/2)$ gate. The plotted controls include the two plunger gates $V_1$ and $V_2$, and the interdot barrier gate, $W_1$. The smooth pulse envelope defines the intended gate operation, while the small fluctuations represent the RTN signals added directly to the control waveforms.

Figure~\ref{fig:rtn_psd}(a) shows example composite RTN traces generated from ensembles, while Fig.~\ref{fig:rtn_psd}(b) displays the corresponding PSDs for different values of $\tau$. Varying $\tau$ shifts the spectral knee $f_c$ across frequency bands: short switching times push $f_c$ into the MHz--GHz regime, producing fast fluctuations that are largely averaged out over the $\sim 10~\mu\mathrm{s}$ gate times considered here, whereas longer switching times shift the spectrum toward lower frequencies and yield increasingly slow, quasi-static noise on the scale of a single quantum-circuit execution.

\subsection{Process Characterization: Fidelity and Tomography}
\label{sec:Process-Characterization}
\subsubsection{Process Fidelity}
\label{sec:process_fidelity}

To quantitatively assess the performance of noisy quantum gates, we use the \emph{average gate fidelity}~\cite{Nielsen2002AverageGateFidelity}
\begin{equation}
    F
    = \frac{d \cdot \left| \frac{1}{d^2} \operatorname{Tr}\!\left(S^{\dagger} S'\right) \right| + 1}{d + 1},
    \label{eq:fidelity}
\end{equation}
where $S$ and $S'$ denote the ideal and noisy superoperators in Liouville space, and $d$ is the Hilbert-space dimension of the $n$-qubit system ($d = 2^n$). This figure of merit compactly quantifies how closely the implemented quantum process approximates its target, with $F = 1$ corresponding to a perfectly realized gate. Throughout this paper, we assume perfect spin-state measurements, restricting our attention to imperfect control during gate operations and reserving measurement noise for future work. 

\subsubsection{Quantum process tomography}
\label{sec:coherent_angle}

To resolve not only the fidelity but also the structure of gate errors induced by voltage noise, we use QPT to reconstruct the full quantum channel implemented by each noisy gate. QPT represents the process via a matrix $\chi$ (or, equivalently, a set of Kraus operators) and therefore retains information about both coherent control imperfections and incoherent noise contributions~\cite{o2004quantum,mohseni2008quantum,Nielsen2010}.

In our two-qubit simulations, we work on a Pauli operator basis $\{\tilde{E}_m\}$ constructed from tensor products of single-qubit Pauli operators. We construct a tomographically complete set of 16 linearly independent input operators
\[
    \rho_j = T_a \ket{00}\!\bra{00}\, T_b ,
\]
with $T_a, T_b \in \{I\otimes I,\, I\otimes X,\, X\otimes I,\, X\otimes X\}$, propagate each $\rho_j$ through the noisy gate, and invert the resulting linear system to obtain the process matrix $\chi$ for that noise realization. The reconstruction procedure, together with the conversion between $\chi$ and the Kraus representation, is described in detail in Appendix~\ref{app:qpt_details}.

From $\chi$ we then extract two derived objects. First, we form an error process matrix $\chi^{\mathrm{err}}$ by factoring out the ideal unitary $U$, so that $\chi^{\mathrm{err}}$ describes only the residual noise channel; in this representation, $\chi^{\mathrm{err}}_{00}$ encodes the process fidelity, while the remaining diagonal elements give the dominant Pauli error weights. Second, we convert $\chi$ to the corresponding Pauli transfer matrix and compute the coherence angle $\Theta$ introduced in Ref.~\cite{Iverson2020Coherence}, which quantifies the coherent (unitary-like) component of the noise channel in the small-error regime. The explicit expressions used to construct $\chi^{\mathrm{err}}$, to obtain the Kraus operators, and to evaluate $\Theta$ are collected in Appendix~\ref{app:qpt_details}.

\section{Effects of noise on quantum gate operations}\label{sec:single_gate}

\subsection{Impact of voltage miscalibration on gate fidelity}\label{subsec:miscalibration}
\subsubsection{Numerical simulation trends}

We first consider miscalibration of the control voltages, a key source of systematic error in silicon spin qubits. Miscalibration is modeled as stochastic scaling and offset errors on each control waveform,
\begin{equation}
    V'(t) - V_{\mathrm{idle}} = a \bigl(V(t) - V_{\mathrm{idle}}\bigr) + b,
    \label{eq:miscalibrated_pulse}
\end{equation}
where $V(t)$ is the ideal waveform, $V'(t)$ is the miscalibrated waveform, and $V_{\mathrm{idle}}$ is a fixed idle operating point. The idle point is chosen such that all qubits share the same effective $g$-factor, all pairwise exchange couplings are negligible, and the ESR drive is off $(\Omega = 0)$. The parameter $a$ thus represents a stochastic amplitude rescaling and $b$ a residual DC offset about $V_{\mathrm{idle}}$, with $a \sim \mathcal{N}(1,\sigma_a^2)$ and $b \sim \mathcal{N}(0,\sigma_b^2)$.

Single-qubit operations in our architecture rely on limited Stark-shift tunability of the $g$-factor (on the order of $10^{-4}$), so non-target qubits cannot be fully detuned from the global ESR drive. Instead, our pulse-design routine constructs control waveforms such that the target qubit executes the desired rotation while non-resonant qubits undergo an effective $2\pi$ rotation about their local quantization axes. In the noiseless case, non-resonant qubits return to their initial states despite experiencing the same global drive and only slightly different $g$-factors.

We benchmark three representative operations:
\begin{itemize}
    \item $RX(\pi/2)$: a single-qubit rotation by $\pi/2$ about the $X$ axis on qubit~1, with qubit~2 undergoing a $2\pi$ rotation;
    \item Hadamard: a single-qubit $\pi$ rotation around the $(1,0,1)$ axis on qubit~2, with qubit~1 undergoing a $2\pi$ rotation;
    \item $\sqrt{\mathrm{SWAP}}$: an exchange-based two-qubit partial-SWAP gate.
\end{itemize}

\begin{figure*}[htbp]
    \centering
    \begin{minipage}[b]{0.32\textwidth}
        \makebox[0pt][l]{\raisebox{16.0\height}{\textbf{(a)}}}%
        \includegraphics[width=\textwidth]{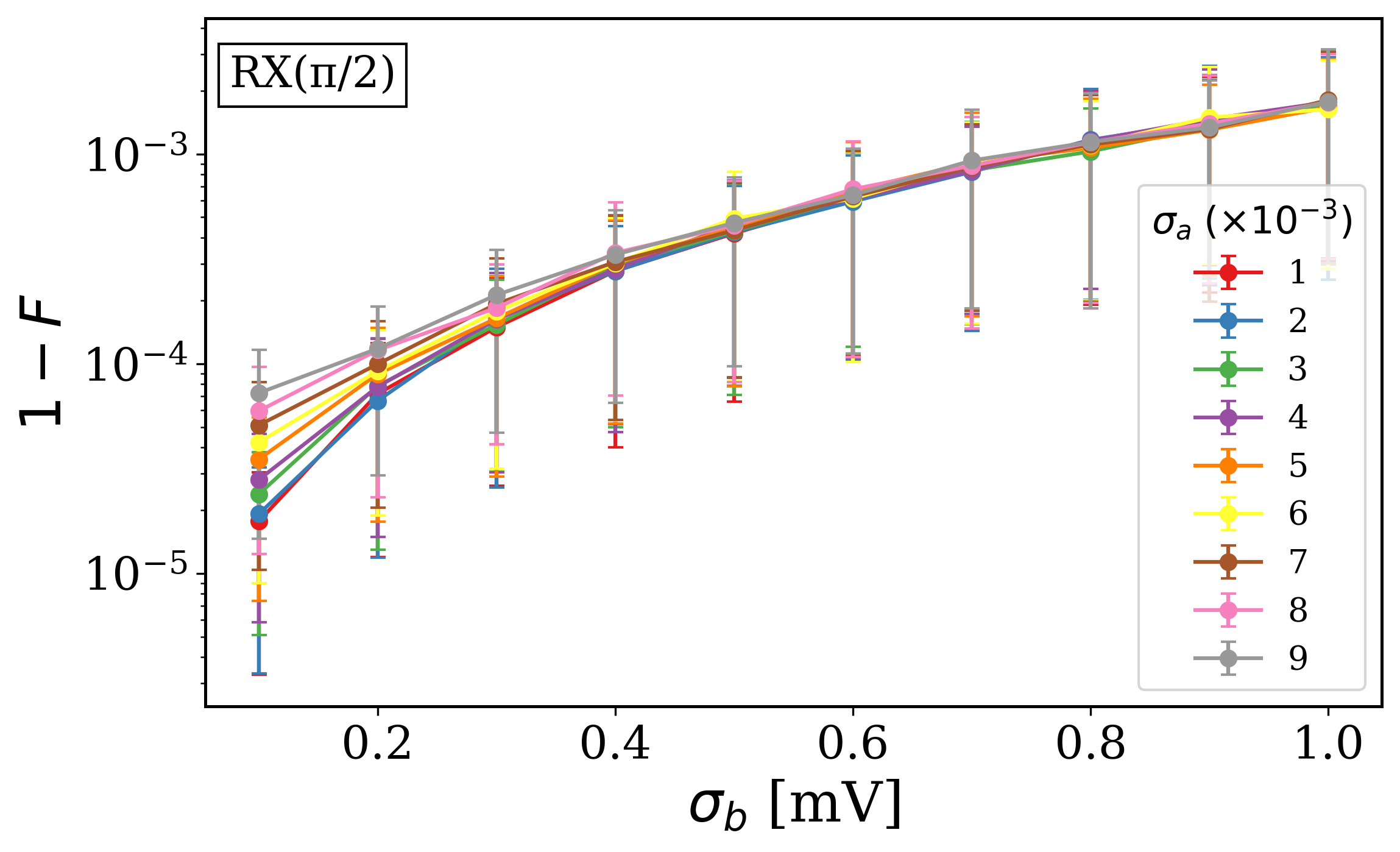}
        \label{fig:combined_rx}
    \end{minipage}
    \begin{minipage}[b]{0.32\textwidth}
        \makebox[0pt][l]{\raisebox{16.0\height}{\textbf{(b)}}}%
        \includegraphics[width=\textwidth]{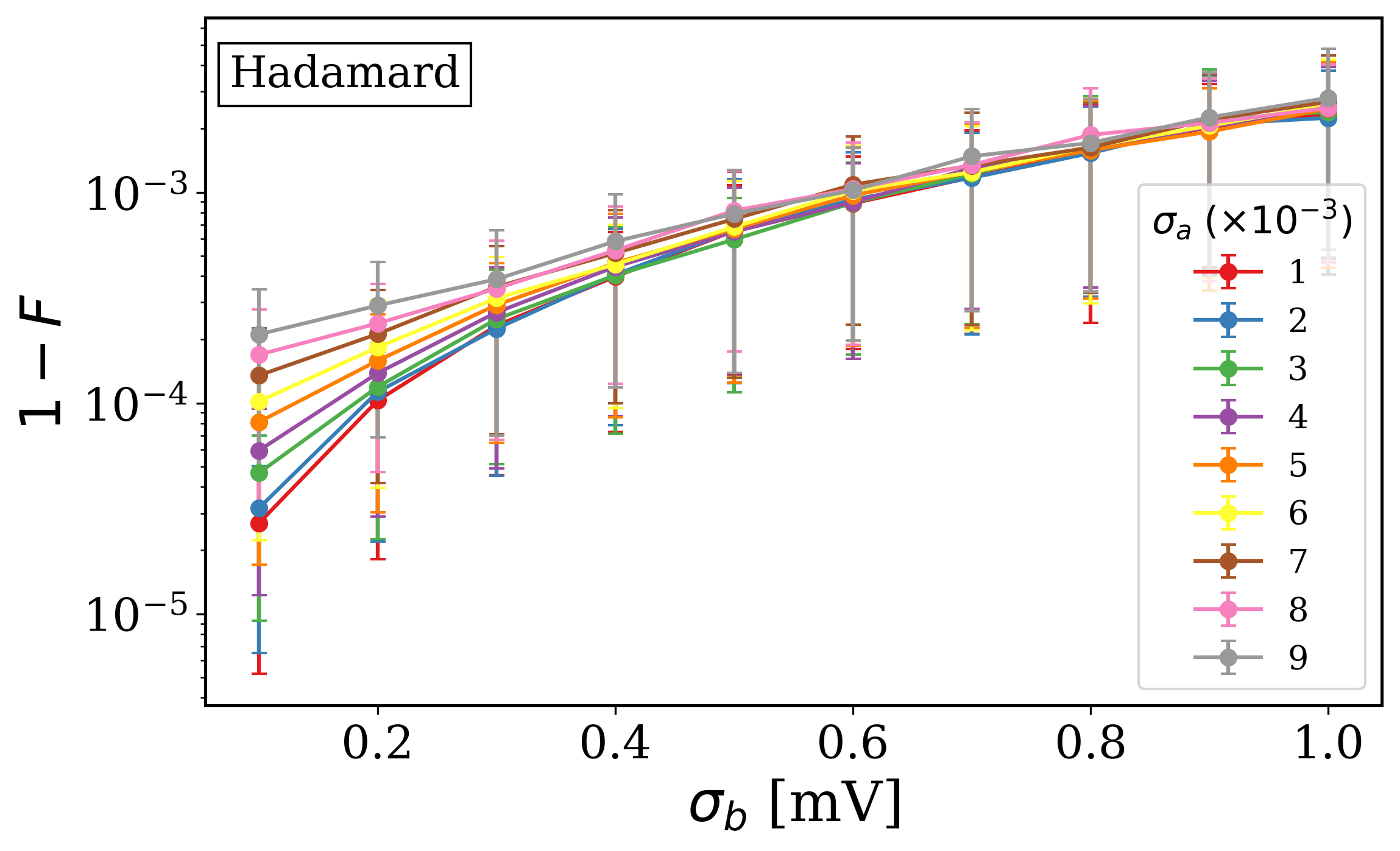}
        \label{fig:combined_h}
    \end{minipage}
    \begin{minipage}[b]{0.32\textwidth}
        \makebox[0pt][l]{\raisebox{16.0\height}{\textbf{(c)}}}%
        \includegraphics[width=\textwidth]{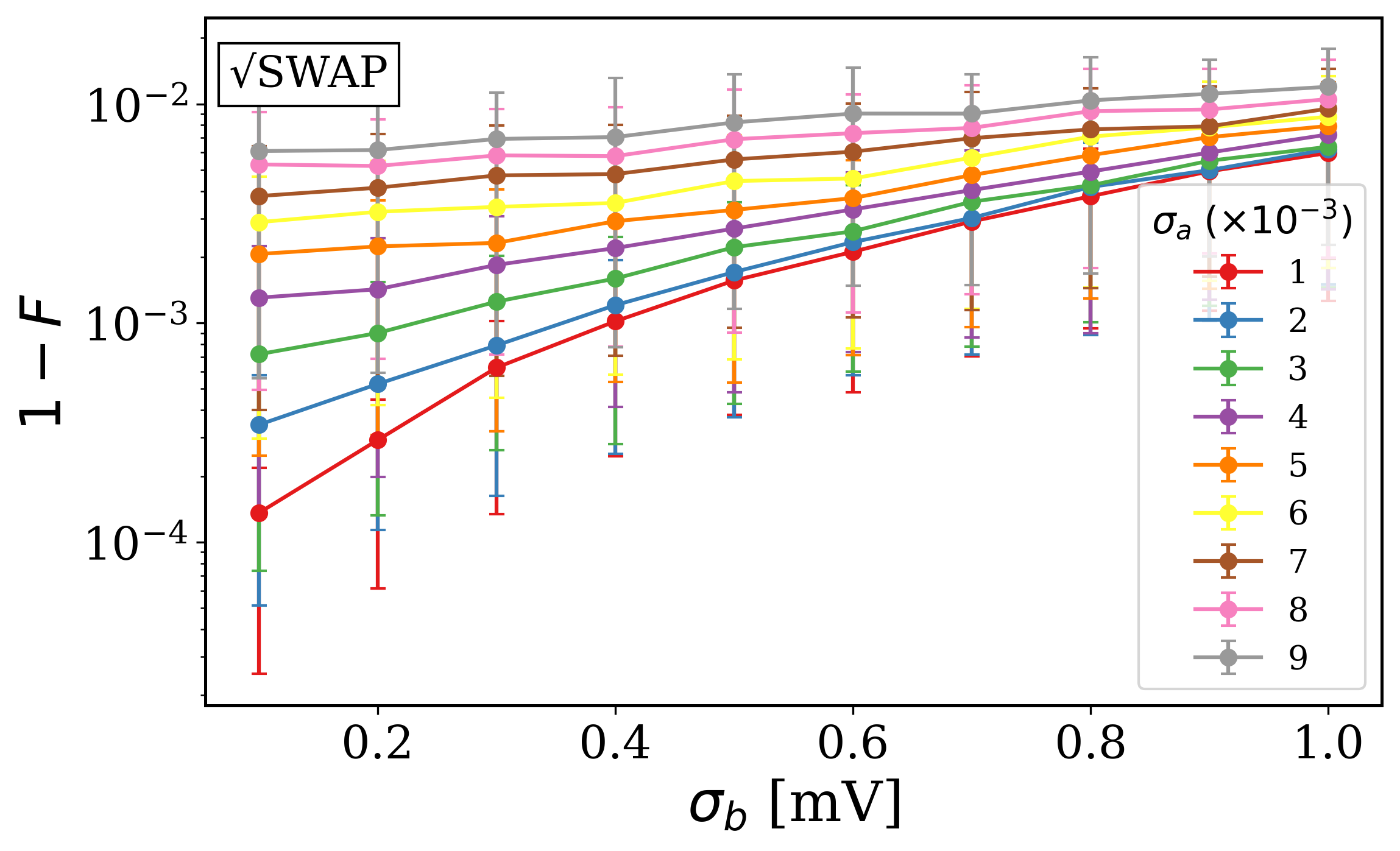}
        \label{fig:combined_rootswap}
    \end{minipage}
    \caption{
Simulated gate infidelity under static voltage miscalibration for (a) $RX(\pi/2)$ (\(t_g = 10\,\mu\mathrm{s} \)), (b) Hadamard (\( t_g = 10\,\mu\mathrm{s} \)), and (c) $\sqrt{\mathrm{SWAP}}$ (\( t_g = 10\,\mu\mathrm{s} \)). Each point shows $1-\langle F\rangle$ averaged over 800 samples of voltage pulses with normally distributed scaling $a$ and offset $b$. Error bars denote the 20th--80th percentile interval of $1-F$.}
    \label{fig:combined_miscal}
\end{figure*}

Each operation uses Gaussian-shaped $g$-factor and ESR control pulses of duration $t_g = 10~\mu\mathrm{s}$. For a given $(\sigma_a,\sigma_b)$, we draw $a \sim \mathcal{N}(1,\sigma_a^2)$ and $b \sim \mathcal{N}(0,\sigma_b^2)$ independently for each of the three control channels $(V_1,V_2,W_1)$ and for each of 800 realizations. For each realization we compute the average gate fidelity $F$ defined in Eq.~\eqref{eq:fidelity} and then form the empirical distribution of $1-F$. The mean $\langle F\rangle(\sigma_a,\sigma_b)$ is used as the average gate fidelity, and the 20th and 80th percentiles of $1-F$ define the error bars shown in Fig.~\ref{fig:combined_miscal}. In this setting, miscalibration acts as a quasi-static \emph{coherent} control error: averaging is performed over fidelities rather than over output states.

Figure~\ref{fig:combined_miscal} summarizes the resulting infidelity $1-\langle F\rangle$ for the three gates. The ESR-driven single-qubit operations [Figs.~\ref{fig:combined_miscal}(a),(b)] show a weak dependence on scaling noise $\sigma_a$: curves at different $\sigma_a$ are tightly clustered, while the dominant trend is the smooth growth of error with offset noise $\sigma_b$. This reflects the approximately linear dependence of the local $g$-factor on plunger voltage in the operating regime, so that amplitude rescaling produces an almost proportional, and relatively mild, change in the effective drive.

By contrast, the exchange-based $\sqrt{\mathrm{SWAP}}$ gate [Fig.~\ref{fig:combined_miscal}(c)] is markedly more sensitive to both scaling and offset miscalibration. Over the same $(\sigma_a,\sigma_b)$ range, its infidelities are typically an order of magnitude (or more) larger than for the single-qubit gates, and the curves fan out visibly with increasing $\sigma_a$. This enhanced susceptibility arises from the strongly nonlinear voltage dependence of the exchange coupling: in the relevant regime $J(V)$ is approximately exponential, so modest changes in pulse amplitude or bias shift lead to disproportionately large changes in the integrated exchange $\int J(t)\,dt$ that sets the effective SWAP angle.

\subsubsection{Comparison of numerical and analytical trends}

\begin{figure}[htbp]
    \centering
    \begin{minipage}[b]{0.45\textwidth}
        \makebox[0pt][l]{\raisebox{22.0\height}{\textbf{(a)}}}%
        \includegraphics[width=\textwidth]{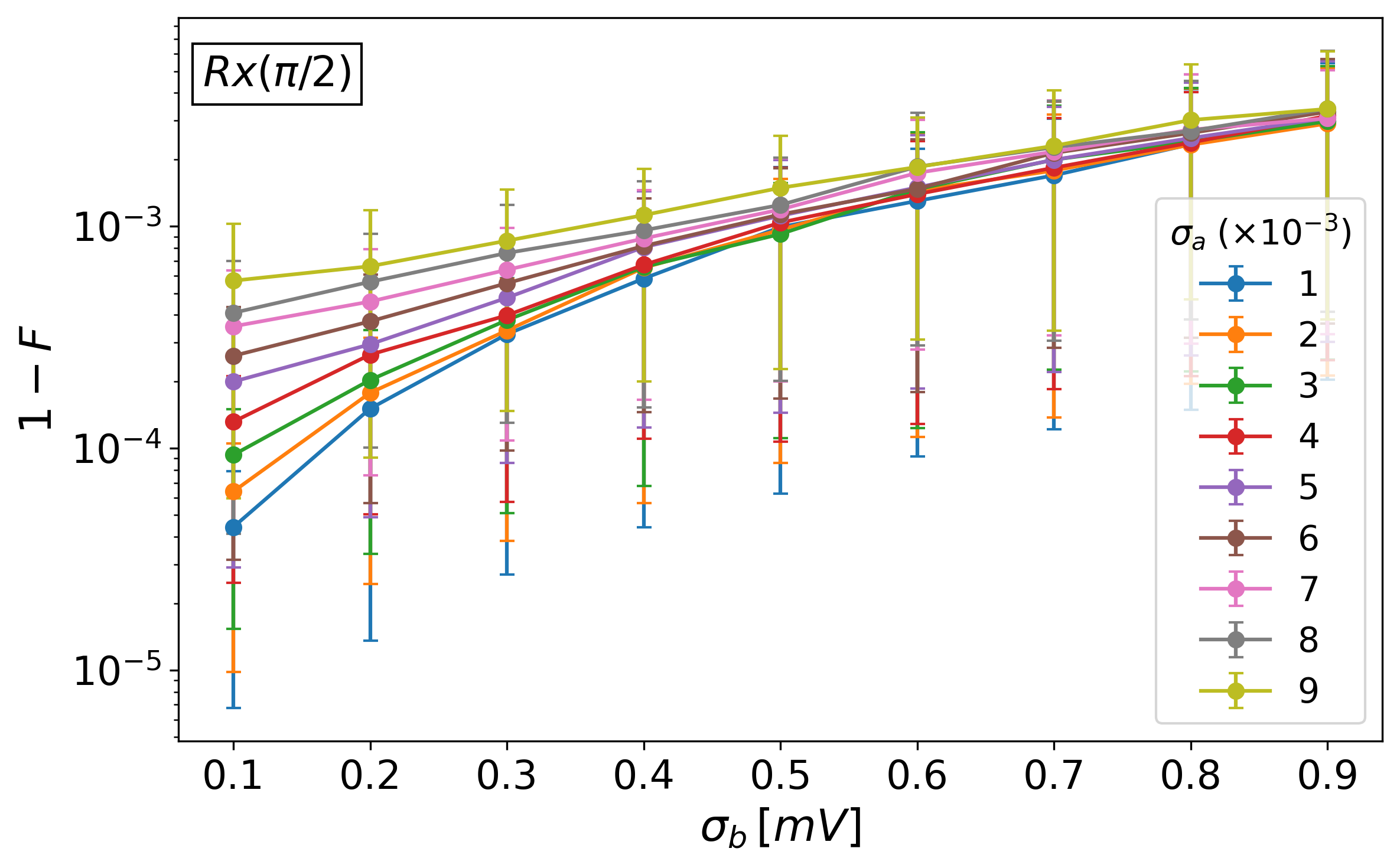}
    \end{minipage}
    \hfill
    \begin{minipage}[b]{0.45\textwidth}
        \makebox[0pt][l]{\raisebox{22.0\height}{\textbf{(b)}}}%
        \includegraphics[width=\textwidth]{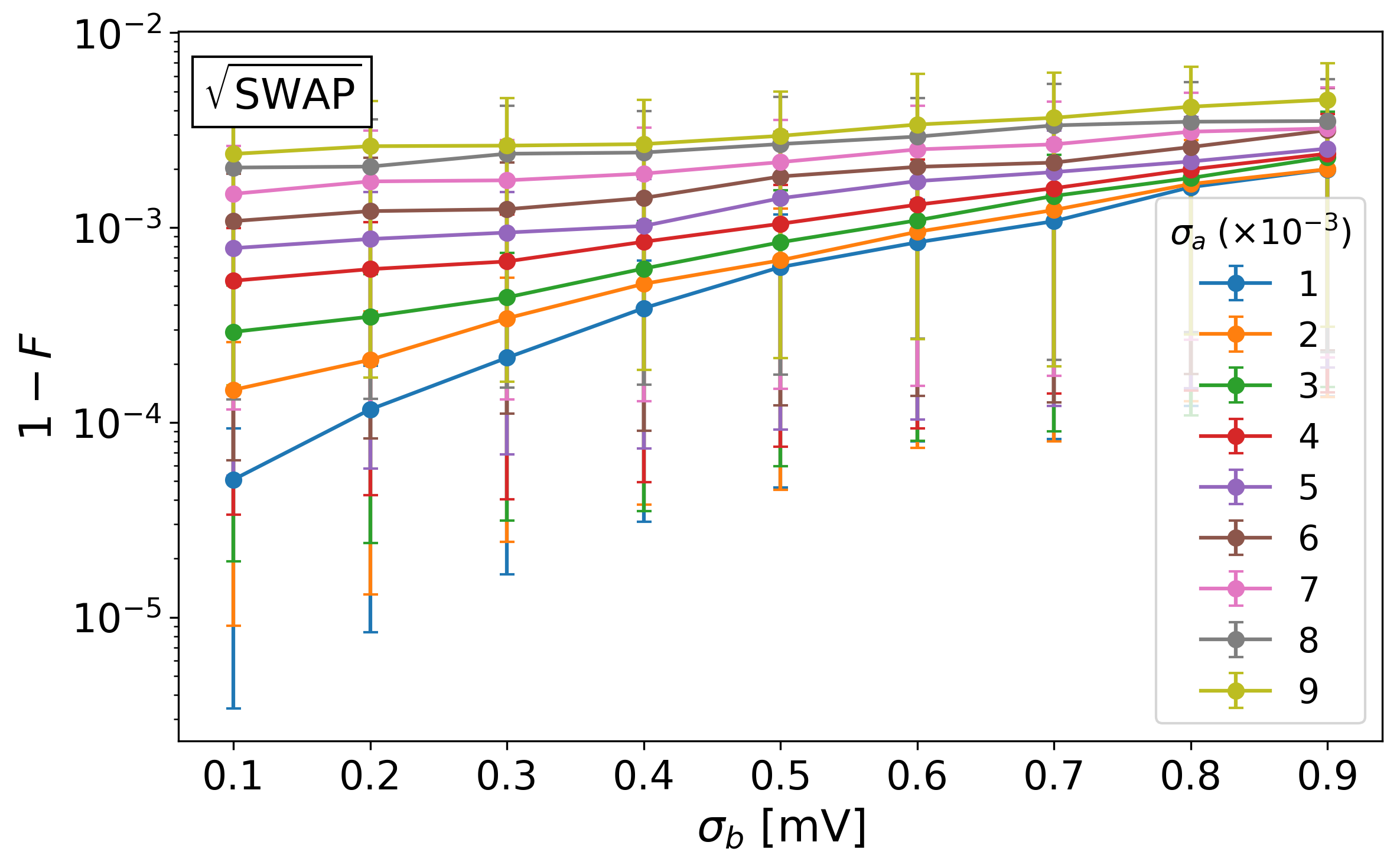}
    \end{minipage}
    \caption{Analytically calculated infidelities of the $RX(\pi/2)$ (a) and $\sqrt{\mathrm{SWAP}}$ (b) gates under static voltage miscalibration for several fixed values of the scaling-noise parameter $\sigma_a$.}
    \label{fig:theory_plots}
\end{figure}

To disentangle the roles of scaling and offset errors, we derive closed-form expressions for the gate fidelities based on the fitted voltage dependences $\delta g_{1,2}(V)$ and $J(V)$ under miscalibrated square pulses. The full derivation is given in Appendix~\ref{app:theory}; the resulting analytical curves for $RX(\pi/2)$ and $\sqrt{\mathrm{SWAP}}$ are shown in Fig.~\ref{fig:theory_plots}. They closely track the numerical simulations in Fig.~\ref{fig:combined_miscal}, with small deviations arising from the fitted parameterizations and from the use of square versus Gaussian pulses.

To quantify sensitivities in a compact form, we fit the small-noise behavior of the simulated infidelity,
\begin{equation}
    r(\sigma_a,\sigma_b)
    \equiv 1 - \langle F\rangle
    \approx c_a \sigma_a^{2} + c_b \sigma_b^{2},
    \label{eq:miscal_quad_model}
\end{equation}
using a global least-squares procedure restricted to the regime $r \ll 1$. Here $\sigma_a$ is the dimensionless scaling-noise amplitude, $\sigma_b$ is the offset-noise amplitude (in volts), and $c_a$ and $c_b$ capture the leading quadratic response to each miscalibration parameter. The extracted coefficients are listed in Table~\ref{tab:miscalibration_quad_fit}. The exchange-driven $\sqrt{\mathrm{SWAP}}$ gate exhibits a much larger $c_a$ than the ESR-driven $RX(\pi/2)$ gate, confirming its substantially higher sensitivity to scaling miscalibration, consistent with the strongly nonlinear $J(V)$ dependence.

\begin{table}[h]
    \centering
    \caption{Quadratic-fit coefficients for the average infidelity $r(\sigma_a,\sigma_b)$ under static voltage miscalibration. Here, $\sigma_a$ is dimensionless and $\sigma_b$ is measured in volts.}
    \label{tab:miscalibration_quad_fit}
    \begin{tabular}{lcc}
        \hline\hline
        Gate & $c_a$ & $c_b$ [V$^{-2}$] \\
        \hline
        $\mathrm{RX}(\pi/2)$       & 0.6423   & $1.718\times10^{3}$ \\
        Hadamard                   & 2.5432   & $2.453\times10^{3}$ \\
        $\sqrt{\mathrm{SWAP}}$     & 80.568   & $5.783\times10^{3}$ \\
        \hline\hline
    \end{tabular}
\end{table}

\subsection{Impact of Random Telegraph Noise on Gate Fidelity}

In Sec.~\ref{sec:citeref}, we discussed how charge noise with a $1/f$-like power spectral density is a leading source of spin-qubit dephasing in silicon devices. Rather than directly simulating idealized $1/f$ noise, our approach models the
environment as an ensemble of two-level fluctuators (TLFs) with fixed switching rates, giving rise to RTN characterized by a Lorentzian power spectral density. This model captures the contribution of individual fluctuators to the noise spectrum and enables a controlled study of how noise switching times affect gate performance.

Building on this noise model, we investigate how RTN degrades quantum gate performance by superimposing the RTN voltage sequences generated in Sec.~\ref{sec:citeref} onto the control pulses of representative single- and two-qubit operations. As in the miscalibration study, we focus on the $RX(\pi/2)$, Hadamard, and $\sqrt{\mathrm{SWAP}}$ gates. For each gate, RTN amplitude $A$, and switching time $\tau$, we propagate the dynamics under 200 independent RTN realizations, average the resulting output density matrices, and compute the average gate fidelity $F$ of the effective noisy channel. In this way, RTN is treated as a stochastic noise process, in contrast to the quasi-static coherent errors used for the miscalibration analysis.

\begin{figure*}[htbp]
    \centering

    \begin{minipage}[b]{0.32\textwidth}
        \makebox[0pt][l]{\raisebox{17.0\height}{\textbf{(a)}}}%
        \includegraphics[width=\textwidth]{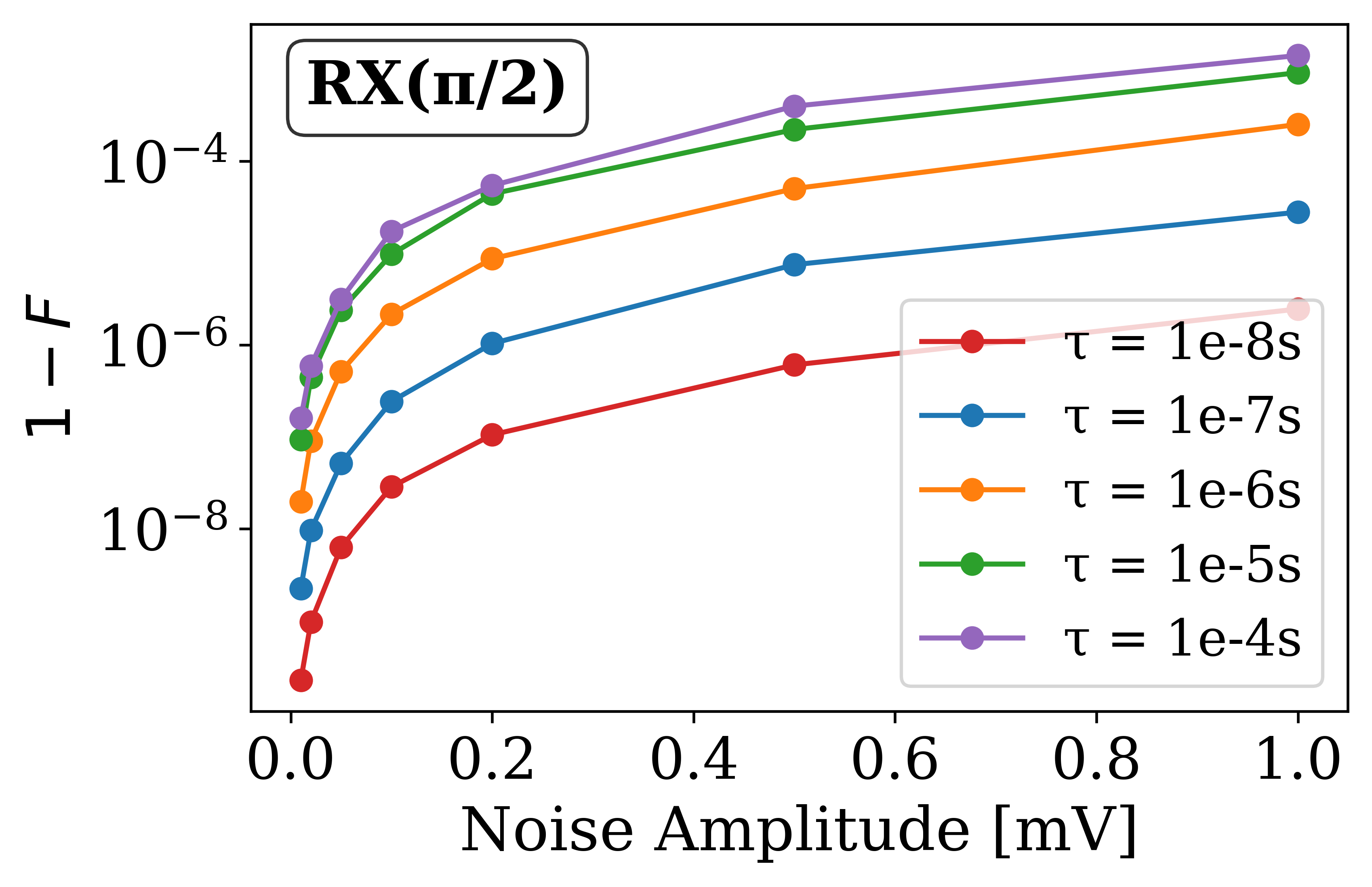}
        \label{fig:d}
    \end{minipage}
    \begin{minipage}[b]{0.32\textwidth}
        \makebox[0pt][l]{\raisebox{17.0\height}{\textbf{(b)}}}%
        \includegraphics[width=\textwidth]{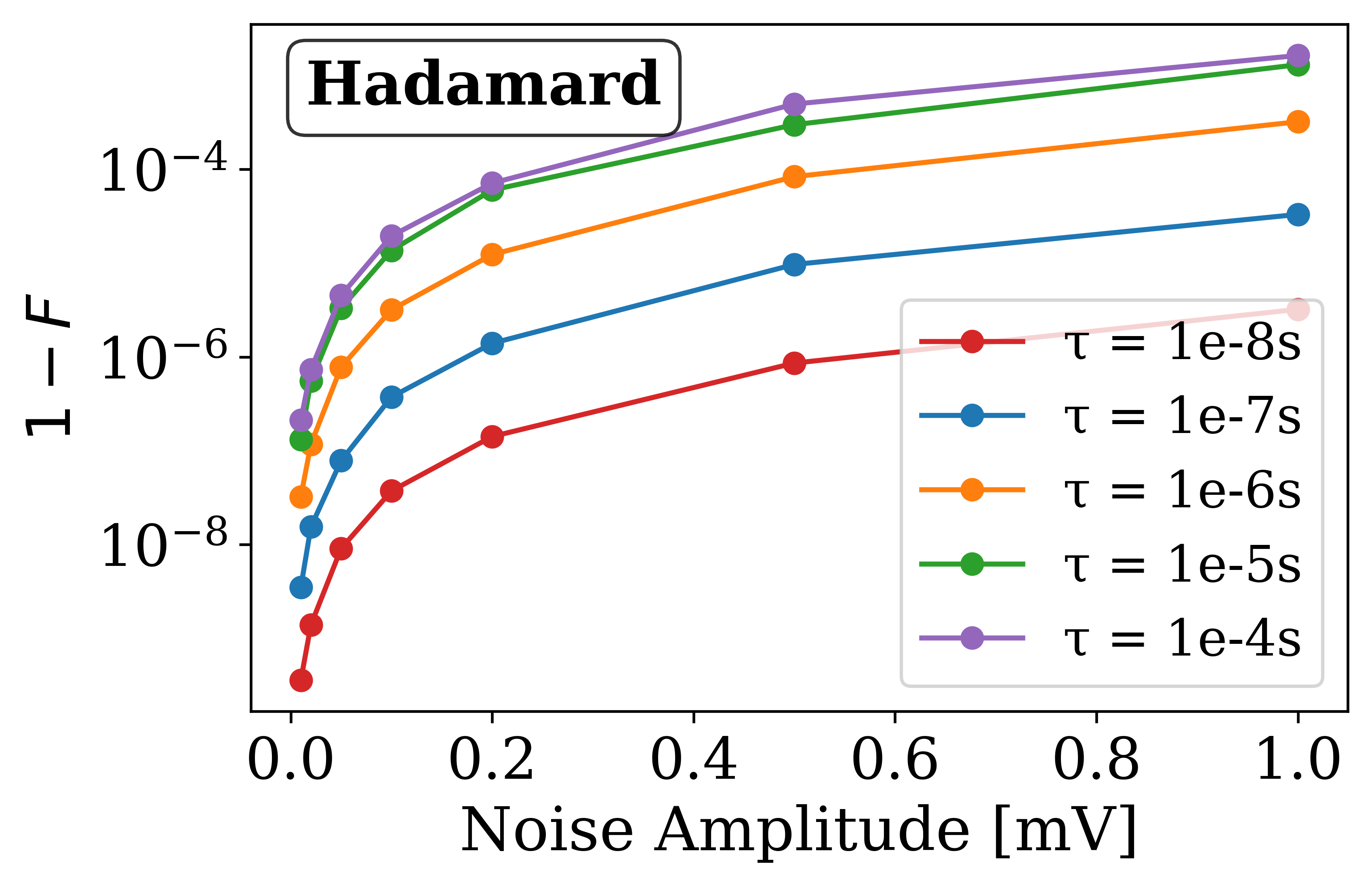}
        \label{fig:e}
    \end{minipage}
    \begin{minipage}[b]{0.32\textwidth}
        \makebox[0pt][l]{\raisebox{17.0\height}{\textbf{(c)}}}%
        \includegraphics[width=\textwidth]{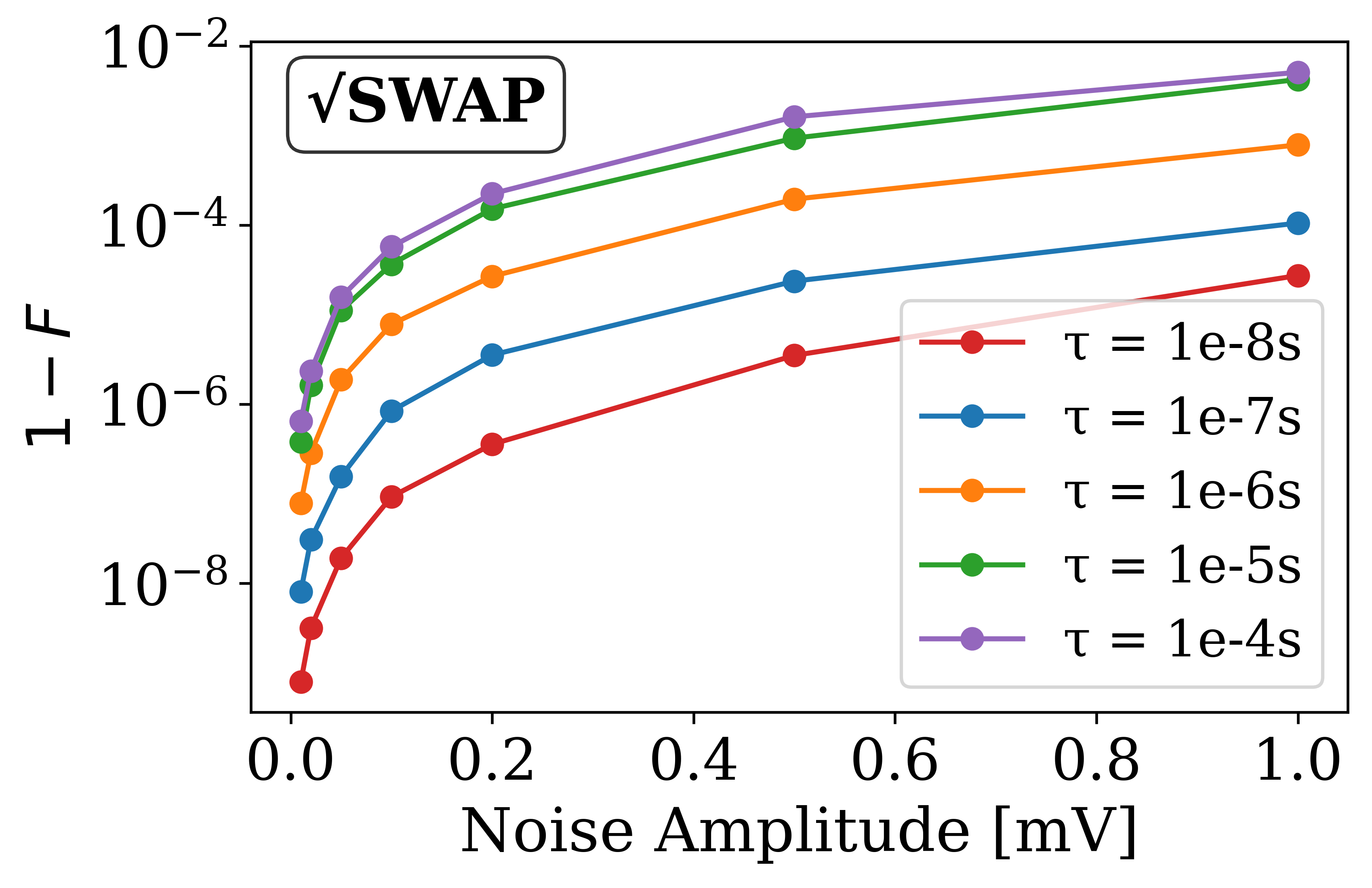}
        \label{fig:f}
    \end{minipage}

    \caption{
       Average gate infidelity under random telegraph noise. (a) $RX(\pi/2)$ gate on qubit 1 (\( t_g = 10\,\mu\mathrm{s} \)); (b) Hadamard gate on qubit 2 (\( t_g = 10\,\mu\mathrm{s} \)); (c) $\sqrt{\mathrm{SWAP}}$ gate (\( t_g = 10\,\mu\mathrm{s} \)).
Each data point shows the average gate fidelity over 200 stochastic noise realizations.
    }
    \label{fig:combined}
\end{figure*}

The resulting trends are summarized in Fig.~\ref{fig:combined}. In this figure, the vertical axis shows the gate infidelity $(1 - F)$ on a logarithmic scale, while the horizontal axis gives the RTN amplitude in millivolts. Each colored curve corresponds to a fixed switching time $\tau$ in the range $10^{-8}\,\mathrm{s}$ to $10^{-4}\,\mathrm{s}$, illustrating how both the strength and characteristic timescale of the charge-noise-equivalent RTN influence the robustness of the three gate types.

Across all three gates, the infidelity exhibits an approximately quadratic dependence on the RTN amplitude $A$:
\[
\log_{10}(1 - F) \propto \log_{10}(A^2)
\quad\Longleftrightarrow\quad
1 - F \approx a(\tau)\,A^2,
\]
where $a(\tau)$ is a gate- and switching-time--dependent prefactor. The fitted values of $a(\tau)$ are listed in Table~\ref{tab:rtn_fit_three}. This quadratic scaling is consistent with a weak-noise expansion, in which the leading contribution to the gate error is governed by the variance of the time-integrated noise and therefore scales as $A^2$, with higher-order terms in $A$ neglected. For fast switching ($\tau \ll t_g$, with $t_g = 10\,\mu\mathrm{s}$), the fluctuations are rapid on the gate timescale and are partially averaged out (motional narrowing), resulting in smaller $a(\tau)$ and reduced infidelity. As $\tau$ approaches and exceeds $t_g$, the RTN becomes effectively quasi-static during the gate, and the curves for different $\tau$ converge, reflecting the fact that slow noise primarily acts as a random but time-independent bias over the duration of a single gate.

\begin{table}[htbp]
\centering
\caption{Quadratic-fit coefficients \(a\) in \(1-F \approx a\,A^2\) for three gates under RTN at different \(\tau\). All gates use the same duration \(t_g=10\,\mu\mathrm{s}\).}
\label{tab:rtn_fit_three}
\begin{tabular}{cccc}
\toprule
\(\tau\) (s) & RX\((\pi/2)\) & Hadamard & \(\sqrt{\mathrm{SWAP}}\) \\
\midrule
\(1\times 10^{-4}\) & \(1.419\times 10^{-3}\) & \(1.624\times 10^{-3}\) & \(5.096\times 10^{-3}\) \\
\(1\times 10^{-5}\) & \(9.190\times 10^{-4}\) & \(1.291\times 10^{-3}\) & \(4.227\times 10^{-3}\) \\
\(1\times 10^{-6}\) & \(2.412\times 10^{-4}\) & \(3.199\times 10^{-4}\) & \(7.889\times 10^{-4}\) \\
\(1\times 10^{-7}\) & \(2.779\times 10^{-5}\) & \(3.300\times 10^{-5}\) & \(1.055\times 10^{-4}\) \\
\(1\times 10^{-8}\) & \(2.459\times 10^{-6}\) & \(3.209\times 10^{-6}\) & \(2.735\times 10^{-5}\) \\
\bottomrule
\end{tabular}
\end{table}

Across these three gate operations, the $\sqrt{\mathrm{SWAP}}$ operation exhibits notably larger fitted coefficients $a$, indicating a substantially stronger sensitivity to RTN of the same parameters. This trend is consistent with our miscalibration study in Subsec.~\ref{subsec:miscalibration}: exchange-based two-qubit gates inherit an intrinsically enhanced susceptibility because the underlying interaction strength responds nonlinearly to variations in the control voltages.

Beyond the relative gate sensitivities, a second clear feature emerges. For switching times comparable to or exceeding the gate duration ($\tau \gtrsim t_g = 10\,\mu\mathrm{s}$), the fidelity curves for the largest $\tau$ values level off onto a plateau for each of the three gates. On the level of a single realization, such slow RTN is effectively quasi-static over the gate and acts like a random offset of the control voltages, akin to drawing a static miscalibration from a distribution. In our RTN analysis, however, we average the output density matrices over many realizations before computing $F_{\mathrm{avg}}$, so slow RTN appears as a predominantly incoherent noise channel, in contrast to the miscalibration study where we average fidelities of coherent unitary errors. For fixed noise amplitude $A$, once $\tau$ reaches $\sim(1\text{--}10)\,t_g$, the infidelity is largely determined by the variance of these quasi-static voltage offsets and depends only weakly on the precise value of $\tau$.

\subsection{Error Analysis via Quantum Process Tomography}

To further elucidate how charge noise manifests at the quantum-channel level, we perform QPT on representative single- and two-qubit gate operations in the presence of RTN. Specifically, we consider an $RX(\pi/2)$ gate applied to qubit~1 and a Control-Z (CZ) gate acting on the two-qubit system. For each gate, we reconstruct both the full process
matrix $\chi$ and the corresponding error process matrix $\chi^{\mathrm{err}}$, as introduced in Sec.~\ref{sec:Process-Characterization}.

\begin{figure*}[htbp]
    \centering

    \begin{minipage}{0.37\textwidth}
        \centering
        \makebox[0pt][l]{\raisebox{150pt}{\textbf{(a)}}}%
        \includegraphics[width=\linewidth]{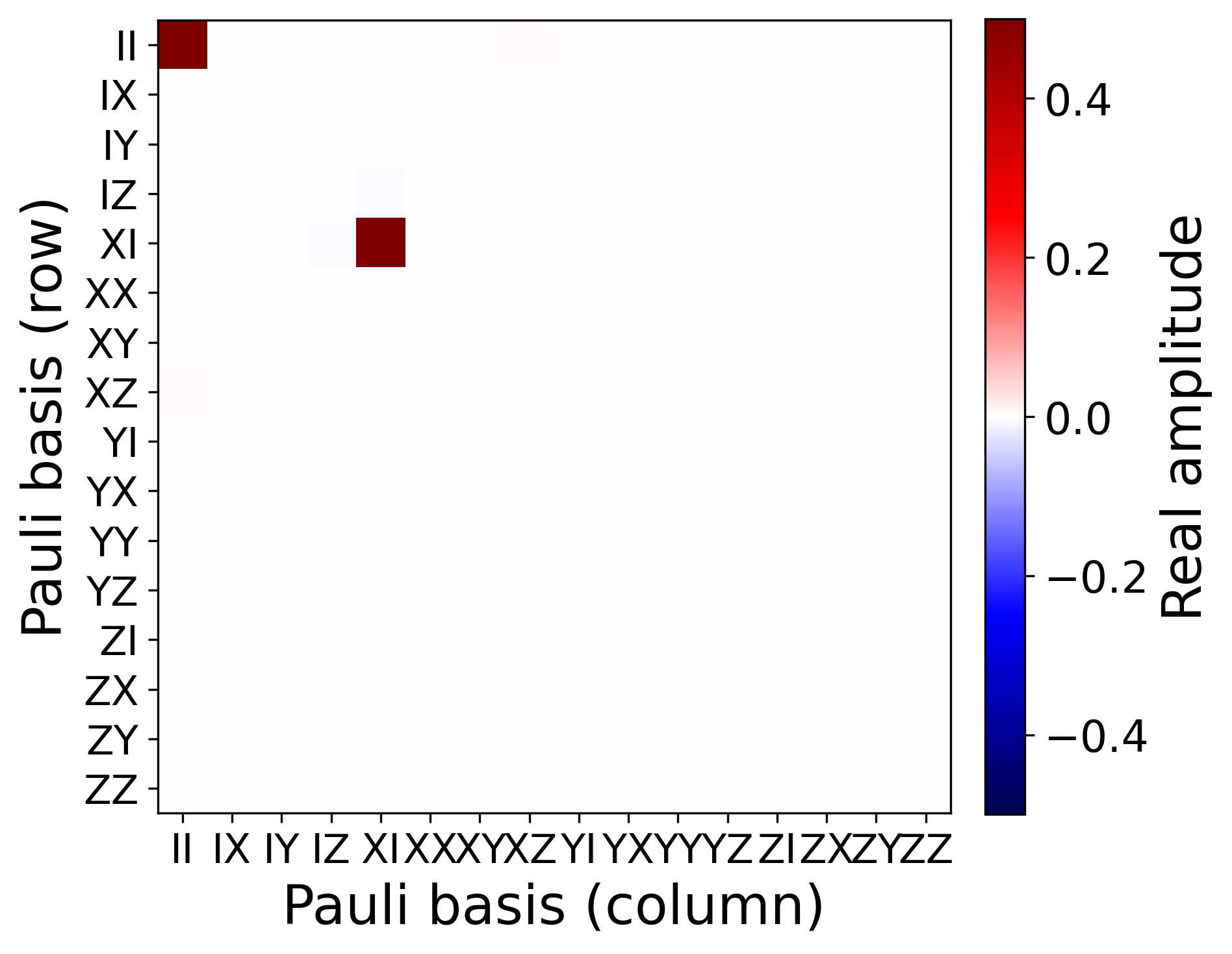}
    \end{minipage}
    \hspace{0.1\textwidth}
    \begin{minipage}{0.37\textwidth}
        \centering
        \makebox[0pt][l]{\raisebox{150pt}{\textbf{(b)}}}%
        \includegraphics[width=\linewidth]{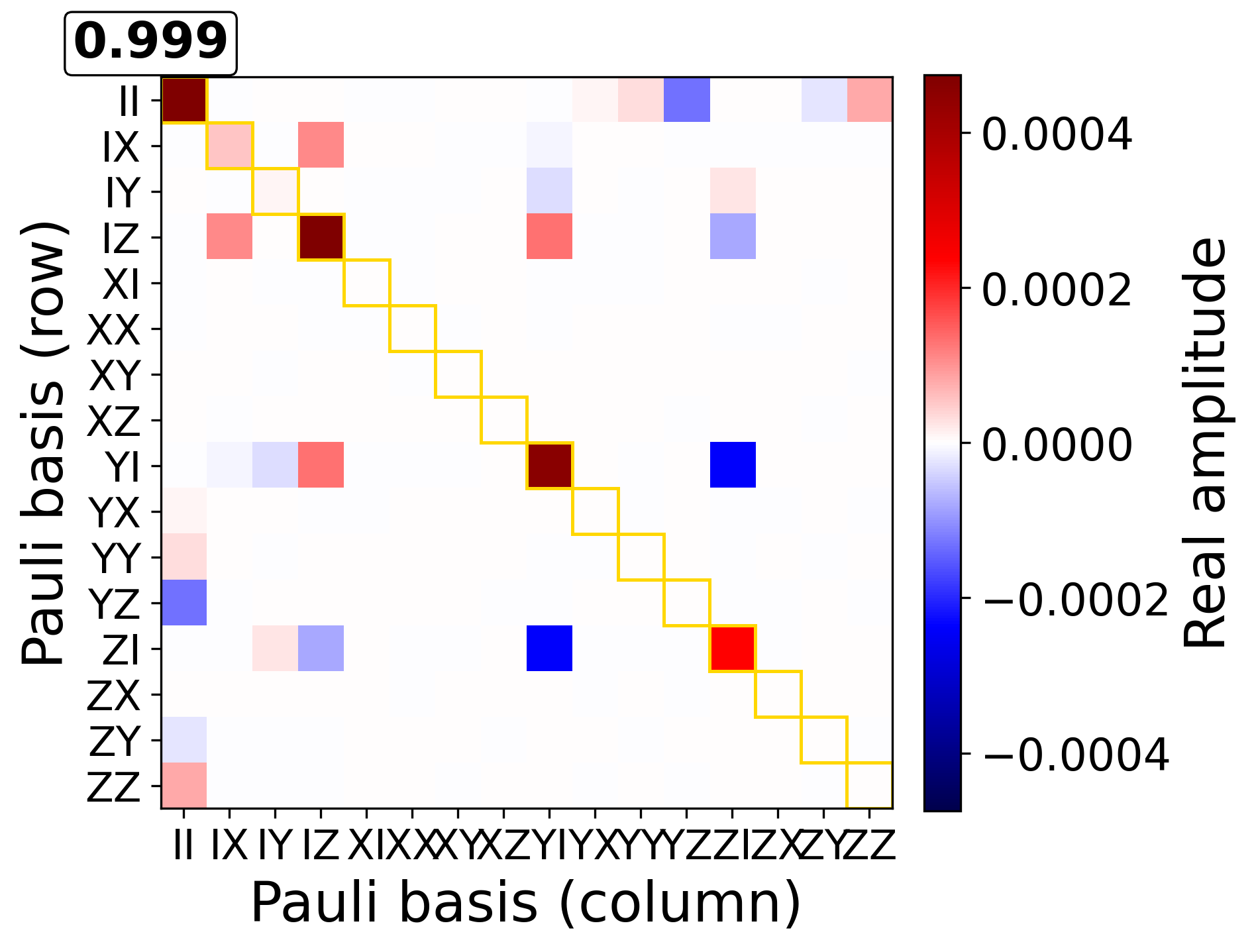}
    \end{minipage}
    \hfill
    \begin{minipage}{0.37\textwidth}
        \centering
        \makebox[0pt][l]{\raisebox{150pt}{\textbf{(c)}}}%
        \includegraphics[width=\linewidth]{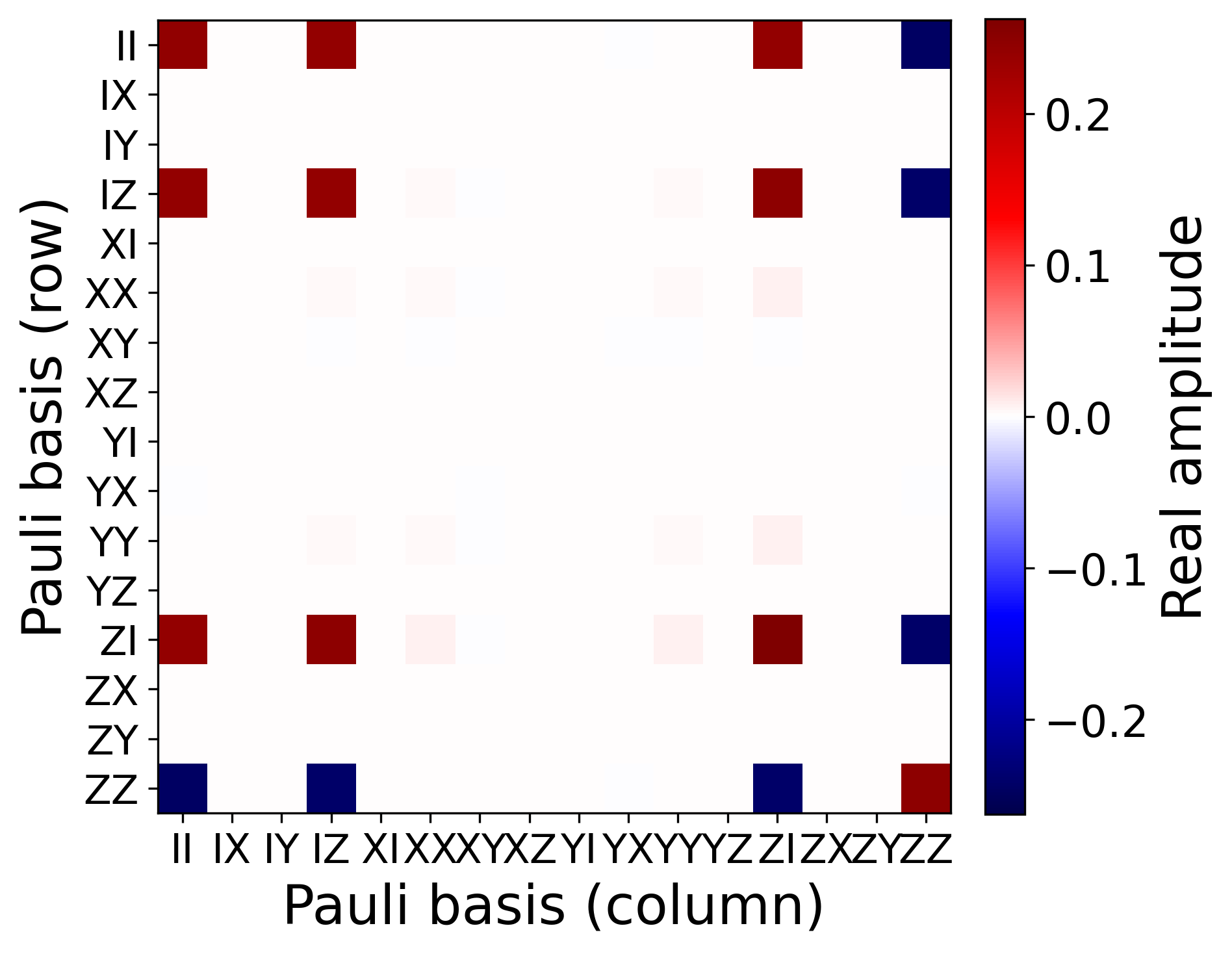}
    \end{minipage}
    \hspace{0.1\textwidth}
    \begin{minipage}{0.37\textwidth}
        \centering
        \makebox[0pt][l]{\raisebox{150pt}{\textbf{(d)}}}%
        \includegraphics[width=\linewidth]{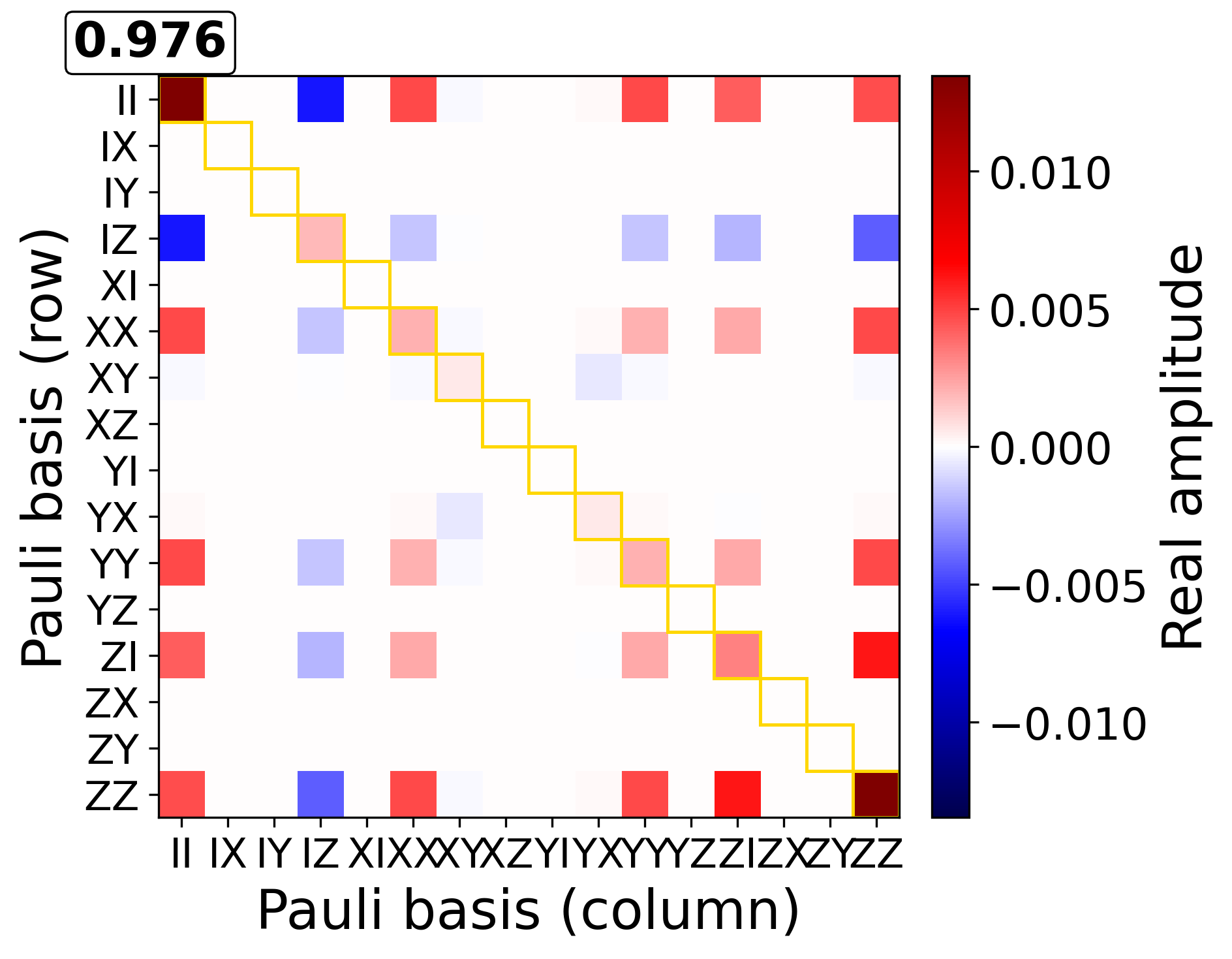}
    \end{minipage}

    \caption{
        Real parts of the reconstructed process matrices and error process matrices for single- and two-qubit gates under RTN with amplitude $A = 3~\mathrm{mV}$ and $\tau = 10^{-7}\,\mathrm{s}$.
        Panels~(a,b) correspond to the $RX(\pi/2)$ gate on qubit~1, while panels~(c,d)
        correspond to the CZ gate. All matrices are expressed in the two-qubit Pauli
        basis and averaged over 200 noise realizations. For visualization of the error
        process matrices, the identity component $\chi^{\mathrm{err}}_{II,II} = \chi^{\mathrm{err}}_{00}$ is excluded from the color scale, which is set by the remaining matrix elements.
        The diagonal elements (highlighted in yellow) of the error process matrices
        represent the Pauli error probabilities within the Pauli-twirling approximation.
    }
    
    \label{fig:chi_rx_cz_comparison}
\end{figure*}

Figure~\ref{fig:chi_rx_cz_comparison} summarizes the real parts of the reconstructed
process matrices and error process matrices for the two gates. Panels~(a) and~(b)
show the process matrix $\chi$ and the error process matrix $\chi^{\mathrm{err}}$
for the $RX(\pi/2)$ gate, while panels~(c) and~(d) show the corresponding quantities
for the CZ gate. All matrices are expressed in the two-qubit Pauli basis and are
averaged over 200 independent RTN realizations with fixed noise amplitude
$A = 3~\mathrm{mV}$ and switching time $\tau = 10^{-7}\,\mathrm{s}$.

The diagonal elements of the error process matrix $\chi^{\mathrm{err}}$ provide information about the dominant Pauli error structure. In the Pauli-basis representation, the diagonal elements are the error probabilities in the so-called Pauli twirling approximation~\cite{Korotkov2013ErrorMatrices}.

For the $RX(\pi/2)$ gate, the leading diagonal contributions are dominated by $IZ$ and $ZI$ terms, consistent with dephasing-type errors and weak crosstalk between the two qubits during an ESR-driven single-qubit rotation. In our model, RTN applied to the plunger gates primarily modulates the effective $g$-factors, leading to residual fluctuations in the local Zeeman splittings that naturally appear as $Z$-type contributions in the Pauli expansion. Smaller but discernible weights in the $IX$ and $IY$ components are indicative of deviations from an ideal rotation about the $X$ axis, consistent with rotation-axis misalignment and amplitude fluctuations induced by control noise.\\
\indent In contrast, the diagonal structure of $\chi^{\mathrm{err}}$ for the CZ gate exhibits substantial weight in $ZZ$, $IZ$, and $ZI$, as well as correlated terms such as $XX$ and $YY$. This structure is consistent with the native two-qubit Hamiltonian containing Zeeman and exchange interactions, which in our model takes the form (in angular frequency units)

\begin{equation}
\begin{aligned}
\mathscr{H}_{\mathrm{CZ}}(t)
=&\;
\frac{\mu_B B_z}{2\hbar}
\left[
\delta g_1(t)\,ZI
+
\delta g_2(t)\,IZ
\right]
\\
&+
\frac{J(t)}{4\hbar}
\left(
XX + YY + ZZ
\right).
\end{aligned}
\end{equation} 

Here, the local $Z$ terms arise from $g$-factor differences, while the exchange pulse $J(t)$ implements an isotropic Heisenberg coupling. In the presence of RTN on the relevant control gates, fluctuations in $\delta g_{1,2}$ and $J$ couple directly to the $IZ$, $ZI$, and $(XX + YY + ZZ)$ components of the Hamiltonian, and the corresponding Pauli structures appear naturally in the reconstructed error channel. Because the two qubits are operated with similar Larmor frequencies and the exchange pulse is not in the extreme dispersive limit, the dynamics cannot be approximated by a purely Ising $ZZ$ interaction; instead, the effective noise channels retain significant $XX$ and $YY$ weight alongside $ZZ$, $IZ$, and $ZI$.

While the diagonal elements of the error process matrix $\chi^{\mathrm{err}}$ provide a clear picture of the dominant Pauli error weights, additional insight into the underlying error mechanisms is obtained from the \emph{off-diagonal}
structure of the channel. To quantify this coherent contribution at the channel level, we evaluate the \emph{coherence angle} $\Theta$, introduced in Sec.~\ref{sec:coherent_angle}.

\begin{figure}[t]
    \centering
    \includegraphics[width=0.95\linewidth]{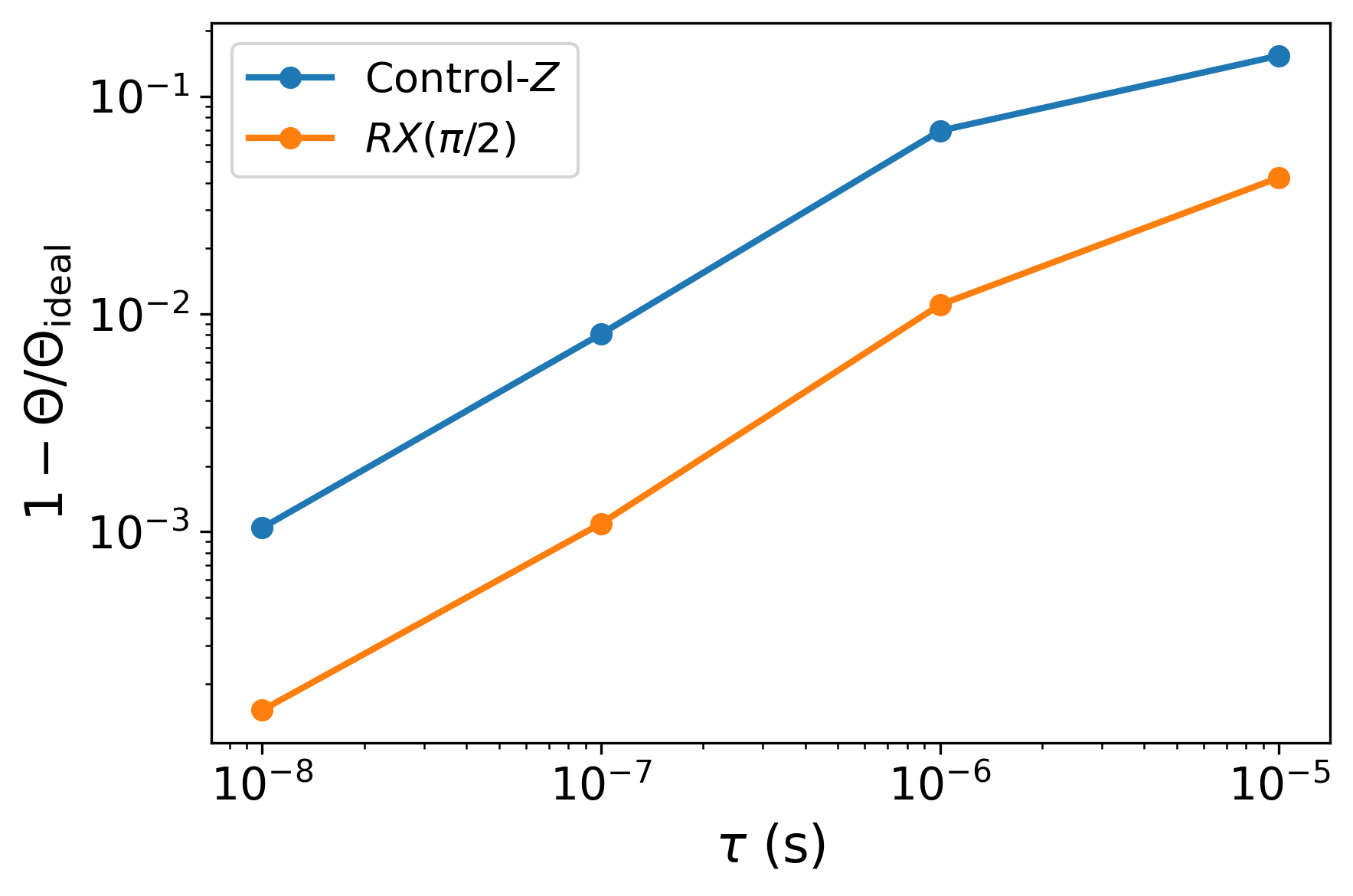}
    \caption{
    Normalized coherence angle $\Theta / \Theta_{\mathrm{ideal}}$ as a function of the noise switching time $\tau$ for the CZ and  $RX(\pi/2)$ gates. The RTN amplitude is fixed at $A=3~\mathrm{mV}$.}
    \label{fig:coherence_angle_tau}
\end{figure}

Figure~\ref{fig:coherence_angle_tau} shows the normalized coherence angle, $\Theta/\Theta_{\mathrm{ideal}}$, as a function of the RTN switching time $\tau$ for the CZ and $RX(\pi/2)$ gates. For both gates, the coherence angle decreases monotonically with increasing $\tau$, as reflected by the systematic increase of $1-\Theta/\Theta_{\mathrm{ideal}}$. This trend indicates that the coherence of the error channel is progressively suppressed as the noise evolves from the fast-fluctuating regime toward the intermediate regime. Across the full range of switching times, the CZ gate exhibits a larger relative suppression of the coherence angle than the single-qubit $RX(\pi/2)$ gate, consistent with its increased sensitivity to correlated control noise.

This interpretation is consistent with the dephasing/rotation channel model discussed in Ref.~\cite{Iverson2020Coherence}, where a quantum channel consists of a mixture of incoherent noise and coherent (unitary over-rotation) error processes. In such mixed channels, the coherence angle provides a quantitative measure of the strength of the coherent component, increasing as unitary, systematic errors become more dominant relative to stochastic noise contributions. Within our RTN model, variations in the noise switching time $\tau$ modify the balance between these two contributions. The observed suppression of the coherence angle with increasing $\tau$ therefore indicates a crossover from error channels dominated by coherent Pauli mixing at short switching times to increasingly incoherent, effectively stochastic noise in the slow-noise limit.

\subsubsection{Coherent Error}\label{sec:coherent_extraction}

The presence of off-diagonal and correlated components in the $\chi$-matrix suggests that the error channel contains a non-negligible coherent contribution. To identify the leading coherent component in the small-noise regime, we examine the Kraus decomposition and focus on the dominant, most unitary-like Kraus operator. Using the procedure described in Appendix~\ref{app:qpt_details}, we extract Kraus representations for three representative gates: $RX(\pi/2)$, Hadamard, and CZ. Throughout this analysis we consider RTN with a fixed switching time $\tau = 10^{-6}\,\mathrm{s}$, chosen as a representative value in the noise regime.

To isolate noise-induced effects from the intended gate action, we work in the error frame by factoring out the target unitary,
\[
    K_i' = U_\mathrm{gate}^\dagger K_i ,
\]
so that the transformed operators $K_i'$ describe only the residual noise channel associated with each gate. We now focus on the dominant Kraus operator $K'_{\mathrm{dom}}$, which carries the largest weight in the decomposition for the cases considered here and is closely related to the leading eigenvector of the process matrix $\chi$. In the weak-noise regime, $K'_{\mathrm{dom}}$ provides an effective description of the dominant, approximately unitary component of the error channel.

Since $K'_{\mathrm{dom}}$ is not, in general, exactly unitary, we approximate its coherent content by projecting it onto the closest unitary. For this, we define an effective error Hamiltonian $H_{\mathrm{err}}$ via a Hermitian projection of the matrix logarithm,
\begin{equation}
    H_{\mathrm{err}}
    = \frac{i \log(K'_{\mathrm{dom}}) + \bigl(i \log(K'_{\mathrm{dom}})\bigr)^\dagger}{2},
\end{equation}
where $\log(\cdot)$ denotes the principal branch of the matrix logarithm. For a perfectly unitary $K'_{\mathrm{dom}}$, the term $i\log(K'_{\mathrm{dom}})$ would already be Hermitian; the above symmetrization ensures that $H_{\mathrm{err}}$ remains Hermitian even when small non-unitary components are present, and thus defines a physically meaningful generator of unitary evolution.

The corresponding approximate unitary error operator is then
\begin{equation}
    U_{\mathrm{err}} = e^{-i H_{\mathrm{err}}}.
\end{equation}
Acting in the gate frame, $U_{\mathrm{err}}$ represents the dominant coherent deviation of the implemented operation from the ideal target.

Given a noisy output state $\rho_{\mathrm{out}}$ generated by the noisy channel $\mathcal{E}$, we define a ``corrected'' state by applying the inverse of this unitary error,
\begin{equation}
    \rho_{\mathrm{corr}} = U_{\mathrm{err}}^\dagger \, \rho_{\mathrm{out}} \, U_{\mathrm{err}}.
\end{equation}
While such a correction could, in practice, be implemented through updated control pulses or a redefinition of the computational frame, here we apply it numerically to assess how much of the observed error arises from coherent contributions.

\subsubsection{Unitary Correction of Coherent Errors}
\label{sec:qpt_rtn}

We now apply the coherent-error extraction method of Sec.~\ref{sec:coherent_extraction} to quantify how much of the RTN-induced gate error can be removed by a unitary correction. When RTN is added to the control voltages, the key observation is that a zero-mean voltage fluctuation $\delta V(t)$ need not remain zero-mean once mapped through a nonlinear dependence of the Hamiltonian parameters on $V$. For exchange-based gates, where $J(V)$ is approximately exponential in the operating regime, noise samples $V(t)+\delta V(t)$ explore regions where $J$ increases more significantly from its noiseless value than it decreases, so that
\[
\langle J(V+\delta V)\rangle > J(V)
\]
even when $\langle \delta V \rangle = 0$. This systematic bias in the time-integrated exchange leads to an average overrotation---i.e., a coherent, unitary error. By contrast, for ESR-driven single-qubit gates the relevant $g$-factor dependence $\delta g(V)$ is nearly linear in $V$ around the working point, so positive and negative fluctuations tend to cancel in the Hamiltonian domain as well, and the net coherent shift is strongly suppressed. In that case, RTN primarily broadens the distribution of gate outcomes rather than shifts their mean.

Superimposed on these coherent distortions are incoherent errors that arise from ensemble averaging over many RTN realizations with the same switching time $\tau$ and amplitude $A$. Each realization corresponds to a different stochastic voltage trajectory and hence to a slightly different unitary evolution. Averaging these at the density-matrix level produces an effective non-unitary channel. If the noise is fast ($\tau \ll t_g$), the trajectories are similar in their time-averaged effect and this averaging induces only modest decoherence. In the opposite limit of slow RTN ($\tau \gtrsim t_g$), different realizations can lead to substantially different gate actions, and the ensemble average produces a channel dominated by dephasing-like, irreducible incoherent error.

\begin{figure*}[htbp]
    \centering
    \begin{minipage}{0.45\linewidth}
        \centering
        \makebox[0pt][l]{\raisebox{150pt}{\textbf{(a)}}}%
        \includegraphics[width=\linewidth]{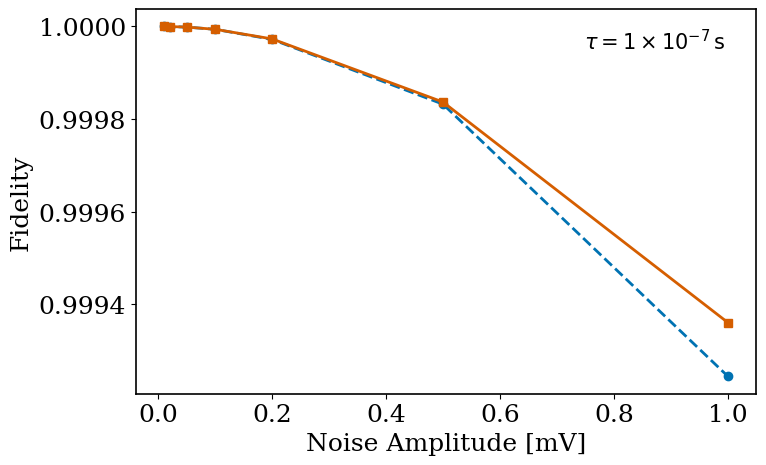}
        \label{fig:cz_unitary_correction_tau1e-7}
    \end{minipage}
    \hfill
    \begin{minipage}{0.45\linewidth}
        \centering
        \makebox[0pt][l]{\raisebox{150pt}{\textbf{(b)}}}%
        \includegraphics[width=\linewidth]{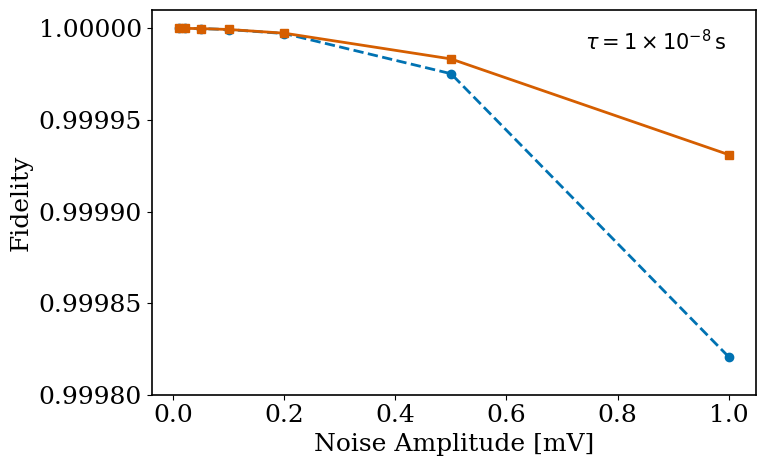}
        \label{fig:cz_unitary_correction_tau1e-8}
    \end{minipage}

    \caption{CZ gate fidelity versus RTN amplitude, illustrating the effect of applying a correction unitary. Blue: before correction; orange: after correction. (a) RTN switching time $\tau = 10^{-7}$ s = $t_g \times 10^{-2}$. (b) $\tau = 10^{-8}$ s = $t_g \times 10^{-3}$.  }
    \label{fig:cz_unitary_correction}
\end{figure*}

Figure~\ref{fig:cz_unitary_correction} illustrates these effects for the CZ gate. For each choice of RTN amplitude $A$ and switching time $\tau$, we first compute the raw average gate fidelity under RTN and then apply the unitary correction $U_{\mathrm{err}}$ extracted from $K'_{\mathrm{dom}}$ to obtain a ``corrected'' fidelity. As expected, the raw fidelity decreases monotonically with amplitude $A$. For small $\tau$, much of this loss is coherent: the corrected fidelity is significantly higher than the uncorrected one. For larger $\tau \sim t_g$ and beyond, the total fidelity drops further, but the gap between raw and corrected fidelities shrinks, signaling that a larger fraction of the error is incoherent and cannot be removed by a unitary correction.

To quantify how effective the coherent correction is, we define the relative \emph{infidelity} improvement as
\begin{equation} 
\mathrm{Improvement} = \frac{r_{\mathrm{bef}} - r_{\mathrm{aft}}}{r_{\mathrm{bef}}} \times 100\%, \qquad r = 1 - F, 
\end{equation} 
where $r_{\mathrm{bef}}$ and $r_{\mathrm{aft}}$ denote the gate infidelities before and after applying the unitary correction, respectively. This quantity measures the fractional reduction in infidelity achieved by the correction procedure, expressed as a percentage.

Table~\ref{tab:cz_improvement} summarizes the relative infidelity improvements for the CZ gate at a fixed RTN amplitude of $A = 1$ mV for different switching times $\tau$, with each value obtained by averaging over 200 independent RTN realizations.

\begin{table}[htbp]
\centering
\caption{Relative improvement of infidelity for the CZ gate for RTN with 1 mV amplitude, versus $\tau$.}
\label{tab:cz_improvement}
\begin{tabular}{cc}
\toprule
$\tau$ (s) & Improvement (\%) \\
\midrule
$1\times 10^{-5}$ & 0.67 \\
$1\times 10^{-6}$ & 1.30 \\
$1\times 10^{-7}$ & 15.36 \\
$1\times 10^{-8}$ & 61.54 \\
\bottomrule
\end{tabular}
\end{table} 

The strong improvement at small~$\tau$ reflects the fact that, in this regime, the leading RTN-induced error manifests as a nearly identical phase overrotation across noise realizations. As a result, a single compensating unitary extracted from $K'_{\mathrm{dom}}$ is able to remove a large fraction of the infidelity. This behavior is consistent with the nonlinear voltage dependence of the exchange interaction discussed earlier in this section.

For long switching times ($\tau \gtrsim t_g$), RTN behaves more like a quasistatic voltage offset: each realization produces a different CZ phase, and averaging over many realizations yields a mixture of distinct unitaries. The resulting channel is predominantly incoherent and cannot be corrected by any single unitary operation.

For single-qubit gates such as $RX(\pi/2)$ and the Hadamard gate, the improvement from coherent correction is negligible. Their approximately linear response $\delta g(V)$ causes positive and negative fluctuations in $V(t)$ to largely cancel in the Hamiltonian domain, so the net coherent distortion is very small.

Overall, these results show that coherent-error correction is effective only when the voltage-to-Hamiltonian mapping is sufficiently nonlinear and the RTN switching time is short compared to the gate duration.

\subsection*{Summary of Gate-Level Results}

The response of all three gates to voltage-level imperfections is well captured by two simple quadratic models. For static miscalibration, the infidelity follows
\[
    r \approx c_a \sigma_a^{2} + c_b \sigma_b^{2},
\]
with coefficients listed in Table~\ref{tab:miscalibration_quad_fit}. Comparing these values shows that the ESR-driven single-qubit gates (RX and Hadamard) are substantially less sensitive to both scaling and offset errors than the exchange-mediated two-qubit gate $\sqrt{\mathrm{SWAP}}$.

For dynamical charge noise modeled as RTN, the infidelity scales as
\[ 
r \approx a(\tau) A^{2},
\]
with coefficients $a(\tau)$ given in Table~\ref{tab:rtn_fit_three}. Here, $a(\tau)$ is smallest for $RX(\pi/2)$ and largest for $\sqrt{\mathrm{SWAP}}$ across all switching times, establishing a consistent hierarchy of robustness. As the RTN switching time increases, $a(\tau)$ for a given gate saturates, reflecting the crossover to a quasi-static noise regime.

Taken together, these fitted models provide a unified quantitative picture of how different spin-qubit gates respond to static miscalibration and dynamical charge noise: the single-qubit gates are least sensitive to voltage-level imperfections, while the $\sqrt{\mathrm{SWAP}}$ gate is the most sensitive. In the next section we apply the same voltage-noise models at the circuit level in a VQE simulation and compare the resulting algorithm-level error trends with these gate-level susceptibilities.

\section{Noise Analysis in a Variational Quantum Eigensolver}
\label{sec:VQE}
\subsection{Variational Quantum Eigensolver Principle}\label{sec:level2}

\begin{figure*}[htbp]
    \centering
    \includegraphics[width=0.9\textwidth]{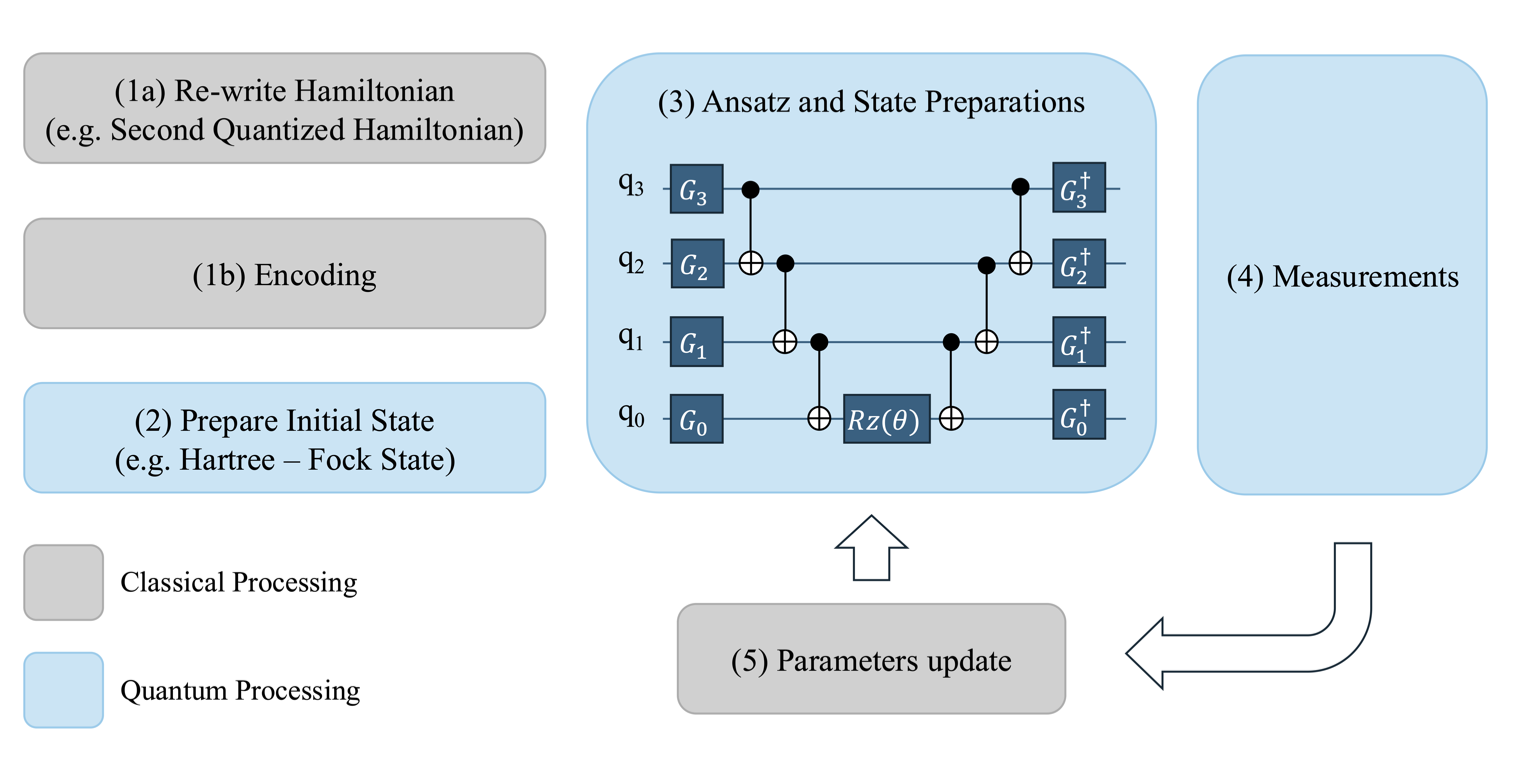}
    \caption{VQE pipeline: (1) Hamiltonian reformulation; (2) initialization, often using the Hartree--Fock state; (3) ansatz preparation with a curated quantum circuit; (4) measurements; (5) classical parameter optimization.}
    \label{fig:vqe_pipeline}
\end{figure*}

Building on the gate-level error characterization above, we now examine how these control imperfections affect the performance of a variational quantum eigensolver. VQE is a hybrid quantum--classical algorithm designed to approximate the ground-state energy of quantum systems. Its workflow is summarized in Fig.~\ref{fig:vqe_pipeline}. The procedure begins with a reformulation of the system Hamiltonian, followed by an initialization in a reference state, such as the Hartree--Fock state. A parameterized ansatz is then applied to construct trial wavefunctions, 
\[
    \ket{\phi(\theta)} = U(\theta)\ket{\phi_0}, 
\]
where $U(\theta)$ is realized by an appropriate quantum circuit.
The expectation value of the Hamiltonian is measured on a quantum processor,
\[
    C(\theta) = \bra{\phi(\theta)} H \ket{\phi(\theta)},
\]
and a classical optimizer updates the parameters $\theta$ to iteratively minimize $C(\theta)$. The process is repeated until convergence, yielding an approximate ground state and corresponding energy,
\[
    E_0 \approx \min_{\theta} C(\theta).
\]

Further details on the Hamiltonian construction, fermion-to-qubit mappings, and ansatz choices are provided in Appendix~\ref{appendix:vqe}.

\subsection{Hydrogen molecule and ansatz}
\label{subsec:H2_ansatz}

\begin{figure}[htbp]
    \centering
    \includegraphics[width=0.25\textwidth]{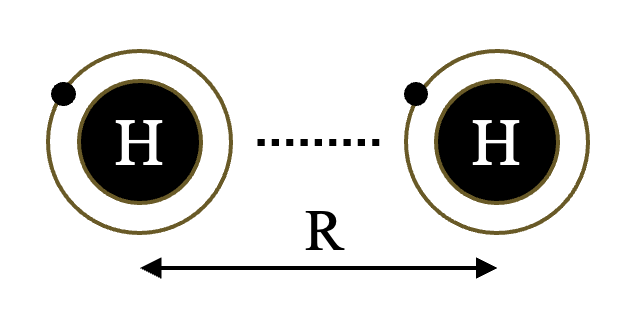}
    \caption{Schematic of the hydrogen molecule with nuclei separated by distance $R$.}
    \label{fig:h2_molecule}
\end{figure}

The hydrogen molecule $\mathrm{H}_2$ (Fig.~\ref{fig:h2_molecule}) is the simplest neutral molecule, consisting of two hydrogen atoms. Our goal is to compute its ground-state energy as a function of the internuclear distance $R$ using the VQE framework described above.

Within the minimal STO-3G basis, the electronic-structure problem is mapped to a 4-qubit Hamiltonian by applying the Jordan--Wigner (JW) transformation to the fermionic Hamiltonian:
\begin{align}
H = &\, h_0 I + h_1 Z_0 + h_2 Z_1 + h_3 Z_2 + h_4 Z_3 + h_5 Z_0 Z_1 \notag \\
    &  + h_6 Z_0 Z_2+ h_7 Z_1 Z_2 + h_8 Z_0 Z_3 + h_9 Z_1 Z_3 \notag \\
    & + h_{10} Z_2 Z_3 + h_{11} Y_0 Y_1 X_2 X_3 + h_{12} X_0 Y_1 Y_2 X_3 \notag \\
    & + h_{13} Y_0 X_1 X_2 Y_3 + h_{14} X_0 X_1 Y_2 Y_3 ,
    \label{eq:Hamiltonian}
\end{align}
where the coefficients $h_i$ depend on $R$ and are obtained from electronic-structure calculations. In our study, these coefficients are taken from restricted Hartree--Fock calculations in the STO-3G basis and a subsequent JW mapping, performed using the open-source quantum chemistry package Psi4~\cite{Smith2020PSI4}. Their numerical values are listed in Appendix~\ref{appendix:vqe_coeff}.

The reference state for our ansatz is
\[
    \ket{\psi_0} = \ket{0011},
\]
corresponding to the Hartree--Fock configuration of two occupied and two virtual canonical orbitals.

\begin{figure}[htbp]
    \centering
    \includegraphics[width=\linewidth]{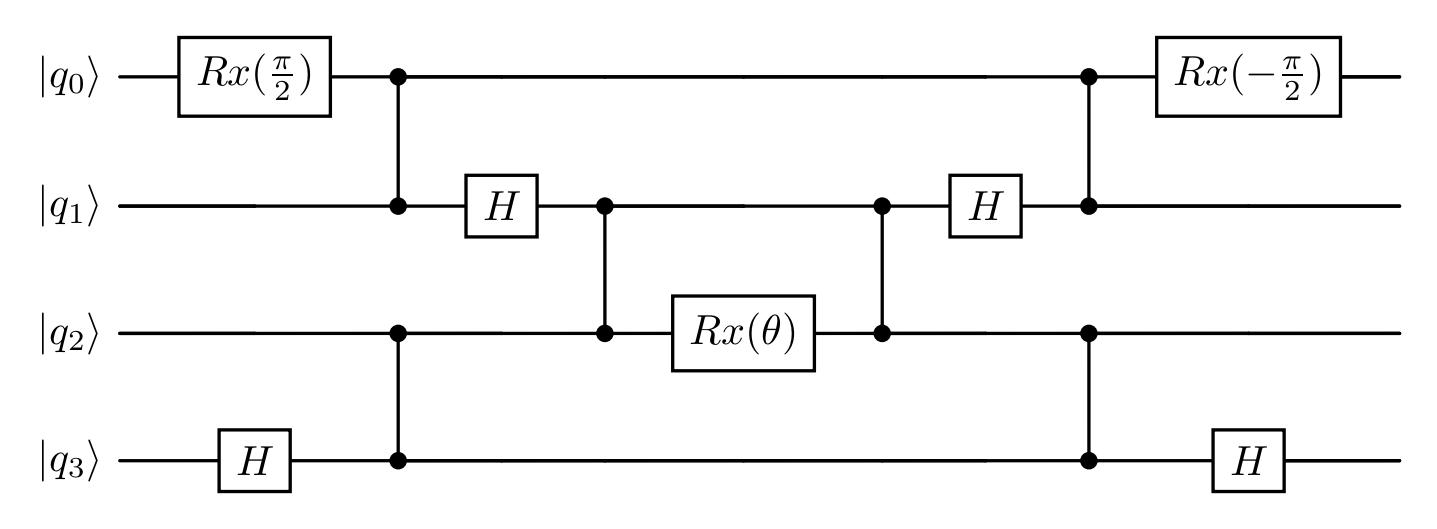}
    \caption{Circuit for the UCC ansatz in Eq.~\eqref{eq:ucc-final-main} used in the VQE simulations of the $\mathrm{H}_2$ ground-state energy.}
    \label{fig:ucc_circuit}
\end{figure}

For this system we use a single-parameter unitary coupled-cluster (UCC) doubles ansatz. In second-quantized form it contains one double-excitation operator, which after JW mapping and restriction to the relevant excitation subspace reduces to a single four-qubit Pauli string. The resulting qubit ansatz can be written as
\begin{equation}
    U(\theta) = \exp\!\left( -i\, \theta\, X_3 X_2 X_1 Y_0 \right),
    \label{eq:ucc-final-main}
\end{equation}
where $\theta$ is the single variational parameter. The corresponding circuit, compiled into single-qubit rotations and CZ gates compatible with our spin-qubit architecture, is shown in Fig.~\ref{fig:ucc_circuit}. A derivation of Eq.~\eqref{eq:ucc-final-main} from the underlying fermionic double excitation and the associated circuit construction are given in Appendix~\ref{appendix:vqe}.

\begin{figure*}[htbp]
    \centering
    \begin{minipage}{0.45\textwidth}
        \centering
        \makebox[0pt][l]{\raisebox{22.0\height}{\textbf{(a)}}}%
        \includegraphics[width=\linewidth]{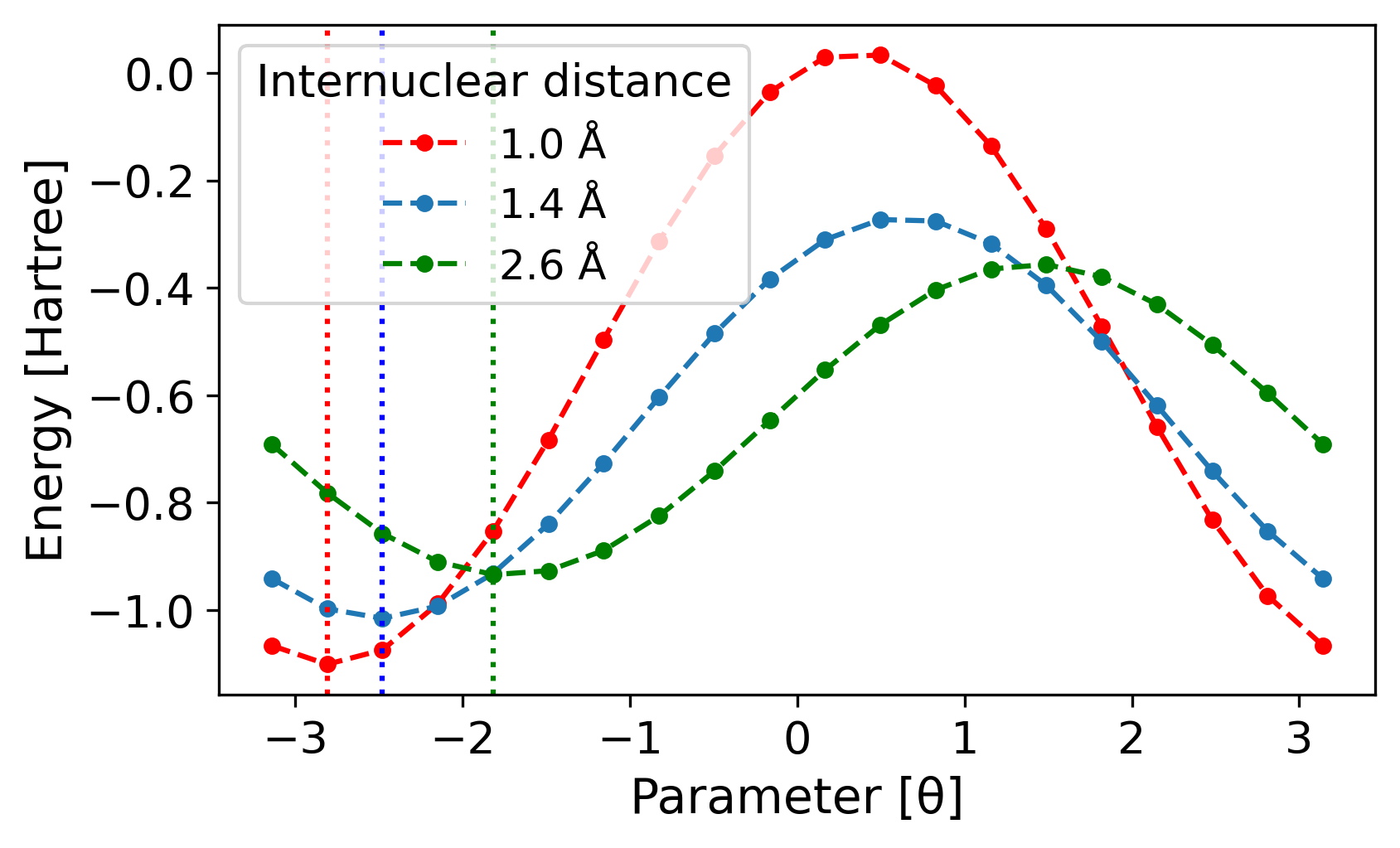}
        \label{fig:vqe_optimization}
    \end{minipage}
    \hfill
    \begin{minipage}{0.45\textwidth}
        \centering
        \makebox[0pt][l]{\raisebox{22.0\height}{\textbf{(b)}}}%
        \includegraphics[width=\linewidth]{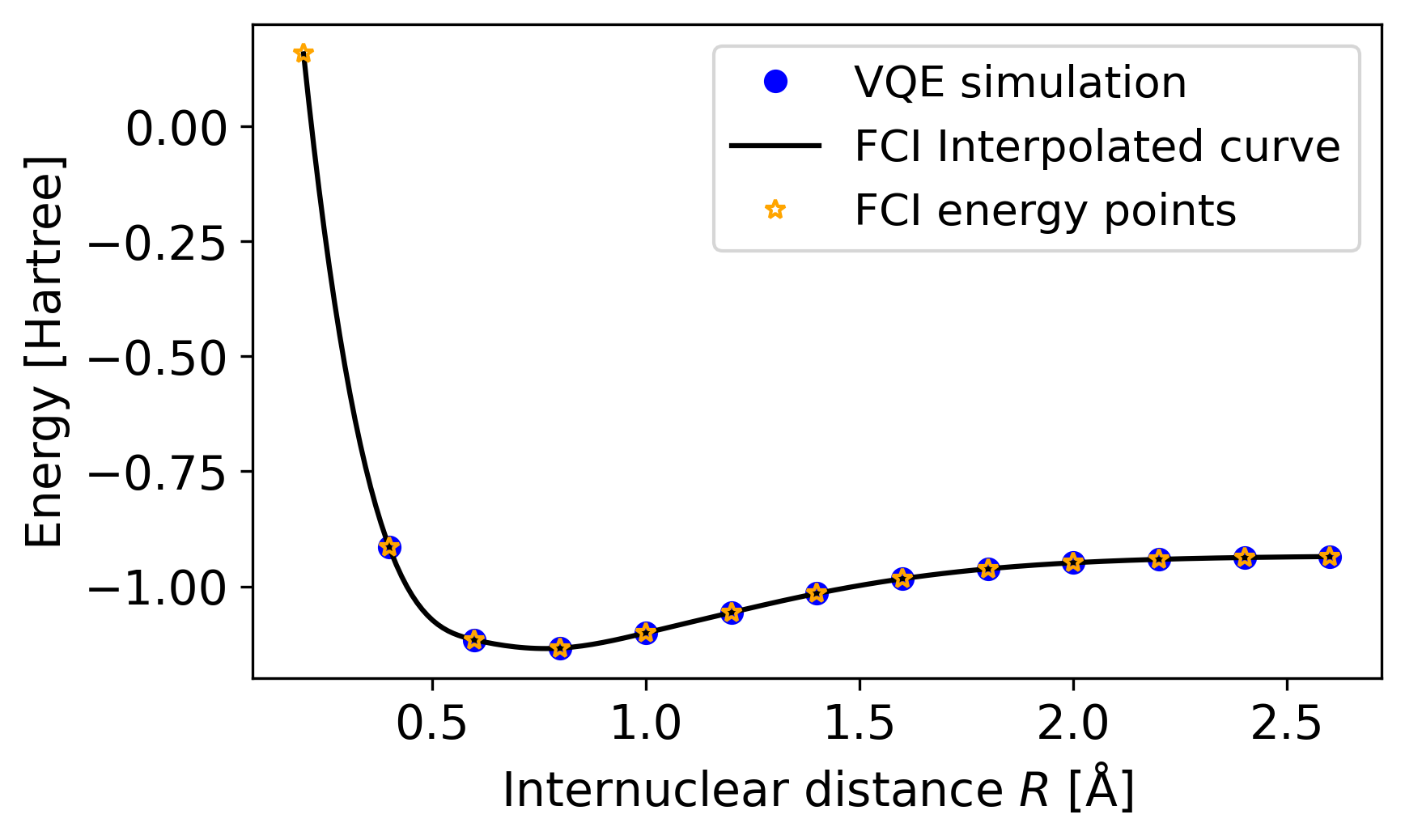}
        \label{fig:vqe_energy_curve}
    \end{minipage}
    \caption{VQE simulation results. (a) Energy landscape of the variational optimization for selected internuclear distances $R$. (b) Ground-state energy curve from VQE simulations (blue dots), compared with exact FCI reference values (black line).}
    \label{fig:vqe_results}
\end{figure*}

Using this ansatz, we perform VQE simulations for various bond lengths $R$ and obtain optimized ground-state energies. Figure~\ref{fig:vqe_results}(a) shows the VQE energy landscape as a function of the variational parameter $\theta$ for three representative internuclear distances. Each curve exhibits a well-defined minimum, corresponding to the optimized ground-state energy for that geometry. The vertical dashed lines mark the optimal values of $\theta$, illustrating that the VQE procedure reliably identifies the minimum across different bond lengths. Figure~\ref{fig:vqe_results}(b) displays the resulting energy curve obtained by extracting the minimum energy from landscapes of this form over a range of $R$. The VQE energies (blue dots) reproduce the exact full configuration interaction (FCI) curve (black line) to within a small numerical error, indicating that the ansatz and optimization procedure are sufficient to accurately capture the ground-state energy of $\mathrm{H}_2$ at the level required for this work. As noted in Sec.~\ref{sec:process_fidelity}, all VQE simulations assume ideal spin-state readout. 

\subsection{Impact of Noise on Ground-State Energy Estimation}

\subsubsection{VQE with Voltage Miscalibration}

 Building upon our analysis of gate fidelity under voltage miscalibration, we now investigate its cumulative impact on a complete quantum algorithm. In Sec.~\ref{sec:single_gate}, we showed that different quantum gates display distinct sensitivities to two basic forms of voltage distortion: a scaling factor $a$, sampled from a normal distribution with standard deviation $\sigma_a$, and an offset $b$, sampled from a normal distribution with standard deviation $\sigma_b$. These parameters model typical calibration imperfections in the control electronics. The differing responses of single- and two-qubit gates to such distortions motivate a more detailed examination of how miscalibration affects the performance of VQE circuits.

To this end, we simulate the VQE procedure for estimating the ground-state energy of the hydrogen molecule in the presence of stochastic voltage miscalibration. For each VQE trial, we generate seven independent pairs of miscalibration parameters $(a_i, b_i)$, with $a_i \sim \mathcal{N}(1, \sigma_a^2)$ and $b_i \sim \mathcal{N}(0, \sigma_b^2)$, and apply them separately to all seven voltage electrodes in the device. These channels comprise four plunger gate voltages and three tunnel barrier voltages in our model of a four-dot linear array. This setup captures the realistic scenario in which calibration errors vary independently across different electrodes in the hardware.

\begin{figure*}[t]
    \centering
    \begin{minipage}{0.45\textwidth}
        \centering
        \includegraphics[width=\linewidth]{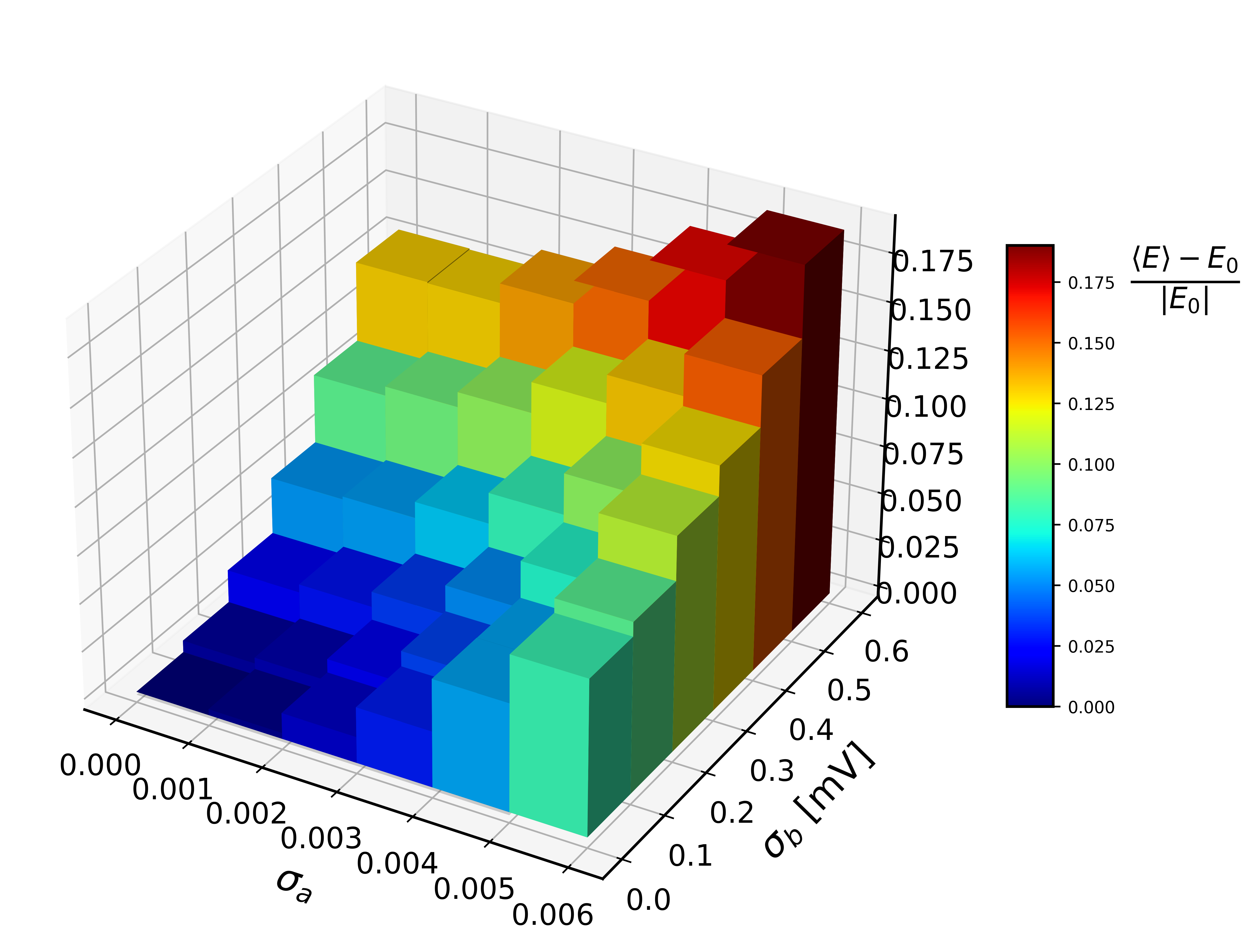}
        \par\smallskip\textbf{(a)} Average energy $\langle E \rangle$
        \label{fig:main_a}
    \end{minipage}
    \hfill
    \begin{minipage}{0.45\textwidth}
        \centering
        \includegraphics[width=\linewidth]{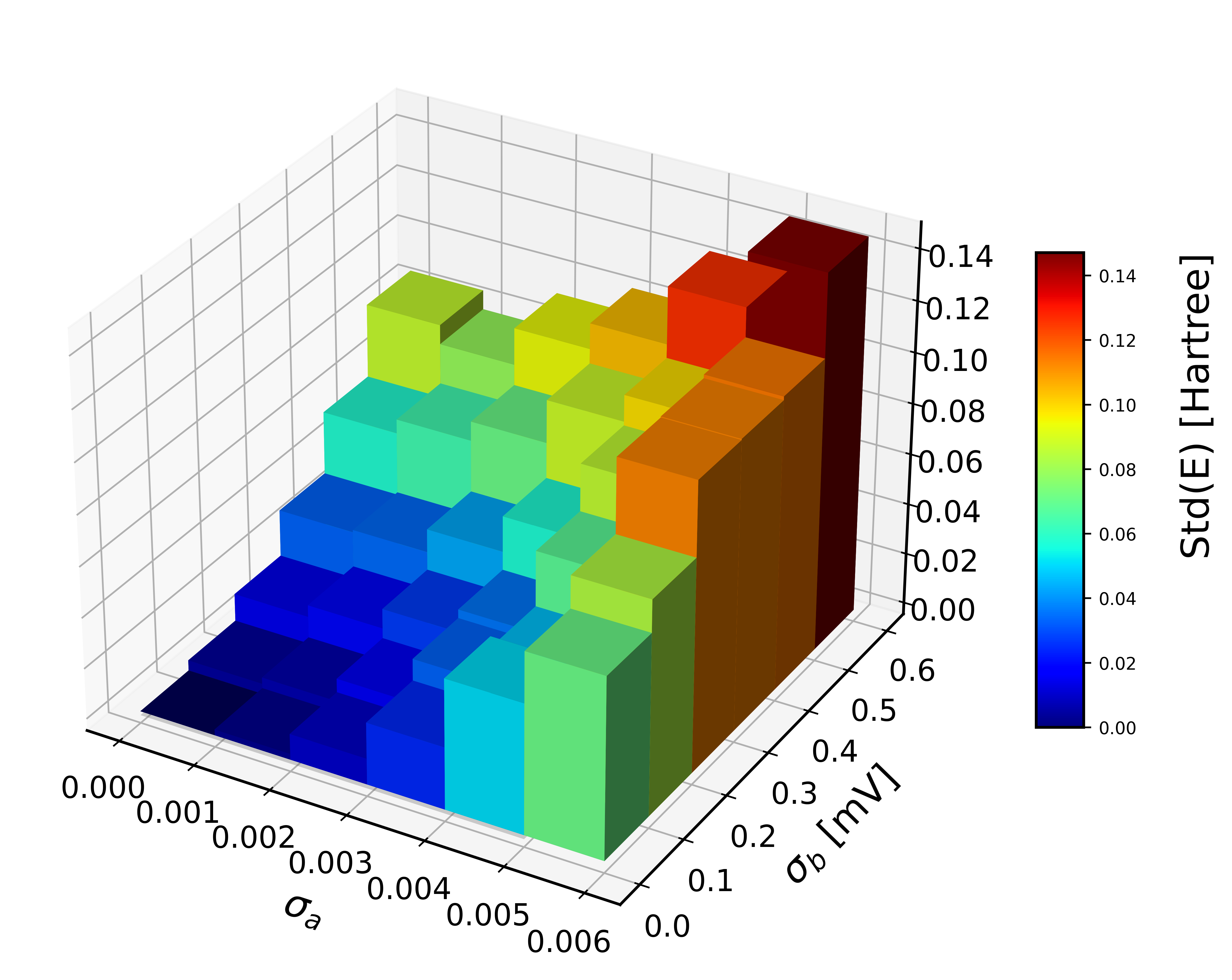}
        \par\smallskip\textbf{(b)} Standard deviation $\sigma_E$
        \label{fig:main_b}
    \end{minipage}

    \vskip\baselineskip

    \begin{minipage}{0.45\textwidth}
        \centering
        \includegraphics[width=\linewidth]{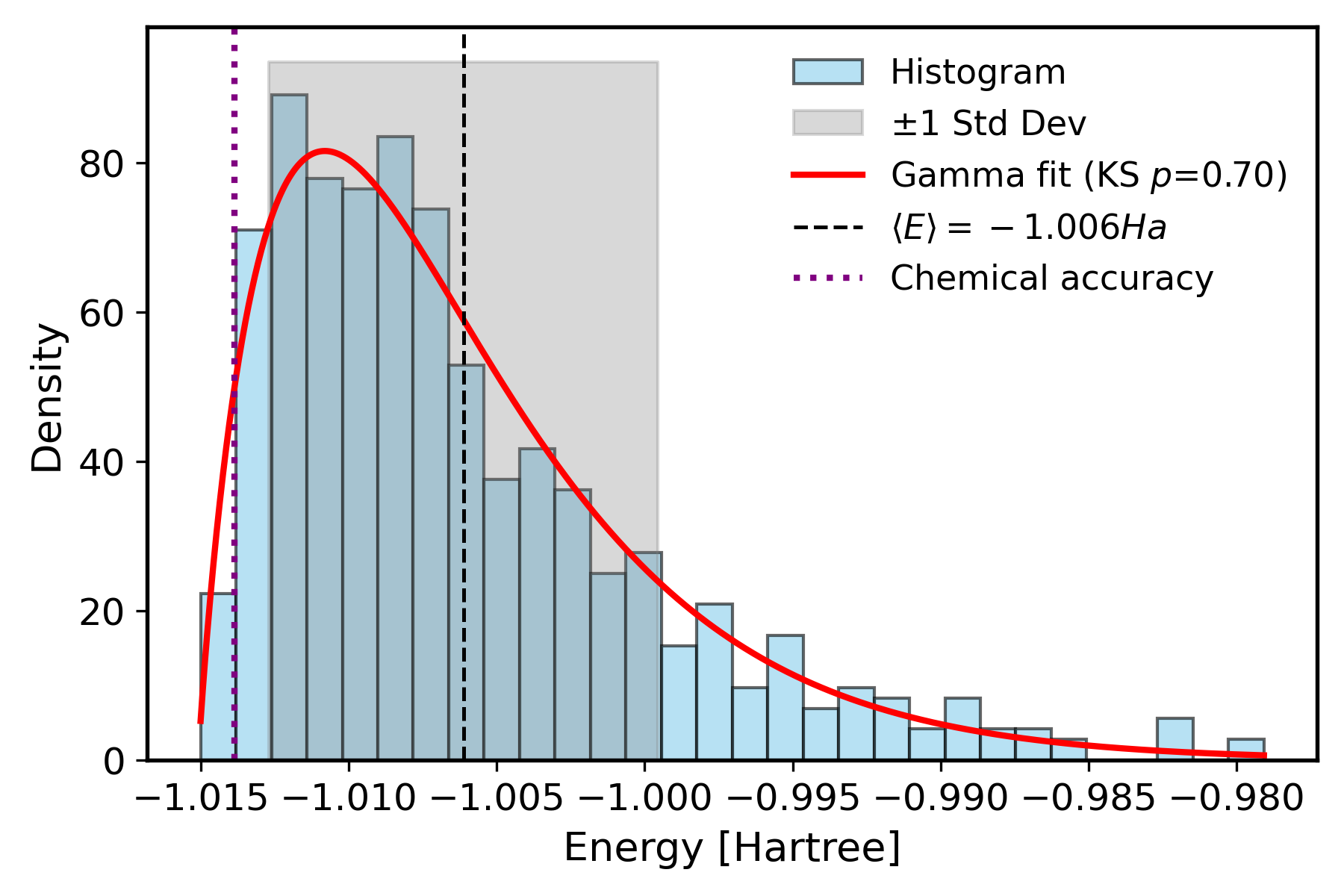}
        \par\smallskip\textbf{(c)} Energy distribution histogram
        \label{fig:main_c}
    \end{minipage}
    \hfill
    \begin{minipage}{0.48\textwidth}
        \centering
        \includegraphics[width=\linewidth]{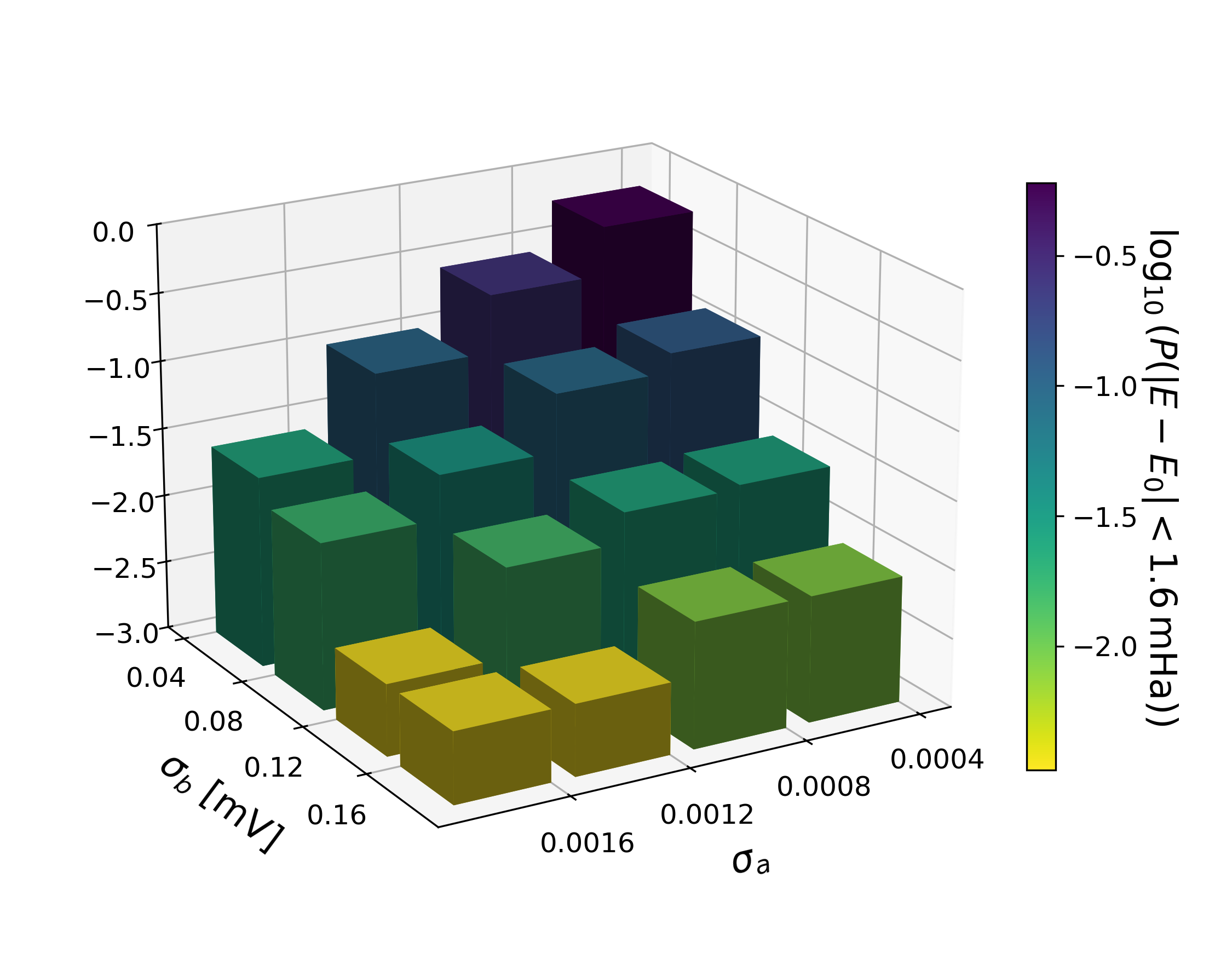}
        \par\smallskip\textbf{(d)} Chemical accuracy probability
        \label{fig:main_d}
    \end{minipage}

   \caption{Effect of static voltage miscalibration on VQE ground-state energy estimation, based on 600 independent VQE trials for each miscalibration setting.
Panels (a) and (b) summarize the mean and standard deviation of the estimated energies as functions of the scaling and offset miscalibration strengths $(\sigma_a,\sigma_b)$. Panel~(c) shows the distribution of VQE energy estimates at a representative miscalibration point $(\sigma_a = 0.001,\ \sigma_b = 0.1~\mathrm{mV})$, together with a fitted Gamma distribution. Vertical lines indicate the sample mean, one standard deviation, and the chemical accuracy threshold. The reported Kolmogorov--Smirnov (KS) $p$-value quantifies the goodness of the fit. Panel~(d) reports the logarithm of the probability of achieving chemical accuracy,
$\log_{10} P(|E - E_0| < 1.6~\mathrm{mHa})$,
computed as the fraction of VQE trials whose energies fall within the chemical-accuracy window. Results are shown for $\mathrm{H}_2$ at $R = 1.4~\mathrm{\AA}$.
}
    \label{fig:main}
\end{figure*}

As shown in Fig.~\ref{fig:main}(a), static voltage miscalibration induces a systematic upward bias in the VQE ground-state energy estimate. The average energy $\langle E \rangle$ increases smoothly with both the scaling mismatch $\sigma_a$ and the offset mismatch $\sigma_b$, without exhibiting sharp threshold behavior over the explored parameter range.
This gradual degradation suggests that the impact of miscalibration on the mean VQE performance can be characterized in terms of calibration tolerances on $(\sigma_a,\sigma_b)$, rather than a sudden failure beyond a critical noise level.

Figure~\ref{fig:main}(b) shows that the standard deviation $\sigma_E$ of the VQE energy estimates also increases with increasing miscalibration strength. To examine the full distribution beyond its first two moments, Fig.~\ref{fig:main}(c) displays the energy histogram  obtained at a representative miscalibration point $(\sigma_a = 0.001,\ \sigma_b = 0.1~\mathrm{mV})$. The distribution is right-skewed and is consistent with a Gamma distribution, reflecting the variational structure of VQE: energy estimates are bounded from below by the exact ground-state energy $E_0$, but may overshoot to substantially higher values. As a result, miscalibration gives rise to rare but large upward fluctuations rather than a symmetric broadening around $E_0$.

Figure~\ref{fig:main}(d) reports the probability of achieving chemical accuracy, $P(|E - E_0| < 1.6~\mathrm{mHa})$, across the $(\sigma_a,\sigma_b)$ parameter space. This probability directly answers the experimentally relevant question of how often a VQE run returns an energy within an acceptable tolerance, and
thereby defines a calibration `tolerance window' for $(\sigma_a,\sigma_b)$ that yields satisfactory algorithm-level performance. As miscalibration increases, the success probability decreases due to the combined effect of a systematic upward bias and an increasingly heavy-tailed distribution of energy estimates.

When combined with the gate-level analysis in Sec.~\ref{sec:single_gate}, these trends suggest a consistent connection between circuit-level performance and individual gate sensitivities. In regions of small $\sigma_a$ and larger $\sigma_b$, the observed energy bias is consistent with the sensitivity of single-qubit ESR gates to voltage-induced $g$-factor offsets. As $\sigma_a$ increases, the more nonlinear voltage dependence of the effective interaction strength in CZ gates correlates with a faster growth in both the mean energy error and its variance. In this sense, the VQE energy error inherits the hierarchy of gate sensitivities identified at the single-gate level.

Overall, static voltage miscalibration induces both a systematic upward bias and a skewed, heavy-tailed distribution in VQE energy estimates. By reporting the chemical-accuracy success probability, Fig.~\ref{fig:main}(d) provides a concrete bridge between device-level calibration imperfections and an experimentally meaningful notion of algorithmic acceptability, enabling one to
specify practical tolerances on $(\sigma_a,\sigma_b)$.

\subsubsection{VQE with Random Telegraph Noise}

We now examine the effect of RTN on VQE performance. In our simulations, RTN is modeled as arising from multiple independent charge fluctuators distributed around the quantum-dot array. For each electrode, the effective gate-voltage noise is given by the averaged RTN signal in Eq.~\eqref{eq:rtn_sum}. Independent $\mathrm{RTN}_{\mathrm{avg}}(t)$ sequences are generated and applied to every plunger and tunnel gate, so that each control channel experiences its own realization of charge-noise-equivalent voltage fluctuations.

\begin{figure}[h]
    \centering
    \includegraphics[width=0.45\textwidth]{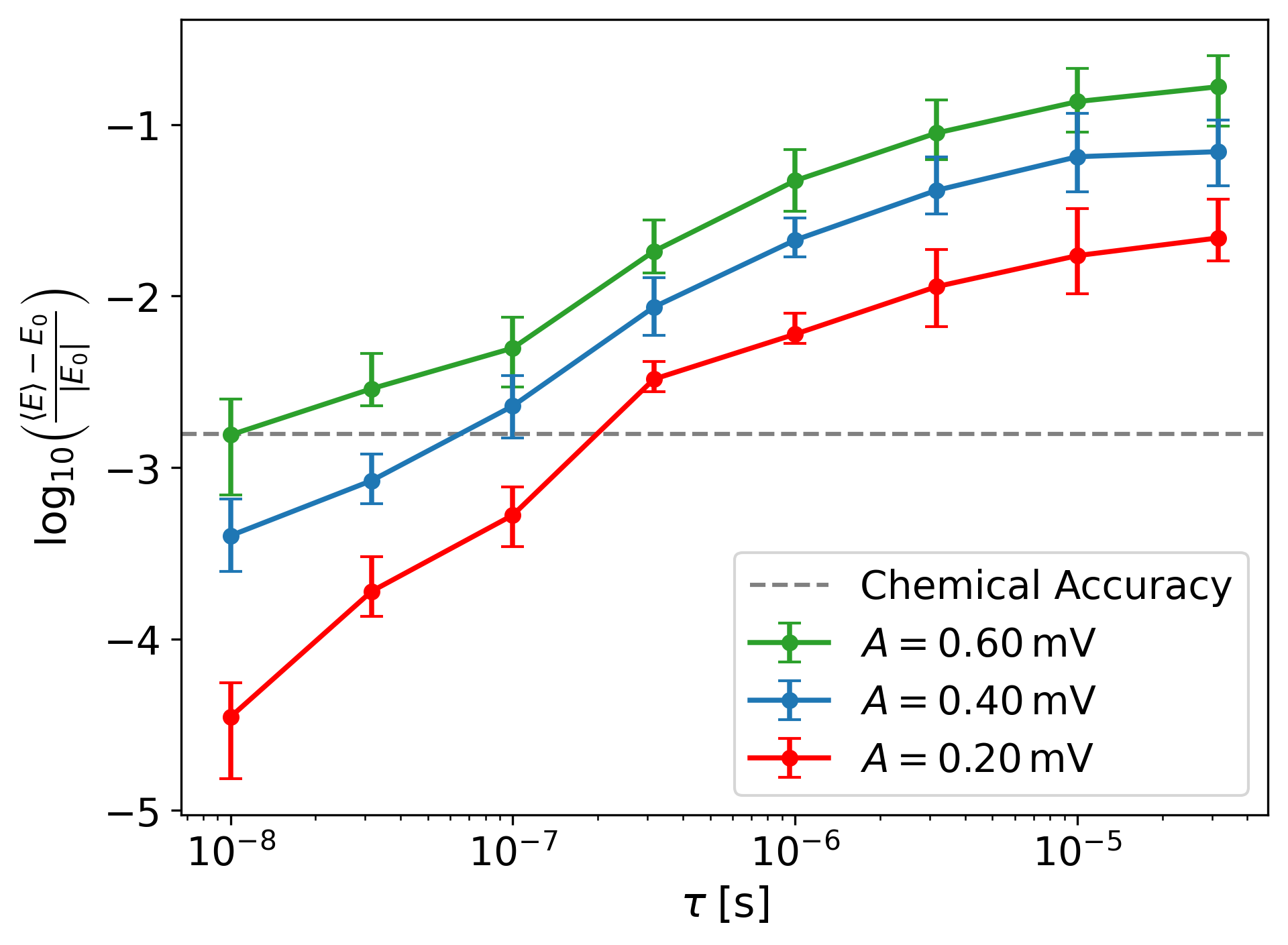}
    \caption{
    Relative error in the VQE-estimated ground-state energy of the hydrogen molecule as a function of the RTN switching time~$\tau$ for three noise amplitudes~$A$.
    The dashed line marks the chemical-accuracy threshold ($1.6\times10^{-3}$~Hartree).
    Error bars denote the standard deviation over 40 independent VQE runs. Results are shown for $\mathrm{H}_2$ at $R = 1.4~\mathrm{\AA}$.
    }
    \label{fig:vqe-rtn}
\end{figure}

The resulting relative error in the VQE-estimated ground-state energy of $\mathrm{H}_2$ as a function of RTN switching time $\tau$ and noise amplitude $A$ is shown in Fig.~\ref{fig:vqe-rtn}. We define a logarithmic relative energy error
\[
    \epsilon = \log_{10}\!\left(\frac{\langle E\rangle - E_0}{|E_0|}\right),
\]
where $\langle E\rangle$ is the VQE-estimated ground-state energy and $E_0$ is the exact (noise-free) ground-state energy.

In the absence of noise, the VQE converges rapidly on our simulator. When RTN is included, the optimizer requires more iterations on average, and the total runtime increases with the switching time $\tau$, reflecting the more rugged optimization landscape induced by slow noise.

As shown in Fig.~\ref{fig:vqe-rtn}, the VQE energy error exhibits a clear dependence on $\tau$. For small switching times (high-frequency RTN, $\tau \lesssim 10^{-7}\,\mathrm{s} = 10^{-2} t_g $), the error remains low and relatively insensitive to further decreases in $\tau$. This is consistent with dynamical averaging: rapid switching causes the RTN to average out over the duration of individual gate operations, so the variational algorithm retains high accuracy despite the presence of noise.

As $\tau$ enters the intermediate regime ($10^{-7}\,\mathrm{s} \lesssim \tau \lesssim t_g = 10^{-5}\,\mathrm{s}$), the relative error grows most rapidly. In this range, the noise is neither fast enough to average out over individual gate operations nor slow enough to be treated as a quasistatic offset. Instead, each RTN trajectory produces voltage fluctuations that persist over several consecutive gates before switching, making the accumulated VQE phase maximally sensitive to the underlying noise. This partially correlated regime leads to a strong, smooth increase of the energy error with~$\tau$, and represents the switching times where VQE is most strongly impacted by dynamical charge noise.

For large $\tau$ ($\tau \gtrsim 10^{-5}\,\mathrm{s}$), the error saturates, consistent with the quasistatic limit in which each VQE run experiences an approximately fixed random voltage offset. Different RTN realizations then act like draws from a static miscalibration distribution, and further increases in $\tau$ have little additional effect on the average error.

While RTN provides a convenient way to isolate the role of a single switching time, realistic charge noise in semiconductor devices is well approximated by a $1/f$-like spectrum. To benchmark our RTN-based conclusions against this more realistic case, we repeated the VQE simulations using synthesized $1/f$ gate-voltage noise with the same rms amplitude $A$ as in Fig.~\ref{fig:vqe-rtn}. The $1/f$ traces were generated by combining RTN components spanning switching times $\tau \in [10^{-8},10^{-4}]~\mathrm{s}$ (i.e., covering the full range explored in Fig.~\ref{fig:vqe-rtn}),
so that the resulting spectrum contains fluctuations on timescales both much faster and much slower than the gate duration.

Despite the broader spectral content, we find that the VQE energy bias induced by $1/f$ noise at a given amplitude is quantitatively close to that produced by RTN with an intermediate switching time, approximately $\tau \sim 10^{-6}~\mathrm{s}$, across the amplitudes considered. This indicates that, although $1/f$ noise contains very slow components, its net impact on the VQE objective is well captured by an effective RTN switching time in the partially correlated regime, providing a useful mapping between the two models.

To complement the trend analysis in Fig.~\ref{fig:vqe-rtn}, we also examine two representative operating points: fast-switching RTN ($\tau = 2\times 10^{-8}$~s) and quasistatic RTN ($\tau = 2\times 10^{-5}$~s), both at fixed amplitude $A = 0.4$~mV. For each condition, the VQE ground-state energy is averaged over 200 independent RTN realizations. We then apply the coherent-error correction procedure of Sec.~\ref{sec:coherent_extraction} to all six CZ gates in the VQE circuit and recompute the energies over a new ensemble of 200 noisy runs. The results are summarized in Table~\ref{tab:vqe_correction}.

\begin{table}[H]
\centering
\caption{VQE-estimated ground-state energies for two representative RTN conditions at fixed amplitude 
$A = 0.4$ mV. Each value is averaged over 200 independent RTN realizations ($E_0 = -1.015467 ~Ha$).}
\label{tab:vqe_correction}
\begin{tabular}{ccc}
\toprule
Condition & Energy (uncorrected) & Energy (corrected) \\
\midrule
$\tau = 2\times 10^{-8}$ s & $-1.015072$ & $-1.015091$ \\
$\tau = 2\times 10^{-5}$ s & $-0.945909$ & $-0.951635$ \\
\bottomrule
\end{tabular}
\end{table}

In both cases, the coherent correction yields a systematic improvement in the estimated ground-state energy, with a larger absolute shift for the quasistatic RTN case where the bias is more pronounced. These results are consistent with Fig.~\ref{fig:vqe-rtn}: fast RTN has only a modest impact on VQE performance, while slow, quasistatic RTN produces a stronger upward bias that can be partially mitigated by coherent post-correction but not fully removed.

The transition between high-frequency and quasi-static regimes underscores the dual nature of RTN as both a stochastic and systematic error source. In the high-frequency regime, the RTN-induced errors are effectively stochastic, introducing random fluctuations in gate fidelities. In contrast, in the quasi-static regime, RTN acts like a slowly varying miscalibration, shifting the variational landscape and degrading convergence.


Taken together, these results show that, for a given amplitude $A$, VQE performance on our quantum-dot architecture is most vulnerable to slow, low-frequency RTN. As seen in Fig.~\ref{fig:vqe-rtn}, the energy error grows rapidly as $\tau$ enters the intermediate and quasistatic regimes and, for the larger amplitudes considered, drives the VQE result beyond chemical accuracy ($1.6\times 10^{-3}$~Hartree). Using this threshold as a benchmark, the VQE energy remains within an acceptable range either when the noise is sufficiently fast ($\tau \lesssim 10^{-7}$--$10^{-6}$~s) or when the RTN amplitude is small (e.g., $A \approx 0.20~\mathrm{mV}$).

The coherent-error correction study refines this picture. At both fast and slow RTN operating points, the correction yields a small but systematic improvement in the VQE energy (Table~\ref{tab:vqe_correction}), demonstrating that a coherent component is present across all $\tau$. However, in the slow-noise regime this improvement accounts for only a modest fraction of the total error, indicating that quasi-static voltage offsets generate a predominantly incoherent bias that cannot be undone by a single unitary. Thus, while coherent correction is beneficial, achieving robust VQE performance ultimately requires hardware-level suppression of slow charge fluctuations.

\subsection*{Summary of algorithm-level results}

Incorporating our hardware-informed noise models into a chemically motivated VQE for the $\mathrm{H}_2$ ground state shows how voltage-level imperfections translate into algorithmic error. Static gate-voltage miscalibration yields a systematic upward bias in the estimated ground-state energy and a strongly right-skewed distribution across repeated runs, well captured by a Gamma-like form with a hard lower bound at the exact energy. Using the chemical-accuracy target of $1.6\times 10^{-3}$~Hartree as a benchmark, we find that achieving chemically accurate estimates with high probability ($>60\%$) requires miscalibration levels
\[
    \sigma_a \lesssim 4\times10^{-4}, \qquad \sigma_b \lesssim 0.04~\mathrm{mV},
\]
for scaling and offset noise on the gate-electrode voltages, respectively.

For RTN, the VQE error is controlled primarily by $\tau/t_g$: fast noise averages out, intermediate noise produces the largest growth in error, and slow noise approaches the static-miscalibration limit.

To assess whether the gate-level quadratic error laws persist at the circuit level, we also fitted the full VQE circuit infidelity under RTN to $1-F_{\mathrm{circ}} = cA^n$. For intermediate and slow RTN, the fitted exponent remains close to $n=2$, consistent with the weak-noise quadratic scaling observed for individual gates. For very fast RTN, however, the apparent exponent increases, indicating that rapid temporal averaging suppresses the leading error contribution and makes higher-order terms in the voltage-to-Hamiltonian mapping more visible. Thus, the quadratic fits should be interpreted as weak-noise phenomenological susceptibility models within the explored parameter regime, rather than universal asymptotic scaling laws for arbitrary gate speeds, circuit depths, or noise spectra.

\section{Discussion and Outlook}
\label{sec:discussion}

The impact of voltage-level noise on silicon spin qubits depends sensitively on how gate voltages map onto the effective spin Hamiltonian. In the Si/SiO$_2$ MOS devices considered here, Stark-shifted $g$-factors vary almost linearly with gate voltage over the operating range, while the exchange coupling $J(V)$ is steep and close to exponential. Because this mapping is obtained from full device-level electrostatics and used directly in the pulse-design routine, cross-talk between electrodes is automatically built into the trajectories $g_i(\vec{V}(t))$ and $J_i(\vec{V}(t))$. As a result, a given pattern of static miscalibration or charge-noise--equivalent voltage fluctuations produces relatively modest distortions of local Zeeman terms but can drive large fractional changes in exchange. This asymmetry underlies the hierarchy observed throughout the paper: ESR-driven single-qubit rotations tolerate moderate voltage error, whereas exchange- and CZ-based entangling gates amplify the same disturbances into much larger coherent over- and under-rotations. In practice, voltage-stability requirements should therefore be set in a gate-dependent way: what is acceptable for single-qubit gates can be insufficient for high-fidelity entangling operations on the same hardware.\\
\indent The quadratic-response model for static miscalibration,
$r \approx c_a \sigma_a^2 + c_b \sigma_b^2$, provides a compact way to encode this difference. The fitted coefficients $c_a$ and $c_b$ act as susceptibility metrics that translate specifications on amplitude scaling and DC offsets in the control stack into expected gate infidelities. Combined with the VQE results in Sec.~\ref{sec:VQE}, they support a simple conclusion: tolerances for chemically accurate VQE are set predominantly by the entangling gates, which impose significantly tighter constraints than ESR-driven single-qubit rotations.\\
\indent A second theme is the role of temporal correlations in the voltage noise. Modeling charge noise as RTN shows that the key parameter is the dimensionless ratio $\tau/t_g$ between the switching time and the gate time. For $\tau \ll t_g$, fluctuations average over each gate and behave as a weak perturbation: gate infidelities scale as $A^2$ with small prefactors, and the induced VQE energy bias remains modest. As $\tau$ enters the intermediate regime ($\tau \sim 0.01$--$1\,t_g$), noise remains correlated across several consecutive gates and is most damaging: gate errors grow rapidly and the VQE output becomes strongly history-dependent. In the quasistatic limit $\tau \gtrsim t_g$, each circuit run samples an effectively static offset, so RTN becomes indistinguishable from stochastic miscalibration at the level of averaged channels. The QPT-based coherence-angle analysis sharpens this picture: increasing $\tau$ drives a crossover from channels with a significant coherent component to ones dominated by incoherent, shot-to-shot variation that cannot be removed by any single compensating unitary.\\
\indent At the algorithm level, static miscalibration produces a strongly right-skewed distribution of VQE energy estimates, well described by a Gamma-like law with a hard lower bound at the true ground-state energy. In this regime the sample mean is a fragile estimator, as a small number of high-energy outliers can bias it far more than they affect the mode or median. This suggests that noisy near-term experiments could benefit from estimators that account for the full energy distribution, such as shifted-distribution fits or mode-seeking procedures. Slow RTN leads to a similar behavior: when $\tau \gtrsim t_g$, each circuit execution effectively samples a different static voltage offset, making the resulting energy distribution resemble that produced by static miscalibration.\\
\indent Our noise model is deliberately focused on voltage-level disturbances that are charge-noise--equivalent: static miscalibration of the control electronics and stochastic fluctuations in gate voltages that map into $g$-factor and exchange noise. Several important silicon-specific mechanisms are not yet included. These include valley physics (device-dependent valley splittings, valley-orbital leakage, and hot spots), hyperfine noise from residual nuclear spins, spin--orbit-mediated relaxation and phonon processes beyond simple $T_1/T_2$ phenomenology, and leakage into higher orbital or multi-electron states. We also do not model the readout stack (spin-to-charge conversion, charge-sensor back-action, and measurement infidelity), nor devices employing micromagnets and electrically driven spin resonance (EDSR), where magnetic-field gradients and spin--orbit coupling introduce additional control channels and noise pathways. For microsecond-scale gates in isotopically enriched Si MOS devices, hyperfine-induced dephasing is substantially suppressed, while spin--orbit- and phonon-mediated relaxation processes are generally slower than the gate durations considered here under typical operating conditions. Available experimental evidence further suggests that low-frequency electrostatic fluctuations are among the most important decoherence mechanisms in these systems. Consequently, although the omitted mechanisms may contribute quantitatively to the overall error budget, they are not expected to alter the central conclusion of this work that low-frequency electrostatic fluctuations remain a dominant limitation for voltage-controlled Si MOS spin-qubit architectures. Nonetheless, extending the co-simulation pipeline to include valley dynamics, micromagnet-induced gradients, more realistic charge-trap ensembles capable of generating $1/f$-like noise spectra, and explicit readout models represents an important direction for future work.\\
\indent While all the numerical thresholds reported here are specific to the Si MOS global-ESR architecture considered in this work and should therefore be viewed as semi-quantitative design guidelines, the main qualitative conclusions are expected to be more broadly applicable. In particular, exchange-based gates remain substantially more voltage-sensitive than ESR-driven single-qubit gates, and the impact of stochastic charge noise is governed primarily by the ratio between the noise switching time and the gate duration.\\
\indent The coherent-error extraction based on QPT and Kraus operators should similarly be viewed as a diagnostic and an upper bound on what calibration could, in principle, achieve, not as a practical protocol. Full process tomography for every gate and bias point is unrealistic on large devices; in practice one would rely on partial characterization, interleaved benchmarking, or model-based calibration to approximate the dominant coherent error. Our analysis quantifies how much of the observed infidelity \emph{could} be removed by such coherent compensation, and where the remaining error is intrinsically incoherent and must be addressed by improved hardware, encoding, or error-mitigation strategies.\\
\indent These results point to several mitigation directions for spin-qubit control hardware and methods. On the hardware side, the inferred tolerance windows for $(\sigma_a,\sigma_b)$ and for RTN amplitudes and switching times provide concrete targets for voltage stability and drift control. Using representative $1/f$ charge-noise spectra together with a simulated lever arm, we estimate an effective RMS gate-voltage noise $\sigma_V \sim 25~\mu\mathrm{V}$, comparable to values reported for recent Si MOS and Si/SiGe devices, placing the simulated noise levels in a realistic regime. Suppressing low-frequency components of the electrostatic noise spectrum---through improved materials and interfaces, more stable bias circuitry, and calibration strategies that explicitly track slow drift---is therefore likely to yield disproportionate gains at the algorithm level. On the control side, robust pulse design for exchange-based gates (and the CZ gates built from them) can reduce sensitivity to amplitude and offset errors; for example, composite and shaped exchange pulses can be engineered to compensate the nonlinearity of $J(V)$~\cite{PhysRevA.110.L040602}. On the algorithm and post-processing side, estimators that exploit knowledge of the energy-distribution shape could partially mitigate the heavy-tailed error structure induced by miscalibration.\\
\indent Several extensions would strengthen the experimental relevance and generality of this approach. A natural next step is to calibrate the RTN and $1/f$ noise parameters---amplitudes, switching-rate distributions, and cross-electrode correlations---directly from measured spectra and charge-sensor data on specific devices. In particular, including spatial correlations in the noise is a further step towards more accurate modeling of charge noise. One could also repeat the present analysis for larger molecular systems and deeper ans\"atze, where many layers of entangling gates may alter the error budget. Incorporating realistic state-preparation and readout, as well as valley and EDSR physics, would enable one to extend the same co-simulation framework to a broader class of silicon spin-qubit architectures.\\
\indent Overall, the combination of hardware-specific modeling, gate-level process tomography, and algorithm-level VQE simulations developed here provides a template for analyzing voltage and charge noise in silicon spin qubits beyond abstract depolarizing models. By explicitly linking gate-electrode disturbances---both static miscalibration and charge-noise--equivalent fluctuations---to $g$-factor and exchange noise, and tracing their effects through individual gates into a chemically relevant VQE application, this framework helps identify which noise sources, gate types, and timescales are most critical to tackle on the path toward scalable, chemically accurate silicon spin-qubit processors.

\begin{acknowledgments}
This research was undertaken thanks in part to funding from the Canada First Research Excellence Fund (Transformative Quantum Technologies) and the Natural Sciences and Engineering Research Council (NSERC) of Canada.
\end{acknowledgments}

\appendix

\section{Effective Spin Hamiltonian}\label{appendix:spin_hamiltonian}

In the main text we describe a hardware--algorithm co-simulation framework in which voltage-dependent electrostatic parameters are mapped to an effective spin Hamiltonian governing the qubit dynamics. For completeness, we provide here the explicit form of the spin Hamiltonian used in our simulations, together with the definitions of all control terms and reference frames.

The simulator constructs an effective spin Hamiltonian incorporating voltage-tunable Zeeman and exchange interactions, expressed (in angular frequency units) as
\begin{multline}
    \label{eq:ham_w_dev_g_factor}
    \mathscr{H}(t)
    = \frac{\mu_B B_{z}}{\hbar} \sum_{j=1}^N \frac{g_j\bigl(\vec{V}(t), \vec{W}(t)\bigr) - g_{0}}{2}\, Z_j
    \\
    - \frac{\omega_\textsc{rf}(t) - \omega_\textsc{rf}^{(0)}}{2} \sum_{j=1}^N Z_j
    \\
    + \frac{\mu_B B_\textsc{rf}(t)}{\hbar}
      \left[
        \cos\phi(t)\, \sum_{j=1}^N X_j
        + \sin\phi(t)\, \sum_{j=1}^N Y_j
      \right]
    \\
    + \sum_{j=1}^{N-1}
      \frac{J_j\bigl(\vec{V}(t), \vec{W}(t)\bigr)}{4\hbar}\,
      \bigl(\vec{\sigma}_j \cdot \vec{\sigma}_{j+1}\bigr),
\end{multline}
where
\begin{itemize}
    \item $\mu_B$ is the Bohr magneton;
    \item $B_{z}$ and $B_\textsc{rf}(t)$ are the static Zeeman field and the envelope of the ESR field (in tesla), respectively;
    \item $\omega_\textsc{rf}(t) = 2\pi f_\textsc{rf}(t)$ and $\phi(t)$ are the time-dependent ESR frequency and phase;
    \item $J_j\bigl(\vec{V}(t), \vec{W}(t)\bigr)$ is the voltage-controlled exchange coupling between dots $j$ and $j+1$ (in joules);
    \item $\vec{\sigma}_j = \{X_j, Y_j, Z_j\}$ is the Pauli vector of the $j^{\mathrm{th}}$ spin;
    \item $g_j\bigl(\vec{V}(t), \vec{W}(t)\bigr)$ is the voltage-controlled in-plane $g$-factor of the $j^{\mathrm{th}}$ spin (the out-of-plane $g$-factor is fixed to 2).
\end{itemize}

Here, $g_0$ is the idling $g$-factor, defined as the value of $g_j$ at a voltage configuration $(\vec{V}_{idle}, \vec{W}_{idle})$ within the charge-stability region for singly occupied dots where all qubits share an identical $g$-factor. The corresponding idling ESR frequency is $\omega_\textsc{rf}^{(0)} = g_0 \mu_B B_z / \hbar$, which sets the rotating-frame reference for spin rotations.

\section{Details of the spin-qubit simulator}
\label{appendix:sim_details}

For completeness, we summarize here the device-level simulation pipeline used to obtain the voltage-dependent spin-Hamiltonian parameters employed in the main text.

\subsection{Electrostatics and single-particle states}

The starting point is a three-dimensional electrostatic model of the Si/SiO$_2$/metal stack, including all plunger, barrier, and tunnel gates, together with the global top gate and screening electrodes [Fig.~\ref{fig:device_single}]. For a grid of control-voltage settings $(\vec{V},\vec{W})$, the Poisson equation is solved using a finite-difference solver~\cite{Birner2007_nextnano}, yielding the electrostatic potential $\Phi(\mathbf{r};\vec{V},\vec{W})$ throughout the device. We then extract a two-dimensional horizontal slice of $\Phi$ at the depth of the electron gas beneath the Si/SiO$_2$ interface and interpolate these slices to obtain a smooth dependence
\[
    \Phi(x,y;\vec{V}) \equiv \Phi(\vec{V})
\]
on the gate-voltage vector.

To obtain localized dot states, we apply a masking procedure that isolates the confining potential for a given single dot or dot pair, and solve the 2D single-particle Schr\"odinger equation in that potential. The resulting ground-state orbitals are used both to define the qubit locations and to parameterize the effective spin and Hubbard models.

\subsection{Effective parameters and Hubbard model}

From the single-particle orbitals we extract local in-plane $g$-factors via a Stark-shift model, using the local electric field and confinement anisotropy, and interdot exchange couplings via Heitler--London or Hund--Mulliken approximations~\cite{Burkard_1999,Pedersen_2007}. Sampling these quantities over the voltage grid and fitting to smooth functional forms produces the effective dependencies $g_j(\vec{V})$ and $J_{j}(\vec{V})$ used in the main text and shown in Fig.~\ref{fig:dg_J_two_column}. In addition, the simulator extracts Hubbard parameters---on-site chemical potentials, tunnel couplings, and Coulomb repulsion terms---which enable cross-checks between spin-only and charge-based descriptions of the device.

\subsection{Spin Hamiltonian and pulse design}

The effective spin Hamiltonian used in the simulations is given in Eq.~\eqref{eq:ham_w_dev_g_factor} of Appendix~\ref{appendix:spin_hamiltonian}. Time-dependent control enters through the global ESR field $B_\textsc{rf}(t)$, its frequency and phase $(\omega_\textsc{rf}(t),\phi(t))$, and the local gate voltages $\vec{V}(t)$ that tune $g_j(\vec{V}(t))$ and $J_{j}(\vec{V}(t))$. Spin dynamics are simulated by numerically integrating the corresponding Lindblad master equation, with optional $T_1$ and $T_2$ processes when included.

Gate pulses are constructed in two stages. First, we choose a time-independent generator $H_0$ for the target gate (e.g., an $X$ rotation or a Heisenberg exchange gate) and a smooth shape function $S(t)$, and impose the time-ordered Hamiltonian
\[
    H(t) = H_0\,\frac{S(t)}{T},
\]
for a chosen duration $T$~\cite{10.1007/978-3-031-84869-8_13}. This determines the desired trajectories $g_j(t)$ and $J_{j}(t)$ that implement the target unitary with unit fidelity in the absence of noise. Second, we invert the fitted maps $g_j(\vec{V})$ and $J_{j}(\vec{V})$ to obtain the corresponding gate-voltage waveforms $\vec{V}(t)$ by solving a coupled system of ordinary differential equations for $V(S)$ and composing with $S(t)$. Because this inversion uses the full device model, capacitive cross-talk between electrodes is automatically incorporated into the resulting pulses.

The limited Stark-shift tunability of the $g$-factor in the modeled Si MOS architecture ($\Delta g/g_0 \sim 10^{-4}$~\cite{Veldhorst2014AddressableQubit}) sets single-qubit ESR gate times in the microsecond range for realistic ESR amplitudes. In the main text, we adopt a common duration $t_g = 10~\mu\mathrm{s}$ for both single- and two-qubit gates, corresponding to exchange energies in the neV range, and then superimpose static miscalibration and RTN-like voltage fluctuations on the resulting control waveforms to study their impact on gate and algorithm performance.

\section{Gate-Time Dependence of Exchange-Gate Fidelity Under RTN Voltage Noise}
\label{app:gate_time_scaling}

In the main text, all single- and two-qubit gates are simulated using a common gate duration $t_g = 10~\mu\mathrm{s}$. This choice is motivated by the global-ESR, Stark-shift-addressed Si MOS architecture considered in this work. However, many silicon spin-qubit experiments operate exchange gates on shorter timescales. To assess how our conclusions depend on gate duration, we additionally simulated the $\sqrt{\mathrm{SWAP}}$ gate for $t_g = 1~\mu\mathrm{s}$ and $t_g = 100~\mathrm{ns}$, while keeping the target gate operation fixed.

For each gate duration, the exchange pulse was adjusted to produce the same $\sqrt{\mathrm{SWAP}}$ operation. The RTN voltage-noise amplitude was fixed at $A = 1~\mathrm{mV}$, and the switching time was scaled proportionally with the gate duration,
\[
    \tau = 0.1\,t_g,
\]
so that all three cases probe the same relative noise-timescale regime. The resulting average gate infidelities are summarized in Table~\ref{tab:gate_time_scaling}.

\begin{table}[h]
\centering
\caption{Average $\sqrt{\mathrm{SWAP}}$ gate infidelity under charge-noise-equivalent RTN voltage noise for different gate durations. The RTN amplitude is fixed at $A = 1~\mathrm{mV}$ and the switching time is scaled as $\tau = 0.1t_g$.}
\label{tab:gate_time_scaling}
\begin{tabular}{cccc}
\toprule
Gate duration $t_g$ & RTN switching time $\tau$ & Infidelity $1-F$ \\
\midrule
$10~\mu\mathrm{s}$ & $1~\mu\mathrm{s}$ & $7.88\times10^{-4}$ \\
$1~\mu\mathrm{s}$ & $100~\mathrm{ns}$ & $8.18\times10^{-4}$ \\
$100~\mathrm{ns}$ & $10~\mathrm{ns}$ & $1.14\times10^{-3}$ \\
\bottomrule
\end{tabular}
\end{table}

The shorter exchange gates show a modest increase in infidelity. This is consistent with the fact that implementing the same exchange-mediated operation in a shorter time generally requires a larger exchange coupling, and therefore may move the device to an operating point where the nonlinear function $J(V)$ is more sensitive to voltage fluctuations. At the same time, a shorter gate duration reduces the time over which the noise is integrated. The observed fidelity degradation therefore reflects a competition between increased instantaneous exchange sensitivity and reduced temporal noise accumulation, and is consequently device- and operating-point dependent.

\begin{figure}[t]
\centering
\includegraphics[width=0.45\textwidth]{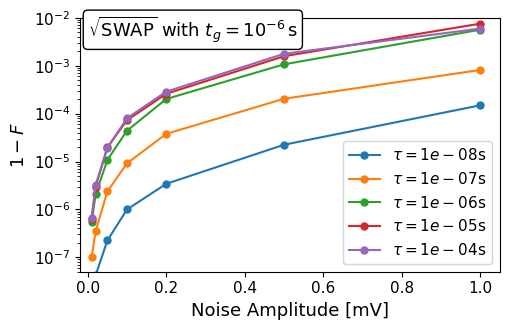}
\caption{Average $\sqrt{\mathrm{SWAP}}$ gate infidelity under charge-noise-equivalent RTN voltage noise for a gate duration $t_g = 1~\mu\mathrm{s}$. Different curves correspond to different RTN switching times $\tau$.}
\label{fig:gate_time_comparison}
\end{figure}

To further examine the role of switching time, we performed a full RTN-amplitude sweep for the $t_g = 1~\mu\mathrm{s}$ case at several values of $\tau$. The resulting average $\sqrt{\mathrm{SWAP}}$ gate fidelity is shown in Fig.~\ref{fig:gate_time_comparison}.

The same qualitative behavior observed in the main text is recovered for the shorter gate. In the fast-switching regime ($\tau \ll t_g$), RTN fluctuations are averaged during the gate evolution and the resulting infidelity remains comparatively small. As $\tau$ approaches $t_g$, the fluctuations become correlated over a significant fraction of the gate, producing larger run-dependent over- or under-rotations and increased infidelity. In the quasi-static regime ($\tau \gg t_g$), the infidelity saturates because the RTN acts as an approximately fixed voltage offset during each gate realization.

These results support the interpretation that the relevant noise timescale is set primarily by the dimensionless ratio $\tau/t_g$. Reducing the gate duration shifts the fast-noise, intermediate, and quasi-static regimes to proportionally shorter absolute switching times. Finally, we emphasize that the numerical fidelities reported for $t_g = 10~\mu\mathrm{s}$ are specific to the Si MOS, global-ESR, Stark-shift-addressed architecture and operating point considered in this work. Although the simulation framework is general, the quantitative noise tolerance depends on the device geometry, electrostatic boundary conditions, voltage-to-Hamiltonian sensitivities, exchange operating regime, control protocol, and the ratio $\tau/t_g$.

\section{Details of quantum process tomography}
\label{app:qpt_details}

In this appendix we summarize the QPT machinery used in the main text to characterize noisy gates and to compute the coherence angle.

\subsection{Process matrix and Kraus representation}

A general quantum operation acting on density operators is described by a completely positive, trace-preserving (CPTP) map
\begin{equation}
    \mathcal{E}: \rho_{\mathrm{in}} \mapsto \rho_{\mathrm{out}}
    = \mathcal{E}(\rho_{\mathrm{in}}).
\end{equation}
Any such map admits a Kraus decomposition~\cite{Nielsen2010}
\begin{equation}
    \mathcal{E}(\rho) = \sum_i K_i \rho K_i^\dagger,
    \qquad
    \sum_i K_i^\dagger K_i = I,
    \label{eq:Kraus_decomposition_app}
\end{equation}
where $\{K_i\}$ are the Kraus operators.

To obtain a convenient representation, we expand the Kraus operators in a fixed operator basis $\{\tilde{E}_m\}$,
\begin{equation}
\begin{aligned}
    K_i &= \sum_m e_{im} \tilde{E}_m, \\
    \mathcal{E}(\rho) &= \sum_{m,n}
    \chi_{mn}\, \tilde{E}_m \rho \tilde{E}_n^\dagger ,
\end{aligned}
\label{eq:chi_matrix_app}
\end{equation}
which defines the process matrix $\chi$, a positive semidefinite Hermitian matrix that fully characterizes the channel. Throughout the QPT procedure, the process matrix is constructed using the operator basis $\{I, X, -iY, Z\}^{\otimes n}$, which provides a convenient representation of the quantum channel. For visualization and for following conversion to Pauli-transfer-matrix and Kraus representations, the reconstructed process matrices are explicitly transformed to the conventional Pauli basis $\{I, X, Y, Z\}^{\otimes n}$ (All process matrices shown in the main text are expressed in this standard Pauli basis).


In standard QPT one prepares a tomographically complete set of $d^2$ linearly independent input operators, applies the unknown process $\mathcal{E}$ to each of them, and reconstructs $\chi$ from tomographically complete measurements on the outputs. In our two-qubit simulations ($d=4$), we follow this procedure in operator form: we generate 16 input operators
\begin{equation}
    \rho_j = T_a \ket{00}\!\bra{00}\, T_b ,
\end{equation}
with
\begin{equation}
    T_1 = I \otimes I,\quad
    T_2 = I \otimes X,\quad
    T_3 = X \otimes I,\quad
    T_4 = X \otimes X ,
\end{equation}
and let the composite index $j$ label the 16 distinct pairs $(T_a,T_b)$. This construction yields a simple, linearly independent operator basis for the two-qubit space.

Each input operator $\rho_j$ is propagated through the noisy gate, producing
\begin{equation}
    \rho_j' = \mathcal{E}(\rho_j)
            = \sum_k \lambda_{jk}\, \rho_k ,
\end{equation}
where the coefficients $\lambda_{jk}$ specify the action of the process in this operator basis. On the other hand, the basis operators act as
\begin{equation}
    \tilde{E}_m \rho_j \tilde{E}_n^\dagger
    = \sum_k \beta^{mn}_{jk}\, \rho_k ,
\end{equation}
relating $\lambda$ and the process matrix via
\begin{equation}
    \sum_{m,n} \beta^{mn}_{jk}\, \chi_{mn}
    = \lambda_{jk}.
\end{equation}
By collecting indices, we view $\beta$ as a $d^4 \times d^4$ matrix and reshape both $\chi$ and $\lambda$ into vectors, which yields the linear system
\begin{equation}
    \beta\, \chi = \lambda ,
\end{equation}
solved numerically to obtain $\chi$ for each noisy gate realization.

Once $\chi$ is known, we also make use of the equivalent Kraus representation. After fixing a vectorization convention $|\!|A\rangle\!\rangle$ for operators $A$, the Choi matrix is
\begin{equation}
    \Lambda
    =
    \sum_{m,n} \chi_{mn}\,
    |\!|\tilde E_m\rangle\!\rangle
    \langle\!\langle \tilde E_n|\!| ,
\end{equation}
which is positive semidefinite for a completely positive map. An eigen-decomposition
\begin{equation}
    \Lambda = \sum_i \lambda_i\,
    |\!|v_i\rangle\!\rangle
    \langle\!\langle v_i|\!|
\end{equation}
then yields the Kraus operators
\begin{equation}
    K_i = \sqrt{\lambda_i}\;
    \mathrm{unvec}\!\left(|\!|v_i\rangle\!\rangle\right),
\end{equation}
where $\mathrm{unvec}$ denotes the inverse of the chosen vectorization map. In our analysis we order the Kraus operators by decreasing eigenvalue weight and identify the \emph{dominant} Kraus operator $K_{\mathrm{dom}}$ as the one associated with the largest $\lambda_i$; after moving to the gate frame via $K'_{\mathrm{dom}} = U^\dagger K_{\mathrm{dom}}$, this operator is used in Sec.~\ref{sec:coherent_extraction} to define an effective error Hamiltonian and corresponding unitary correction.

\subsection{Error process matrix}

To separate the intended unitary evolution from residual noise, we construct an \emph{error process matrix} $\chi^{\mathrm{err}}$ by factoring out the target unitary $U$ from the reconstructed channel~\cite{Korotkov2013ErrorMatrices}. In the Pauli basis this is defined via
\begin{equation}
    \rho_{\mathrm{out}} =
    \sum_{m,n} \chi^{\mathrm{err}}_{mn}\,
    \tilde{E}_m\, U\, \rho_{\mathrm{in}}\,
    U^\dagger\, \tilde{E}_n^\dagger ,
    \label{eq:error_process_def_app}
\end{equation}
so that $\chi^{\mathrm{err}}$ captures only the deviation from the ideal gate. In the absence of noise, the error process reduces to the identity channel,
\begin{equation}
    \chi^{\mathrm{err}}_{mn}
    = \delta_{m0}\, \delta_{n0},
    \qquad
    \chi^{\mathrm{err}}_{00} = 1 ,
\end{equation}
and deviations of $\chi^{\mathrm{err}}_{00}$ from unity directly quantify the process infidelity. The diagonal elements of $\chi^{\mathrm{err}}$ in the Pauli basis can be interpreted, in the Pauli-twirling approximation, as effective Pauli error weights~\cite{Korotkov2013ErrorMatrices}.

\subsection{Pauli transfer matrix and coherence angle}

For some analyses it is convenient to work in the Pauli transfer matrix (PTM) representation. Given a trace-preserving channel $\mathcal{E}$ and an orthonormal Pauli basis $\{\tilde{E}_\mu\}$, the PTM elements are defined as
\begin{equation}
    N_{\mu\nu}
    =
    \frac{1}{d}\,\mathrm{Tr}\!\left[\tilde{E}_\mu\,\mathcal{E}(\tilde{E}_\nu)\right].
\end{equation}
Using the reconstructed process matrix $\chi$, the PTM can be written as
\begin{equation}
    N_{\mu\nu}
    = \frac{1}{d}
    \sum_{i,j}
    \chi_{ij}\,
    \mathrm{Tr}\!\left[
        \tilde{E}_\mu\,\tilde{E}_i\,\tilde{E}_\nu\,\tilde{E}_j
    \right].
\end{equation}

we then extract the unital block $N_u = N_{\mu\nu}[1{:},1{:}]$ corresponding to the action of the channel on traceless Pauli operators. Following Ref.~\cite{Iverson2020Coherence}, we quantify the coherent component of the noise through the \emph{coherence angle} $\Theta$, which in the small-error limit is approximated by
\begin{equation}
    \Theta^2
    \approx
    \frac{1}{d^2 - 1}
    \sum_{i \neq j}
    (N_u)_{ij}^2.
    \label{eq:theta_nu_app}
\end{equation}
This expression directly relates $\Theta$ to the squared weight of off-diagonal elements of $N_u$, and therefore to the amount of coherent mixing between different Pauli components induced by the channel. In the main text we use $\Theta$ (normalized to its ideal value) as a compact scalar diagnostic of the coherent part of the gate error as a function of RTN parameters.


\section{Analytical Derivation of Fidelity Under Voltage Miscalibration}
\label{app:theory}

This appendix derives the analytical fidelity expressions used to interpret
the miscalibration trends in Fig.~\ref{fig:combined_miscal}. We treat square
control pulses and consider separately the Stark-shift--driven $RX(\pi/2)$ gate
and the exchange-driven $\sqrt{\mathrm{SWAP}}$ gate. In both cases we start from
the miscalibrated waveform model
\begin{equation}
    V'(t) - V_{\mathrm{idle}} = a \bigl(V(t) - V_{\mathrm{idle}}\bigr) + b,
\end{equation}
where $V_{\mathrm{idle}}$ is the idle operating point defined in
Sec.~\ref{sec:single_gate} and $a$, $b$ are a dimensionless scaling factor and
voltage offset, respectively. Throughout, $a$ and $b$ are treated as random
variables drawn from Gaussian distributions with variances $\sigma_a^2$ and
$\sigma_b^2$.

\subsection{$RX(\pi/2)$ gate}
\label{app:rx}

We first consider an $RX(\pi/2)$ rotation on the resonant qubit. For the two
plunger gates $V_1,V_2$ and the tunnel barrier $W_1$, miscalibration gives
\begin{align}
    V_i' - V_{\mathrm{idle}} &= a\,(V_i - V_{\mathrm{idle}}) + b, \quad i=1,2, \nonumber\\
    W_1' - W_{\mathrm{idle}} &= a\,(W_1 - W_{\mathrm{idle}}) + b.
\end{align}
The local $g$-factor deviation transforms as
\begin{equation}
    \delta g' = a\,\delta g + b(c_1 + c_2 + c_3),
    \label{eq:dg_prime_main_app}
\end{equation}
where the ideal deviation relative to $V_{\mathrm{idle}}$ is
\[
    \delta g
    = c_1(V_1 - V_{\mathrm{idle}})
    + c_2(V_2 - V_{\mathrm{idle}})
    + c_3(W_1 - W_{\mathrm{idle}}),
\]
and $c_i$ are the linear Stark coefficients.

\subsubsection*{Resonant qubit}

For a square pulse of duration $T$, the resonant-qubit Hamiltonian can be written as
\begin{equation}
    H = \frac{\mu_B B_z}{2\hbar}\,\delta g \, Z + \Omega_x X + \Omega_y Y
      = \frac{1}{2T}\,\theta\, \mathbf{n}\!\cdot\!\boldsymbol{\sigma},
\end{equation}
and, under miscalibration,
\begin{equation}
    H' = \frac{\mu_B B_z}{2\hbar}\,\delta g' \, Z + \Omega_x X + \Omega_y Y
       = \frac{1}{2T}\,\theta' \,\mathbf{n}'\!\cdot\!\boldsymbol{\sigma}.
\end{equation}
Defining
\begin{equation}
    \frac{\mu_B B_z}{\hbar}\,\delta g = \frac{\Gamma}{T},
\end{equation}
the ideal unitary is
\begin{equation}
    U_{\rm res}(\theta,\mathbf n)
    = \cos\!\left(\frac{\theta}{2}\right)I
      - i\sin\!\left(\frac{\theta}{2}\right)(\mathbf n\!\cdot\!\boldsymbol{\sigma}),
\end{equation}
with target values $\Gamma = 0$, $\theta = \pi/2$, and
$\mathbf n = (1,0,0)$ for an ideal $RX(\pi/2)$ gate.

With miscalibration,
\begin{align}
    \Gamma' &= a\Gamma
    + T\frac{\mu_B B_z}{\hbar}\,b(c_1 + c_2 + c_3), \\
    \theta' &= \sqrt{\Gamma'^2 + \theta^2}, \\
    \mathbf n' &= \left(\frac{\theta}{\theta'},\, 0,\, \frac{\Gamma'}{\theta'}\right).
\end{align}
The overlap fidelity between the ideal and miscalibrated single-qubit unitaries
$\mathcal{F} = \tfrac{1}{2}\left|\mathrm{tr}\,U_{\mathrm{ideal}}^\dagger
U_{\mathrm{miscalibrated}}\right|$ is then
\begin{equation}
    \mathcal{F}_{\rm res}
    = \cos\!\Big(\tfrac{\theta}{2}\Big)\cos\!\Big(\tfrac{\theta'}{2}\Big)
      + \sin\!\Big(\tfrac{\theta}{2}\Big)\sin\!\Big(\tfrac{\theta'}{2}\Big)\frac{\theta}{\theta'},
\end{equation}
where we used $\mathbf n\!\cdot\!\mathbf n' = \theta/\theta'$.

\subsubsection*{Non-resonant qubit}

The non-resonant qubit experiences the same global ESR drive but is detuned so that
in the ideal case it undergoes a net $2\pi$ rotation. Let $\theta_n = 2\pi$ be
the ideal rotation angle. For the chosen drive amplitude $\theta$ on the resonant
qubit, the corresponding $z$-component of the non-resonant qubit's effective rotation
vector is
\[
    \Gamma_n = \sqrt{\theta_n^2 - \theta^2} = \sqrt{4\pi^2 - \theta^2}.
\]
Under miscalibration, we write
\[
    \Gamma_n' = a \Gamma_n + \Delta_z,
\]
with
\[
    \Delta_z
    = T\,\frac{\mu_B B_z}{\hbar}\,b(c_1 + c_2 + c_3),
\]
so the miscalibrated non-resonant rotation angle is
\begin{equation}
    \theta_n' = \sqrt{\big(a\sqrt{4\pi^2 - \theta^2} + \Delta_z\big)^2 + \theta^2}.
\end{equation}
The ideal rotation has $\theta_n = 2\pi$, so
$\cos(\theta_n/2) = -1$ and $\sin(\theta_n/2) = 0$. The corresponding
single-qubit overlap fidelity simplifies to
\begin{equation}
    \mathcal{F}_{\rm nonres}
    = -\cos\!\Big(\tfrac{\theta_n'}{2}\Big)
    = \cos\!\Big(\tfrac{\theta_n' - 2\pi}{2}\Big).
\end{equation}

\subsubsection*{Two-qubit fidelity}

Because the total operation factorizes as
$U_{\rm total} = U_{\rm res} \otimes U_{\rm nonres}$, the two-qubit overlap
fidelity for $\theta = \pi/2$ can be written in closed form as
\begin{equation}
    \mathcal{F}_{\mathrm{rot}}
    = -\frac{\sqrt{2}}{2}\cos\!\Big(\tfrac{\theta_n'}{2}\Big)
      \left[
        \cos\!\Big(\tfrac{\theta'}{2}\Big)
        + \sin\!\Big(\tfrac{\theta'}{2}\Big)\frac{\theta}{\theta'}
      \right],
\end{equation}
which is then converted to the average gate fidelity $F$ using
Eq.~\eqref{eq:fidelity},
\begin{equation}
    F_{\mathrm{rot}}
    = \frac{\mathcal{F}_{\mathrm{rot}}^2\,d + 1}{d+1},
\end{equation}
with $d=4$ for two qubits.

\subsection{$\sqrt{\mathrm{SWAP}}$ gate}
\label{app:rswap}

We now consider exchange-driven entangling gates implemented via the isotropic
exchange interaction. The exchange coupling is modeled as
\begin{equation}
    J(V_1,V_2,W_1)
    =
    C\,e^{\gamma_1(V_1-V_{\mathrm{idle}})}
     e^{\gamma_2(V_2-V_{\mathrm{idle}})}
     e^{\gamma_3(W_1-W_{\mathrm{idle}})},
     \label{eq:exchange_coupling_fitting}
\end{equation}
which depends exponentially on the control voltages. Under miscalibration, all
three voltages transform according to the same $(a,b)$ pair, so
\begin{equation}
    J'
    = J^{\,a}\,\exp\!\big[b(\gamma_1+\gamma_2+\gamma_3)\big].
\end{equation}

For a gate of duration $T$, the corresponding exchange-induced phase is
\begin{align}
    \phi  &= \frac{TJ}{2\hbar}, \\
    \phi' &= \frac{TJ'}{2\hbar}
          = \phi\,J^{\,a-1}\exp\!\big[b(\gamma_1+\gamma_2+\gamma_3)\big].
\end{align}
Using the identity
\[
    \mathrm{SWAP} = \tfrac{1}{2}(II + XX + YY + ZZ),
\]
the ideal and realized unitaries can be written as
\begin{equation}
\begin{aligned}
    U_{\rm ideal}
    &= e^{i\phi/2}\Big(II\cos\phi - i\,\mathrm{SWAP}\,\sin\phi\Big), \\
    U_{\rm real}
    &= e^{i\phi'/2}\Big(II\cos\phi' - i\,\mathrm{SWAP}\,\sin\phi'\Big).
\end{aligned}
\end{equation}
The two-qubit overlap fidelity
$\mathcal{F}_{\sqrt{\mathrm{SWAP}}}
 = \tfrac{1}{4}\left|\mathrm{tr}\,U_{\mathrm{ideal}}^\dagger U_{\mathrm{real}}\right|$
evaluates to
\begin{multline}
    \mathcal{F}_{\sqrt{\mathrm{SWAP}}}
    = \Big|
        \cos\phi\,\cos\phi'
        + \sin\phi\,\sin\phi' 
        \\
        + \frac{i}{2}\big(\sin\phi\,\cos\phi' - \cos\phi\,\sin\phi'\big)
      \Big|.
\end{multline}
For small phase error $\delta\phi = \phi' - \phi$, the imaginary contribution
$\tfrac{1}{2}\sin(\phi - \phi') = -\tfrac{1}{2}\sin\delta\phi$ is higher order in
$\delta\phi$ and can be neglected to leading order, giving the approximation
\begin{equation}
    \mathcal{F}_{\sqrt{\mathrm{SWAP}}}
    \approx \cos(\phi - \phi').
\end{equation}
This compact form makes explicit the strong susceptibility of the
$\sqrt{\mathrm{SWAP}}$ gate to scaling miscalibration, since $\phi'$ inherits the
full exponential sensitivity of the underlying coupling $J(V)$.


\section{Details of the VQE Framework}
\label{appendix:vqe}

\subsection{Electronic Hamiltonian}

The molecular Hamiltonian for a system of $K$ nuclei and $N$ electrons is
\begin{align}
H = & -\sum_i \left( \frac{\hbar^2}{2m_e} \nabla_i^2 \right)
      - \sum_I \left( \frac{\hbar^2}{2M_I} \nabla_I^2 \right) \nonumber \\
    & - \sum_{i,I} \left( \frac{e^2}{4\pi \epsilon_0} \frac{Z_I}{|r_i - R_I|} \right)
      + \frac{1}{2} \sum_{i \neq j} \left( \frac{e^2}{4\pi \epsilon_0} \frac{1}{|r_i - r_j|} \right) \nonumber \\
    & + \frac{1}{2} \sum_{I \neq J} \left( \frac{e^2}{4\pi \epsilon_0} \frac{Z_I Z_J}{|R_I - R_J|} \right),
\end{align}
where $r_i$ and $R_I$ denote electronic and nuclear positions, and $Z_I$ and $M_I$ are the nuclear charges and masses.

Within the Born--Oppenheimer approximation~\cite{Brandas2021}, the nuclei are treated as fixed classical point charges and the nuclear kinetic energy is neglected. For a given nuclear configuration, the nuclear--nuclear repulsion term is then a constant energy shift. The electronic Hamiltonian can be written schematically as
\begin{equation}
    H_{\mathrm{el}} = T_e + V_{ne} + V_{ee} + E_{\mathrm{nn}},
\end{equation}
where $E_{\mathrm{nn}}$ is the nuclear repulsion energy. In what follows, we work in atomic units
($\hbar = m_e = e = 4\pi \epsilon_0 = 1$), so these constants are absorbed into the one- and two-electron integrals defined below.

\subsection{Second Quantization}

In second-quantized form, the electronic Hamiltonian is expressed as
\begin{equation}
    H = \sum_{p,q} h_{pq}\, a_p^\dagger a_q
      + \frac{1}{2} \sum_{p,q,r,s} h_{pqrs}\, a_p^\dagger a_q^\dagger a_r a_s,
\end{equation}
where $a_p^\dagger$ and $a_p$ are fermionic creation and annihilation operators acting on spin-orbitals $\{\phi_p(x)\}$, with $x = (r,\sigma)$ denoting the combined spatial and spin coordinate. In this work, the spin-orbitals are taken to be the Hartree--Fock molecular orbitals obtained from a restricted Hartree--Fock calculation in the STO-3G basis. The one- and two-electron integrals are
\begin{equation}
    h_{pq}
    = \int dx\, \phi_p^*(x)
      \left(
        -\frac{\nabla^2}{2} - \sum_I \frac{Z_I}{|r - R_I|}
      \right) \phi_q(x),
\end{equation}
\begin{equation}
    h_{pqrs}
    = \int dx_1\,dx_2\,
      \frac{\phi_p^*(x_1)\,\phi_q^*(x_2)\,\phi_r(x_2)\,\phi_s(x_1)}{|r_1 - r_2|}.
\end{equation}

\subsection{Mapping to Qubits}

The fermionic Hamiltonian can be mapped to a qubit Hamiltonian using transformations such as the Jordan--Wigner (JW) mapping~\cite{Jordan1928}, yielding
\begin{equation}
    H = \sum_k h_k P_k,
\end{equation}
where the $P_k$ are tensor products of Pauli operators acting on qubits (Pauli strings), and the coefficients $h_k$ depend on the molecular geometry and basis set.

\subsection{UCC Ansatz and \texorpdfstring{$\mathrm{H}_2$}{H2} Specialization}

A widely used chemically motivated ansatz is the unitary coupled-cluster (UCC) ansatz. Starting from a reference state $\ket{\psi_0}$ (typically the Hartree--Fock state), the UCC wavefunction is
\begin{equation}
    \ket{\psi(\theta)} = e^{T - T^\dagger} \ket{\psi_0},
\end{equation}
where $T$ is a truncated cluster operator. For a UCCSD (singles and doubles) truncation,
\begin{align}
    \hat{T}_1 &= \sum_{i \in \mathrm{occ}} \sum_{a \in \mathrm{virt}}
                 t_{ai}\, a_a^\dagger a_i, \\
    \hat{T}_2 &= \sum_{i,j \in \mathrm{occ}} \sum_{a,b \in \mathrm{virt}}
                 t_{abij}\, a_a^\dagger a_b^\dagger a_i a_j.
\end{align}
Here ``occ'' and ``virt'' denote occupied and virtual spin-orbitals in the reference configuration, and the $t_{ai}$ and $t_{abij}$ are variational amplitudes. The unitary form $e^{T-T^\dagger}$ ensures a norm-preserving evolution and is particularly well suited for implementation on quantum hardware.

For the $\mathrm{H}_2$ molecule in a minimal STO-3G basis, there are only two occupied and two virtual spin-orbitals, and the UCC ansatz reduces to a single double-excitation parameter. Using the Hartree--Fock reference
\[
    \ket{\psi_0} = \ket{0011},
\]
the corresponding unitary can be written in second quantization as
\begin{equation}
    U = \exp\!\left[\theta_{\mathrm{amp}}
    \left( a_2^\dagger a_3^\dagger a_1 a_0
          - a_0^\dagger a_1^\dagger a_3 a_2 \right)\right],
    \label{eq:ucc-second-quantized-app}
\end{equation}
where $\theta_{\mathrm{amp}}$ is the (single) cluster amplitude.

Applying the JW transformation to the creation and annihilation operators yields a sum of eight commuting four-qubit Pauli strings,
\begin{align}
    U
    &= \exp\!\bigg(
    \frac{i\theta_{\mathrm{amp}}}{8}\sum_{k=1}^{8} s_k P_k
    \bigg),
    \label{eq:ucc-jw-app}
\end{align}
where each $P_k$ is a product of $X$ and $Y$ operators (e.g., $X_3 Y_2 X_1 X_0$, $Y_3 X_2 X_1 X_0$, etc.) and $s_k = \pm 1$. All $P_k$ mutually commute, so
\[
    \exp\!\left(\sum_k c_k P_k\right)
    = \prod_k \exp(c_k P_k).
\]

When these operators act on the two-dimensional subspace spanned by $\{\ket{0011}, \ket{1100}\}$, the Jordan--Wigner $Z$ strings reduce to overall $\pm 1$ phases, and each $P_k$ becomes proportional (up to a sign) to the same effective operator $X_3 X_2 X_1 Y_0$~\cite{Hempel2018QuantumChemistry}. Consequently, the combined action of all eight commuting exponentials is equivalent to a single rotation
\begin{equation}
    U = \exp\!\left( -i\, \theta_{\mathrm{amp}}\, X_3 X_2 X_1 Y_0 \right),
    \label{eq:ucc-final-app}
\end{equation}
on this subspace.

This form leads directly to a standard four-qubit implementation: one encodes the parity of the four computational qubits into a single qubit using a short sequence of CNOT gates, applies a single-qubit $Z$-rotation $R_Z(2\theta_{\mathrm{amp}})$ on the parity qubit in an appropriate basis, and then uncomputes the parity~\cite{Nielsen2010}. In the main text, this construction is adapted to the single- and two-qubit gates natively available in our spin-qubit architecture, yielding the circuit shown in Fig.~\ref{fig:ucc_circuit}. The explicit Pauli strings and coefficients of the $\mathrm{H}_2$ qubit Hamiltonian $H = \sum_k h_k P_k$ used in these simulations are listed in Appendix~\ref{appendix:vqe_coeff}.

\section{Coefficients for \( R = 1.4 \, \text{\AA} \)}
\label{appendix:vqe_coeff}

For completeness, we list here the Pauli coefficients entering the $\mathrm{H}_2$ Hamiltonian of Eq.~\ref{eq:Hamiltonian} at an internuclear distance of $R = 1.4 \,\text{\AA}$ in the STO-3G basis. These
were obtained using the open-source quantum chemistry package Psi4~\cite{Smith2020PSI4}.

\begin{table}[H]
\centering
\begin{tabular}{|c|c|}
\hline
\textbf{Operator} & \textbf{Coefficient} \\
\hline
$I$                    & $-0.473800316$ \\
$Z_0$                  & $0.100535574$  \\
$Z_1$                  & $0.100535574$  \\
$Z_2$                  & $-0.049032364$ \\
$Z_3$                  & $-0.049032364$ \\
$Z_0 Z_1$              & $0.141204681$  \\
$Z_0 Z_2$              & $0.086787499$  \\
$Z_0 Z_3$              & $0.142543021$  \\
$Z_1 Z_2$              & $0.142543021$  \\
$Z_1 Z_3$              & $0.086787499$  \\
$Z_2 Z_3$              & $0.148911897$  \\
$X_0 X_1 Y_2 Y_3$      & $-0.055755522$ \\
$X_0 Y_1 Y_2 X_3$      & $0.055755522$  \\
$Y_0 X_1 X_2 Y_3$      & $0.055755522$  \\
$Y_0 Y_1 X_2 X_3$      & $-0.055755522$ \\
\hline
\end{tabular}
\caption{Qubit-Hamiltonian coefficients for $\mathrm{H}_2$ at $R = 1.4 \, \text{\AA}$,
computed using Psi4 and the STO-3G basis.}
\end{table}

\section{Effect of Correlated Gate-Voltage Noise}
\label{app:correlated_noise}

In the main text, RTN-induced charge noise is modeled as statistically independent voltage fluctuations on each control electrode. Here we estimate how correlated gate-voltage noise would modify the effective exchange-noise spectrum.

We focus on the exchange interaction $J$, since it is the most voltage-sensitive Hamiltonian parameter in our model. Linearizing the fitted exchange dependence around the operating point gives
\begin{equation}
    \frac{\delta J(t)}{J}
    =
    \gamma_1 \delta V_1(t)
    + \gamma_2 \delta V_2(t)
    + \gamma_3 \delta W_1(t),
    \label{eq:correlated_noise_linearized_J}
\end{equation}
where $\gamma_i$ are the sensitivities of $J$ to the corresponding gate voltages. Defining
\[
    \boldsymbol{\gamma}
    =
    \begin{pmatrix}
    \gamma_1 & \gamma_2 & \gamma_3
    \end{pmatrix},
    \qquad
    \delta \mathbf{V}(t)
    =
    \begin{pmatrix}
    \delta V_1(t) & \delta V_2(t) & \delta W_1(t)
    \end{pmatrix}^{T},
\]
the exchange-noise power spectral density can be written as~\cite{Cheng2025CorrelatedNoise}
\begin{equation}
    S_J(f)
    =
    J^2\,\boldsymbol{\gamma}\,K_V(f)\,\boldsymbol{\gamma}^{T},
    \label{eq:correlated_noise_matrix}
\end{equation}
where $K_V(f)$ is the voltage-noise PSD matrix. This form makes explicit that correlations between electrodes enter through the off-diagonal elements of $K_V(f)$.

For statistically independent RTN voltage fluctuations with electrode-dependent amplitudes $\eta_i$, the voltage-noise PSD matrix is diagonal:
\[
    K_V^{\mathrm{ind}}(f)
    =
    K_x(f)\,
    \mathrm{diag}(\eta_1^2,\eta_2^2,\eta_3^2),
\]
where $K_x(f)$ is the unit-amplitude RTN spectral shape,
\[
K_x(f)
=
\frac{\tau}
     {1+(\pi f\tau)^2},
\]
which is the Debye--Lorentzian spectrum of a random telegraph noise process with characteristic switching time $\tau$; and therefore
\begin{equation}
    S_J^{\mathrm{ind}}(f)
    =
    J^2 K_x(f)
    \left(
    \gamma_1^2\eta_1^2+
    \gamma_2^2\eta_2^2+
    \gamma_3^2\eta_3^2
    \right).
    \label{eq:correlated_noise_independent}
\end{equation}

In the opposite limiting case, a single fluctuator couples simultaneously to all three gates, so that
\[
    \delta \mathbf{V}(t) = \boldsymbol{\eta}\,x(t),
    \qquad
    \boldsymbol{\eta}
    =
    \begin{pmatrix}
    \eta_1 & \eta_2 & \eta_3
    \end{pmatrix}^{T},
\]
where the same RTN source $x(t)$ couples to all three control electrodes with coupling strengths $\eta_i$. The voltage-noise PSD matrix is then rank one,
\[
    K_V^{\mathrm{corr}}(f)
    =
    K_x(f)\,\boldsymbol{\eta}\boldsymbol{\eta}^{T},
\]
and the exchange-noise PSD becomes
\begin{equation}
    S_J^{\mathrm{corr}}(f)
    =
    J^2 K_x(f)
    \left(
    \gamma_1\eta_1+
    \gamma_2\eta_2+
    \gamma_3\eta_3
    \right)^2 .
    \label{eq:correlated_noise_correlated}
\end{equation}
Compared with Eq.~\eqref{eq:correlated_noise_independent}, the correlated case contains cross terms proportional to
$2\gamma_i\gamma_j\eta_i\eta_j$. These terms can either enhance or suppress the effective exchange-noise PSD, depending on the relative signs and magnitudes of the voltage sensitivities and fluctuator couplings.

For the operating point considered in this work, the fitted sensitivities are approximately
\[
    \gamma_1 \simeq -10~\mathrm{V}^{-1},
    \qquad
    \gamma_2 \simeq -10~\mathrm{V}^{-1},
    \qquad
    \gamma_3 \simeq 136~\mathrm{V}^{-1}.
\]
Thus the exchange interaction is much more sensitive to the tunnel-gate voltage $W_1$ than to the plunger voltages. As a simple illustrative case, assume equal coupling amplitudes,
$\eta_1=\eta_2=\eta_3=\eta$. For independent noise,
\[
    S_J^{\mathrm{ind}}(f)
    \propto
    \left[2\times(-10)^2+136^2\right]\eta^2K_x(f)
    =
    18696\,\eta^2K_x(f),
\]
whereas for fully correlated noise,
\[
    S_J^{\mathrm{corr}}(f)
    \propto
    (-10-10+136)^2\eta^2K_x(f)
    =
    13456\,\eta^2K_x(f).
\]
In this specific operating regime, the correlated contribution is partially suppressed because the plunger-gate sensitivities have the opposite sign to the tunnel-gate sensitivity. In another device geometry or operating point, the same cross terms could instead enhance the exchange-noise PSD.

The independent-noise approximation used in the main simulations is therefore useful for isolating the effect of local charge fluctuations, but it is not universal. Correlated gate-voltage noise can quantitatively modify the effective exchange-noise spectrum, and should be included in future device-specific noise models when measured voltage-noise covariance data or microscopic trap models are available.

\bibliographystyle{apsrev4-1}

\end{document}